\documentclass[fleqn,usenatbib]{mnras}

\usepackage[T1]{fontenc}

\usepackage{graphicx}	
\usepackage{amsmath}	
\usepackage{amssymb}	
\usepackage{relsize}     
\usepackage[dvipsnames]{xcolor} 
\usepackage[normalem]{ulem} 
\usepackage{wasysym}
\usepackage{newtxtext,newtxmath}
\usepackage{orcidlink}

\DeclareRobustCommand{\VAN}[3]{#2}
\let\VANthebibliography\thebibliography
\def\thebibliography{\DeclareRobustCommand{\VAN}[3]{##3}\VANthebibliography}
\DeclareRobustCommand{\DE}[3]{#2}
\let\DEthebibliography\thebibliography
\def\thebibliography{\DeclareRobustCommand{\DE}[3]{##3}\DEthebibliography}
\DeclareRobustCommand{\DA}[3]{#2}
\let\DAthebibliography\thebibliography
\def\thebibliography{\DeclareRobustCommand{\DA}[3]{##3}\DAthebibliography}

\defcitealias{Vijayan+19}{V19}
\defcitealias{Yates+13}{Y13}

\definecolor{darkgreen}{rgb}{0.0,0.75,0.0}
\definecolor{darkblue}{rgb}{0.0,0.0,0.5}
\definecolor{orange}{rgb}{1.0,0.65,0.0}
\definecolor{grey}{gray}{0.7}

\newcommand{\correc}[1]{\color{black}#1\color{black}{}} 

\newcommand{\eg}[0]{$\textnormal{e.g. }$}
\newcommand{\ie}[0]{$\textnormal{i.e. }$}
\newcommand{\tn}[1]{\textnormal{#1}}
\newcommand{\sub}[1]{_{\textnormal{#1}}}

\newcommand{\logMm}[0]{\tn{log}(M_{*}/\Msun)}

\newcommand{\lgal}[0]{\textsc{L-Galaxies}}
\newcommand{\lgaltt}[0]{\textsc{L-Galaxies 2020}}
\newcommand{\binaryc}[0]{\textsc{binary\_c}}

\newcommand{\Msun}[0]{\,\textnormal{M}_{\textnormal{\astrosun}}}

\newcommand{\Zsun}[0]{\,\textnormal{Z}_{\textnormal{\astrosun}}}

\newcommand{\mumH}{\mbox{$\mu m_\mathrm{H}$}}

\title[Impact of binary stars on dust and metals]{The impact of binary stars on the dust and metal evolution of galaxies}

\author[Robert M. Yates et al.]{Robert M.~Yates$^{\orcidlink{0000-0001-9320-4958}\,1,2}$\thanks{E-mail: \href{mailto:r.yates3@herts.ac.uk}{r.yates3@herts.ac.uk}},
David Hendriks$^{\orcidlink{0000-0002-8717-6046}\,2}$,
Aswin P. Vijayan$^{\orcidlink{0000-0002-1905-4194}\,3,4}$,
Robert G. Izzard$^{\orcidlink{0000-0003-0378-4843}\,2}$,
\newauthor
Peter A. Thomas$^{\orcidlink{0000-0001-6888-6483}\,5}$,
and Payel Das$^{\orcidlink{0000-0002-3239-4222}\,2}$
\\
$^{1}$Centre for Astrophysics Research, University of Hertfordshire, Hatfield, AL10 9AB, UK\\
$^{2}$Department of Physics, University of Surrey, Stag Hill, Guildford, GU2 7XH, UK\\
$^{3}$Cosmic Dawn Center (DAWN)\\ 
$^{4}$DTU-Space, Technical University of Denmark, Elektrovej 327, DK-2800 Kgs. Lyngby, Denmark\\
$^{5}$Astronomy Centre, University of Sussex, Falmer, Brighton BN1 9QH, UK\\
}

\date{Accepted XXX. Received YYY; in original form ZZZ}

\pubyear{2023}

\begin{document}
\label{firstpage}
\pagerange{\pageref{firstpage}--\pageref{lastpage}}
\maketitle

\begin{abstract}
We present detailed implementations of (a) binary stellar evolution (using \binaryc{}) and (b) dust production and destruction into the cosmological semi-analytic galaxy evolution simulation, \lgal{}. This new version of \lgal{} is compared to a version assuming only single stars and to global and spatially-resolved observational data across a range of redshifts ($z$). We find that binaries have a negligible impact on the stellar masses, gas masses, and star formation rates of galaxies \correc{\textit{if}} the total mass ejected by massive stars is unchanged. This is because massive stars determine the strength of supernova (SN) feedback, which in turn regulates galaxy growth. \correc{Binary effects, such as common envelope ejection and novae, affect carbon and nitrogen enrichment in galaxies, however heavier alpha elements are more affected by the choice of SN and wind yields.} Unlike many other simulations, the new \lgal{} reproduces observed dust-to-metal (DTM) and dust-to-gas (DTG) ratios at $z\sim{}0-4$. This is mainly due to shorter dust accretion timescales in dust-rich environments. However, dust masses are under-predicted at $z\gtrsim{}4$, highlighting the need for enhanced dust production at early times in simulations, possibly accompanied by increased star formation. On sub-galactic scales, there is very good agreement between \lgal{} and observed dust and metal radial profiles at $z=0$. A drop in DTM ratio is also found in diffuse, low-metallicity regions, contradicting the assumption of a universal value. We hope that this work serves as a useful template for binary stellar evolution implementations in other cosmological simulations in future.
\end{abstract}



\color{black}
\section{Introduction} \label{sec:Intro}
Stellar evolution plays a critical role in overall galaxy evolution. In particular, it determines the timing, energetics, and chemical composition of the material ejected by stellar winds and supernovae (SNe). This material drives the stellar feedback and chemical enrichment of galaxies, and consequently many other key evolutionary processes, such as gas cooling and star formation. It is therefore crucial for galaxy evolution simulations to take proper account of stellar evolution in their `sub-grid modelling'.

However, a major aspect of stellar evolution that is rarely considered in cosmological-scale simulations is stellar multiplicity. Observations now suggest that over 50 per cent of massive stars (\ie{}those above $\sim{}8\Msun$) evolve in binary or higher-multiple systems (\eg{}\citealt{Sana+12}). This binary stellar evolution (BSE) can have a significant impact on the lifetime, nucleosynthesis, and end state of these stars, due to the complex interactions with their lower-mass counterparts (\eg{}\citealt{Podsiadlowski+92,DeMarco&Izzard17,Moe&DiStefano17}). Given that these massive stars are also the major contributors to ionising photons, stellar feedback, and chemical enrichment in galaxies (\eg{}\citealt{Nomoto+13,Goetberg+19,Kobayashi+20}) it is important to assess how their binarity impacts galaxy evolution as a whole.

Although BSE has been included in some stand-alone galaxy chemical enrichment (GCE) models (\eg{}\citealt{DeDonder&Vanbeveren04,Mennekens&Vanbeveren16,Cote+18}), and individual aspects of massive binaries have been studied on larger scales, such as transient event rates, \correc{compact binary mergers,} and r-process element production \eg{}\citealt{van_de_Voort+15,Schneider+17,Stanway+18,Chruslinska+18,Cote+19,Artale+20,Kobayashi+23,Peron+23}), most cosmological-scale galaxy evolution simulations to date only consider single-star stellar populations in their sub-grid modelling (\eg{}\textsc{Santa Cruz}, \citealt{Arrigoni+10a}; \textsc{L-Galaxies}, \citealt{Yates+13}; \textsc{Illustris}, \citealt{Vogelsberger+13}; \textsc{Gaea}, \citealt{DeLucia+14}; \textsc{Eagle}, \citealt{Schaye+15}; \textsc{Fire-2}, \citealt{Hopkins+18}; \textsc{Sag}, \citealt{Collacchioni+18}; \textsc{Illustris-TNG}, \citealt{Pillepich+18a}; \textsc{Simba}, \citealt{Dave+19}; \textsc{Simba-C}, \citealt{Hough+23}). Consequently, important binary phenomena such as type Ia supernovae (SNe-Ia) are also only treated approximately in these simulations. Typically, an analytic delay-time distribution (DTD) is assumed and the rate of SN-Ia events per stellar population scaled to match the iron abundances observed in early-type galaxies (ETGs) or the solar neighbourhood (\eg{}\citealt{Greggio+05,Matteucci+06,Matteucci+09,Yates+13}). While this approach returns a good match to galaxy-scale observations, it does so essentially by construction. Other binary phenomena, such as common envelope (CE) ejection and novae, are often ignored altogether.

Similarly, dust can play a critical role in galaxy evolution. Observationally, intervening dust attenuates the UV and optical light emitted by galaxies, re-emitting it at longer wavelengths and therefore changing their observed colours or obscuring them completely (for a recent review, see \citealt{Galliano+18}). It can also provide a useful independent tracer of the metal and gas content in galaxies (\eg{}\citealt{Eales+12,Brinchmann+13}). Physically, dust is a source of depletion for refractory elements, thus changing the gas-phase abundances in the interstellar medium (ISM) (\eg{}\citealt{Savage&Sembach96,Jenkins09}) and affecting cooling rates (\eg{}\citealt{Sutherland&Dopita93}).

A number of cosmological-scale galaxy evolution simulations now include dust models (\eg{}\citealt{McKinnon+17,Popping+17,Vijayan+19,Li+19,Hou+19,Triani+20,Graziani+20,Dayal+22,Esmerian&Gnedin22,Parente+22,Parente+23}). However, these simulations produce significantly different predictions for the evolution of the dust-to-metal (DTM) and dust-to-gas (DTG) ratios in galaxies across cosmic time \citep{Popping&Peroux22}, and have therefore been unable to firmly constrain which processes dominate dust production and destruction in the Universe, particularly at high redshift ($z$, \citealt{Mattsson+15,Ferrara+16,Ferrara&Peroux21}).

Therefore, in this work we implement new models for (a) binary stellar populations and (b) dust into the semi-analytic galaxy evolution simulation, \textsc{L-Galaxies 2020} \citep{Henriques+20,Yates+21a}, in order to self-consistently constrain the impact of BSE on the dust and metal evolution of galaxies. In particular, we present global and radially-resolved measurements of gas and stellar masses, chemical abundances, DTM and DTG ratios, SN rates, and cosmic densities, and compare these to the latest multi-redshift observations.

In Section \ref{sec:L-Galaxies}, we describe the \lgal{} base simulation used in this work. Section \ref{sec:Binary stellar population model} presents the \binaryc{} BSE model and its implementation into \lgal{}, while Section \ref{sec:Dust model} presents our new dust production and destruction model. In Section \ref{sec:Results}, we provide results from two versions of \lgal{}, one with BSE included and one without, and make detailed comparisons to observational data. Section \ref{sec:Conclusions} provides our conclusions and future prospects. Throughout, we take a Hubble parameter of $h\equiv{}H_{0}/100=0.673$, log to the base 10, and distances in co-moving units.

\section{Galaxy evolution simulation} \label{sec:L-Galaxies}
\lgal{}\footnote{\url{https://lgalaxiespublicrelease.github.io}} (\eg{}\citealt{Springel+05,Guo+11,Henriques+15,Henriques+20}) is a cosmological-scale, semi-analytic simulation of galaxy evolution, built to run on the merger trees generated from N-body simulations of dark matter halo formation. This work is based on the most recent public release, \lgaltt{} \citep{Henriques+20,Yates+21a}, which we modify and run on both the \textsc{Millennium} ($l\sub{box} = 480.3\, \tn{Mpc}/h$, $m\sub{particle} = 9.61\times10^8\,\Msun/h$, \citealt{Springel+05}) and \textsc{Millennium-II} ($l\sub{box} = 96.1\, \tn{Mpc}/h$, $m\sub{particle} = 7.69\times10^6\,\Msun/h$, \citealt{Boylan-Kolchin+09}) N-body simulations, re-scaled to a \textit{Planck 2013} cosmology \citep{Planck14}: $\Omega_{\Lambda,0}=0.685$, $\Omega\sub{m,0}=0.315$, $\Omega\sub{b,0}=0.0487$, $\sigma_8=0.826$, $n\sub{s}=0.96$, $h=0.673$ (see \citealt{Angulo&Hilbert15}).

Semi-analytic simulations like \lgal{} treat all baryonic processes analytically, in a similar way to sub-grid models in hydrodynamical simulations. These models allow the transfer of mass and energy between `homogeneous spatial components' inside dark matter haloes and galaxies, rather than between computational particles or cells. This method is highly computationally efficient, allowing many millions of galaxies to be modelled from high $z$ to the present day at a computational cost (in terms of CPU time) many thousands of times smaller than for hydrodynamical simulations of comparable size. \lgal{} is able to estimate global properties such as stellar masses and specific star formation rates ($\tn{sSFR} = \tn{SFR}/M_{*}$) to within $\sim{}0.3$ dex of those predicted by fully hydrodynamical simulations run on the same dark matter haloes, for systems where active galactic nucleus (AGN) feedback is not significant \citep{Ayromlou+21a}.

There is a wide range of baryonic physics included in \lgal{}, as described in \citet{Henriques+20} and the \lgaltt{} Model Description\footnote{\url{https://lgalaxiespublicrelease.github.io/Hen20\_doc.pdf}}. Most recently, models for (a) the partition of HI and H$_{2}$ within the ISM \citep{Fu+10}, (b) delayed chemical enrichment from SNe and stellar winds (\citealt{Yates+13}), and (c) radially-resolved ISM and stellar discs \citep{Fu+13} were incorporated. Resolved discs are represented by twelve `radial rings' with outer radii given by $\tn{log}(R_{r}) = \tn{log}(0.44)+r\,\tn{log}(1.5)\ \tn{kpc}/h$, where $r=0-11$ \citep{Fu+13}. \correc{Gas (and dust) is then allowed to flow inwards over time with a velocity given by $v\sub{inflow}=\alpha_{v}R$, where $\alpha_{v}=0.6\tn{ km s}^{-1}\tn{ kpc}^{-1}$.} \lgal{} can therefore resolve the gas and stellar components of galaxies down to sub-kpc scales in their inner $\sim{}1$ kpc, as well as at lower spatial resolution in their outskirts and surrounding subhalo. The following spatial components are considered: central supermassive black hole (SMBH), stellar bulge, radially-resolved stellar disc, radially-resolved cold-gas disc (representing the interstellar medium, ISM), hot-gas reservoir (representing the circumgalactic medium, CGM), stellar halo, and an ejecta reservoir (representing gas ejected out of a galaxy's dark matter subhalo via feedback). 

Star formation histories (SFHs) are also stored for each stellar component \citep{Yates+13,Shamshiri+15}. These comprise 20 bins, structured such that there is higher temporal resolution in the recent past, enabling properties such as rest-frame magnitudes and colours to be optimally calculated on-the-fly within the simulation. These SFHs also allow the timing of chemical element ejection by stars and SNe to be accurately calculated (see Section \ref{sec:Binary stellar population model}).

Finally, following \citet{Yates+21a}, we base our current study on a version of \lgal{} known as the \textit{`modified model'} (hereafter, MM). This version allows the efficient ejection of newly-synthesised metals out of galaxies. 90 per cent of the material ejected by SNe-II and 70 per cent of the material ejected by SNe-Ia is directly deposited into the CGM surrounding galaxies, without first mixing with the ISM. This material can then cool back onto the disc at later times. The remaining ejecta material is allowed to fully mix with the local ISM, before being available for star formation or later removal through entrainment in SN-driven winds. This version of \lgal{} was designed to emulate the metal-rich galactic outflows observed in many real (\eg{}\citealt{Martin+02,Strickland+04,Tumlinson+11}) and simulated (\eg{}\citealt{Emerick+20a,Gutcke+20}) star-forming galaxies, and produces a gradual enrichment of galaxies over cosmic time which is more in line with recent observational inferences from ALMA, Keck, and JWST (\eg{}\citealt{Jones+20,Sanders+21,Schaerer+22,Arellano-Cordova+22,Trump+23,Sanders+23,Curti+23,Brinchmann23}).

\section{Binary stellar evolution (BSE) model} \label{sec:Binary stellar population model}
In this work, we implement the mass- and metallicity-dependent chemical yields, stellar lifetimes, and event rates from synthetic stellar populations generated by the \binaryc{}\footnote{\url{https://binary\_c.gitlab.io/}} code \citep{Izzard+06,Izzard+09,Izzard+18,Izzard&Jermyn22} into the \lgal{} simulation described above. \textsc{Binary\_c} is a BSE framework designed to model the evolution of single and binary stars using semi-analytic methods. Included in this modelling is the calculation of chemical yields from a wide range of ejection processes, including stellar and Wolf-Rayet (WR) winds, Roche-lobe overflow (RLOF), CE ejection, various types of (super)nova explosion, and others. We make use of a recent extension to this framework, \textsc{binary\_c-python}\footnote{\url{https://binary_c.gitlab.io/binary_c-python/}} \citep{Hendriks+23}, to calculate yields and event rates on a ``per population basis'' as a function of time after their formation. Such yields are perfect for implementation into large-scale galaxy evolution simulations, which do not have the mass or time resolution required to resolve individual stars. \correc{\binaryc{} is preferred in this work over other BSE models in the literature (\eg{}SeBa, \citealt{PortegiesZwart+96}; BPASS, \citealt{Eldridge+17}; and MOBSE, \citealt{Giacobbo+18}) because it is specifically designed to rapidly model the impact of (binary) stellar evolution on a very wide range of nucleosynthetic yields on-the-fly.}

Below, we describe the key aspects of \binaryc{} of relevance here. A list of the key \binaryc{} parameters used to generate our stellar populations is also available online in the Supplementary Material.\footnote{\url{https://lgalaxiespublicrelease.github.io/Yates23_doc.pdf}}

\subsection{Binary stellar populations} \label{sec:Binary stellar populations}
In this work, we utilise stellar populations generated by \binaryc{} v2.2.1 using \textsc{binary\_c-python} v0.9.4 for six distinct metallicities ($Z \equiv{} M\sub{Z}/M\sub{b} = 0.0001, 0.001, 0.004, 0.008, 0.01, 0.03$). The standard \binaryc{} set-up is used for this work, which includes a \citet{Kroupa01} initial mass function (IMF) and a maximum star mass of $80\Msun$. The fraction of stars that are in binary systems, as well as the initial orbital period and mass-ratio distribution, are determined following \citet{Moe&DiStefano17}. This mass-dependent prescription predicts that $\sim{}75$ per cent of $8\Msun$ stars are in higher-order (binary, triple, or quadruple) systems, with this fraction reaching $\gtrsim{}90$ per cent at $\gtrsim{}25\Msun$ (see fig. 39 from \citealt{Moe&DiStefano17}).  All multiple systems are treated as binaries here, leaving the higher-order effects caused by triple and quadruple interactions for future work. We also note that in the case of wide binaries, the component stars essentially evolve as single stars even though they are strictly in binary systems. Below, we outline the key prescriptions included in \binaryc{} v2.2.1 which are of most relevance here. The impact of modifying these standard assumptions will be the focus of future work.

\subsubsection{Stellar yields} \label{sec:Stellar yields}
\binaryc{} provides yields for a total of 117 chemical elements (487 isotopes), opening-up a whole new window of investigation into the chemical evolution of galaxies in cosmological-scale simulations. The implementation of \binaryc{} into \lgal{} has been designed to allow the user to freely choose which elements/isotopes are tracked. For the purposes of this work, we only consider the 11 chemical elements commonly used in current cosmological-scale simulations, including previous versions of \lgal{}: H, He, C, N, O, Ne, Mg, Si, S, Ca, and Fe. Follow-up papers in this series will consider additional elements. These 11 elements account for $\sim{}98$\% of the total baryonic mass in the solar photosphere, so also allow for a robust study of the total metallicity in most astrophysical regions.

For AGB winds from both single and binary stars, the stellar yields calculated by \citet[][based on those from \citealt{Karakas+02}]{Izzard+04,Izzard+06} are used. For SNe-Ia, we assume the yields from the DD2 model (\ie{}assuming a delayed detonation at a density of $2.2\times{}10^{7}\,\tn{g cm}^{3}$) from \citet{Iwamoto+99} for Chandrasekhar-mass detonations and from \citet{Livne&Arnett95} for sub-Chandrasekhar-mass detonations. For SNe-II, the metallicity-dependent yields of \citet{Chieffi&Limongi04} are used. For details on the yield sets for less-common transient events, along with information on the full range of ejection processes modelled in \binaryc{}, see the \binaryc{} online documentation.

\begin{figure}
\centering
  \includegraphics[width=0.88\linewidth]{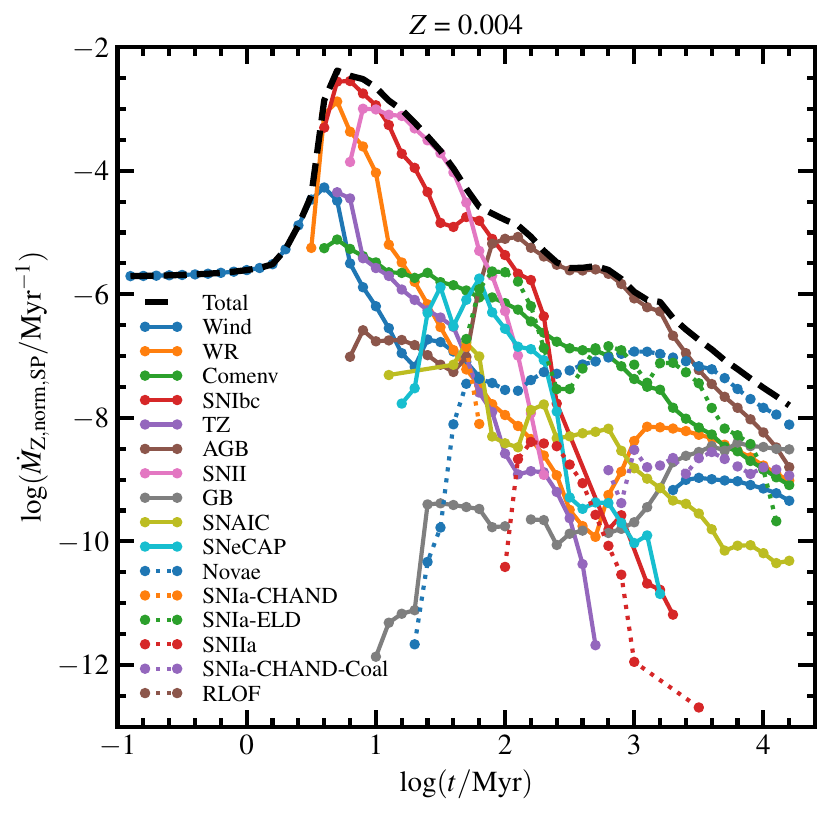}
 \caption{The normalised metal ejection rate ($\dot{M}\sub{Z,norm,SP} = \tn{d}M\sub{Z}/\tn{d}t\,1/\Msun$), for a \binaryc{} stellar population including single and binary stars with an initial metallicity of $Z=0.004$. The 16 different ejection processes modelled in \binaryc{} are shown (see \binaryc{} documentation for details).}
 \label{fig:binaryc_all_yields_z0004}
\end{figure}

Fig. \ref{fig:binaryc_all_yields_z0004} provides an illustration of the normalised metal ejection rate (\ie{}the metal ejection rate per solar mass of stars formed) for a stellar population with birth metallicity $Z = 0.004$, as a function of time after star formation. This is obtained by summing all nine chemical elements heavier than He considered in this work. The full \correc{complement} of ejection processes modelled in \binaryc{} is shown (see legend). Stellar winds dominate enrichment at very early times ($t\lesssim{}6$ Myr), after which prompt ejection events such as WR winds, SNe-Ibc, and SNe-II dominate until $t\sim{}100$ Myr. At later times, AGB winds and finally novae dominate. Enrichment from stellar winds is a strong function of metallicity in \binaryc{}. For example, the normalised metal ejection rate from winds at $t=1$ Myr increases by $\sim{}4.5$ dex from $Z=0.0001$ to $0.03$. This is because metals increase the opacity of the surface layers, thus increasing the radiation pressure on them.

Yield sets such as these represent a significant improvement for GCE models in simulations, which typically just assume single-star yields for SNe-II and AGB winds, along with a simplified prescription for SN-Ia enrichment.

\subsubsection{Common envelope (CE) ejection} \label{sec:Common envelope ejection}
During their evolution, stars in binary systems expand and may overflow their Roche lobe, \ie{}the volume in which material is strictly bound to only that star. This leads to mass transfer which is either stable or unstable. We determine the stability of mass transfer based on the criteria from \citet{Claeys+14}. When mass transfer is unstable, the system undergoes a CE evolution where the transferred material engulfs both stars, and orbital energy and angular momentum is lost through dynamical friction with this envelope. This leads to shrinkage of the orbit, and possibly the merging of the two stars. To determine the final orbital separation we use the $\alpha_{\mathrm{CE}}-\lambda_{\mathrm{CE}}$ prescription of \citet{webbinkDoubleWhiteDwarfs1984} and \citet{Hurley+02} to calculate the energy necessary to expel the envelope, where $\alpha_{\mathrm{CE}}$ is the efficiency with which orbital energy is transferred to the CE and $\lambda_{\mathrm{CE}}$ scales the binding energy of the donor's envelope. We take $\alpha_{\mathrm{CE}}$ = 1, i.e. 100 per cent of the transferred orbital energy is transformed into energy to eject the envelope, and $\lambda_{\mathrm{CE}}=0.5$.

\subsubsection{Remnants of massive stars} \label{sec:Remnants of massive stars}
The \binaryc{} code calculates the remnant mass of a massive star following the prescription of \citet{Hurley+00,Hurley+02}, which depends on the CO core mass at the end of the star's life,
\begin{equation}\label{eqn:remnant_mass}
M\sub{rem} = 1.17 + 0.09M\sub{CO}\ .
\end{equation}
Stars which undergo SNe-II typically have $M\sub{CO} \lesssim{} 35\Msun$ (see \eg{}\citealt{Limongi&Chieffi18}), thus obtaining remnant masses of $M\sub{rem} \lesssim{}4.3\Msun$, meaning that the majority of the star's initial mass is ejected during the SN or in the pre-SN winds. We note that other prescriptions which account for dependencies on metallicity and explosion energy (\eg{}\citealt{Fryer+12,Ertl+16}) tend to predict higher remnant masses, implying lower ejected yields of newly-synthesised material from SNe-II. However, the detailed effects of binary evolution on pre-SN mass loss could counteract these effects to some extent (\eg{}\citealt{Schneider+21,Farmer+23}). For example, \citet{Schneider+21} find that the stripping of a massive star by its binary companion can lead to larger carbon abundances and/or lower $M\sub{CO}$ at core-helium exhaustion, and thus higher explosion energies, lower remnant masses, and greater ejection of heavy elements at core collapse. A detailed investigation into the impact of different remnant and fallback prescriptions on the chemical evolution of galaxies will be the focus of future work.

\subsubsection{SN-Ia progenitors} \label{sec:SN-Ia progenitors}
The evolution of SN-Ia progenitor systems is modelled in \binaryc{} following \citet{Claeys+14}. In this formalism, three SN-Ia pathways are considered; the single-degenerate scenario (SNIa-CHAND), double-degenerate scenario (SNIa-CHAND-Coal), and sub-Chandrasekhar-mass scenario (SNIa-ELD). The relative frequency of the latter two scenarios in \binaryc{} depends on the assumed mass that must be accreted before a white dwarf can ignite as an edge-lit detonation (ELD). Here, this accretion mass is set to $0.15\Msun$.

The double-degenerate scenario is the dominant pathway for late-time (\ie{}$\gtrsim{}200$ Myr) SNe-Ia in the \citet{Claeys+14} formalism, whereas the single-degenerate scenario dominates at earlier times (see their fig. 7). For the canonical CE efficiency of $\alpha{}\sub{ce}=1$, a SN-Ia DTD that roughly traces the power-law shape inferred from observations of SN-Ia rates in local galaxies is obtained (\eg{}\citealt{Maoz+11,Maoz+14}). However, the normalisation of this DTD in \binaryc{} is significantly lower than observed, reflecting the well-known difficulty stellar evolution models have in generating enough SNe-Ia per population to match observed SN-Ia rates and iron abundances (see \eg{}\citealt{Kobayashi+23}). The impact of this discrepancy on the properties of galaxies in \lgal{} is discussed in Section \ref{sec:Results}. Recent attempts to improve the modelling of mass transfer during binary evolution could improve the SN-Ia rate in models like \binaryc{} \citep{Temmink+23}. The impact of such approaches will be investigated using \lgal{} in future work.

\subsection{Implementation into galaxy evolution simulations} \label{sec:Implementation into L-Galaxies}
Traditionally, galaxy evolution simulations utilise stellar yields calculated for individual stars of various initial masses and metallicities. The overall mass ejection rate from a whole stellar population at time $t$ is then calculated using the GCE equation (\eg{}\citealt{Tinsley80,Matteucci+12}), \ie{}the integral over the death rate at time $t$ of stars of mass $M$ multiplied by the mass ejected by each of those stars (see \eg{}section 4 of \citealt{Yates+13}). \correc{However, in this work we utilise the extension to \binaryc{} called \textsc{binary\_c-python}, which generates yields for entire stellar populations. In essence, \textsc{binary\_c-python} calculates the integral of the yield ejected by each star multiplied by the IMF.} Therefore, the GCE equation does not need to be solved again within \lgal{}. Instead, only the synchronisation of these yields with the SFHs of model galaxies (see Section \ref{sec:L-Galaxies}) is required. This process is described below, using the example of a single radial ring within the stellar disc of a model galaxy.

At each simulation timestep, \correc{the upper and lower edges of the $i$th SFH bin ($t\sub{upper,i}$ and $t\sub{lower,i}$) are determined. Then, the normalised yield of element $x$ ejected at time $t$ by stars born in SFH bin $i$ is calculated for each of the six metallicities ($Z$) considered in \binaryc{}, by numerically integrating the \binaryc{} mass ejection rates between these two times,}
\begin{equation}\label{eqn:integral_across_time_bins}
\frac{M\sub{x,i}(Z,t)}{\Msun} = \int^{t\sub{upper,i}}_{t\sub{lower,i}}\left(\frac{\tn{d}M\sub{x}(Z,t)}{\tn{d}t}\frac{1}{\Msun}\right)\,\tn{d}t\ .
\end{equation}
\correc{The true value of $M\sub{x,i}(Z,t)$ is then obtained for radial ring $r$ by interpolating between the six values using the actual metallicity of the stars at their formation in \lgal{}, $Z\sub{r,i}$. This true normalised yield is then multiplied by the SFR in the $i$th SFH bin,} $\psi\sub{r,i}$, to obtain the true mass ejection rate. The mass ejection rates from all SFH bins are then added together to obtain the total mass ejection rate of element $x$ at time $t$ for ring $r$,
\begin{equation}\label{eqn:sum_of_ej_rates}
\dot{M}\sub{x,r}(t) = \mathlarger{\mathlarger{\sum}}_{i}\left(\frac{M\sub{x,i}(Z=Z\sub{r,i},t)}{\Msun}\cdot{}\psi\sub{r,i}\right)\ ,
\end{equation}
and the total mass of element $x$ ejected in that timestep by stars in ring $r$ is then
\begin{equation}\label{eqn:final_mas_ejected}
M\sub{x,r} = \dot{M}\sub{x,r}(t)\,\Delta{}t\ ,
\end{equation}
where $\Delta{}t$ is the simulation timestep width.

We note that carrying-out all these steps for every radial ring of every galaxy within the simulation would be computationally prohibitive. Therefore, we adopt the approach used for the GCE model in previous versions of \lgal{} \citep{Yates+13}, by carrying-out the first part (\ie{}up to Eqn.~\ref{eqn:integral_across_time_bins}) in pre-processing. This takes advantage of the fact that the timestep spacings and SFH bin spacings are identical for all galaxies (and all radial rings) in \lgal{}. Such a uniform time structure allows us to loop through all timesteps just once before the main simulation code is executed, to calculate $M\sub{x,i}(Z,t)/\Msun$ and store it in a look-up table with dimensions of timestep, SFH bin, initial metallicity, and chemical element. This look-up table is then used within \lgal{} to solve Eqns.~\ref{eqn:sum_of_ej_rates} and \ref{eqn:final_mas_ejected}.

\subsection{Comparing single and binary stellar population yields} \label{sec:Comparing single star and binary population yields}
Before assessing the impact of binaries in a complex galaxy evolution simulation, it is informative to first compare the basic ejection rates predicted by \binaryc{} with those from single-star-only stellar populations. To do this, we compare three set-ups: (a) the mixed (\ie{}binary+single star) stellar population from \binaryc{} described in Section \ref{sec:Binary stellar populations}, (b) an equivalent single-star-only stellar population from \binaryc{}, and (c) a single-star-only stellar population including an analytic prescription for SNe-Ia from the previous \lgal{} GCE model, first presented by \citet[][hereafter \citetalias{Yates+13}]{Yates+13}.

This third set-up assumes a \citet{Chabrier03} IMF and includes \citet{Marigo01} yields for AGB stars, \citet{Thielemann+03} yields for SNe-Ia, and \citet{Portinari+98} yields for SNe-II, similar to many other cosmological-scale simulations. Metal enrichment from stars of mass $0.85$ to $120\Msun$ is assumed, along with a power-law SN-Ia delay-time distribution (DTD) of slope -1.12 and a SN-Ia factor of $A\sub{SNIa}=0.035$ (see \citealt{Yates+21a} for details). This $A\sub{SNIa}$ parameter determines the fraction of stellar systems in the mass range $3-16\Msun$ that are SNe-Ia progenitor systems, and essentially corrects for any shortfall in the predicted yield of iron-peak elements by amplifying the expected SN-Ia fraction to compensate. The ejection rates for AGB winds, SNe-II, and SNe-Ia are then calculated using the standard GCE equation (eqn. 5 in \citetalias{Yates+13}). These are comparable to ejection rates from the \binaryc{} set-ups using Eqn. \ref{eqn:sum_of_ej_rates} above.

We note that the different IMF upper-mass limits between the \citetalias{Yates+13} prescription ($120 \Msun$) and that from \binaryc{} ($80 \Msun$) does not have a significant impact on the evolution of model galaxies. When changing this limit from $120 \Msun$ to $80 \Msun$ in the \citetalias{Yates+13} set-up, we find no significant changes to stellar and gas masses of galaxies at $z=0$, and average ISM oxygen abundance only drops by $\sim{}0.04$ dex (due to having slightly fewer core-collapse SNe).

\begin{figure}
 \centering
 \includegraphics[width=0.82\linewidth]{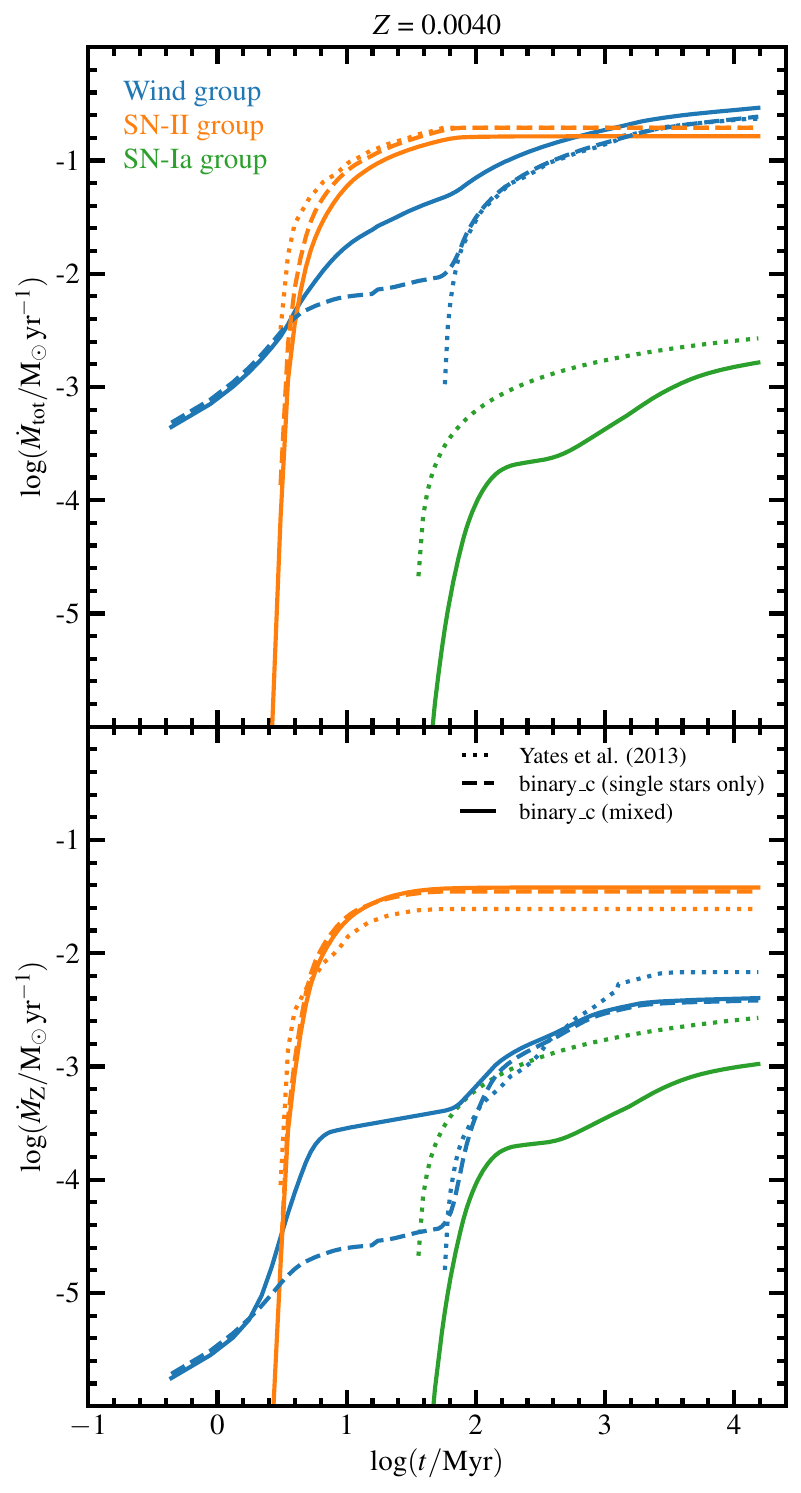}
 \caption{\textit{Top panel:} Total mass ejection rate from stellar winds (blue), SNe-II type enrichment channels (orange), and SNe-Ia + novae (green), assuming a constant SFR of $\psi{}=1\ \Msun/\tn{yr}$ and a fixed metallicity of $Z=0.004$. Dotted lines denote the single-star-only set-up used in previous versions of \lgal{}, dashed lines denote the single-star-only set-up from \binaryc{}, and solid lines denote single+binary stellar populations from \binaryc{}. \textit{Bottom panel:} Same as top panel, but for the metal mass ejection rate, considering C, N, O, Ne, Mg, Si, S, Ca, and Fe.}
 \label{fig:mass_and_SNrate_comp}
\end{figure}

\begin{figure*}
\centering
 \includegraphics[width=0.95\linewidth]{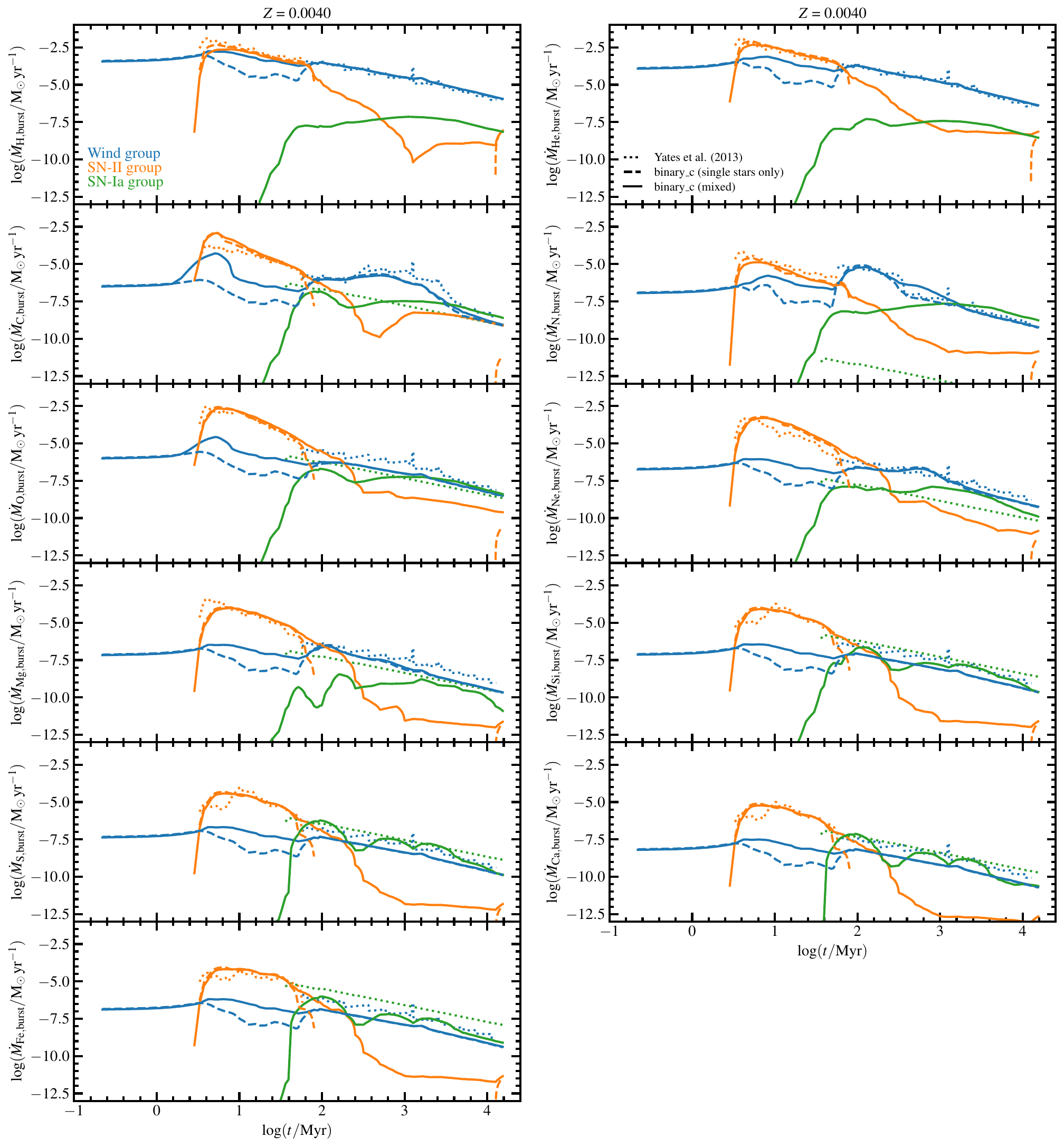}
 \caption{Mass ejection rates from a single $1\Msun/\tn{yr}$ burst of star formation for the 11 chemical elements considered in this work. Dotted lines denote the single-star-only set-up used in previous versions of \lgal{}, dashed lines denote the single-star-only set-up from \binaryc{}, and solid lines denote single+binary stellar populations from \binaryc{}.}
 \label{fig:yields_comp}
\end{figure*}

For the purposes of this comparison, the yields from \binaryc{} have been grouped into the following three broad categories, to roughly match the three enrichment channels previously implemented into \lgal{}:

\begin{itemize}
\item Wind group: AGB, RLOF, Wind, Thorne-Zytkow, GB, CE
\item SN-II group: SNII, SNIIa, SNIbc, WR, SNeCAP, SNAIC
\item SN-Ia group: SNIaELD, SNIaChandCoal, SNIaChand, Novae
\end{itemize}

These set-ups are then run through an arbitrary SFH (here a constant SFR of $1\ \Msun/\tn{yr}$), to obtain the total mass ejection rate as a function of time.

Fig.~\ref{fig:mass_and_SNrate_comp} shows the total mass ejection rates (top panel) and metal mass ejection rates (bottom panel) for the mixed set-up from \binaryc{} (solid lines), the single-star-only set-up from \binaryc{} (dashed lines), and the \citetalias{Yates+13} set-up (dotted lines), at an example metallicity of $Z=0.004$.

We can see that the \textit{total} mass ejected by massive stars (\ie{}from the SN-II group, orange) is quite similar between these three set-ups. Such a result was also found by \citet{DeDonder&Vanbeveren04} when comparing single-star-only vs. binary-only stellar populations. This indicates that the total energy available for SN feedback in \lgal{} will be relatively unchanged when switching from a single-star-only to mixed prescription (see Section \ref{sec:General galaxy properties}). However, the bottom panel of Fig.~\ref{fig:mass_and_SNrate_comp} shows that the \textit{metal} mass ejected by massive stars is slightly greater when using \binaryc{}. This is partly due to the more efficient mass loss that massive stars undergo when in binary systems (comparing the dashed and solid orange lines), but is predominantly due to the different SN-II yield tables used, \ie{}\citealt{Chieffi&Limongi04} in \binaryc{} and \citealt{Portinari+98} in \citetalias{Yates+13}, as discussed in Section \ref{sec:General galaxy properties}.

Fig. \ref{fig:mass_and_SNrate_comp} also shows that CE ejection is an important additional component to the `Wind group' when assuming binary stellar populations, particularly $\sim{}4-64$ Myr after star formation. Such a process is not possible when only considering single stars. It is also clear from Fig. \ref{fig:mass_and_SNrate_comp} that SNe-Ia ejecta is significantly reduced when using \binaryc{}. This is due to the under-estimation of the SN-Ia rate, as discussed in Section \ref{sec:SN-Ia progenitors}, and leads to a deficit in the production of elements such as Fe in \lgal{} (see Section \ref{sec:Radial profiles}).

Ejection from SNe-II is largely independent of metallicity in \binaryc{}. Conversely, ejection rates from the Wind group increase with metallicity, predominantly due to increasing yields for elements heavier than N as opacity increases. There is also a large drop in the early-time ejection rate from SNe-Ia at both very low ($Z=0.0001$) and very high ($Z=0.03$) metallicity in \binaryc{}, predominantly due to a decrease in prompt ($\sim{}100$ Myr) SNe-Ia leading to a drop in the yield of heavy $\alpha$ elements and Fe.

We also note that binaries contribute up to 90 per cent of the elemental ejection from the `SN-II group' in the \binaryc{} mixed populations, mainly via Wolf-Rayet winds and traditional SNe-II. In the single-star-only scenario, most of this material is simply ejected by isolated massive stars instead.

Finally, Fig.~\ref{fig:yields_comp} shows the individual chemical element ejection rates from a single $1\Msun/\tn{yr}$ burst of star formation with $Z=0.004$. There is a reasonable agreement between the AGB yields from \citet[][used by \binaryc{}]{Izzard+04,Izzard+06} and \citet[][used by \citetalias{Yates+13}]{Marigo01} at late times, although the \citet{Marigo01} yields tend to be higher (especially at lower metallicities) and more sensitive to star mass and therefore time. The inclusion of early stellar winds in \binaryc{} extends its `Wind group' ejection to much earlier times, and CE ejection provides a boost to the yields of all elements (particularly carbon) in \binaryc{} when binaries are included. \correc{However, we note that the overall carbon ejection from the \binaryc{} set-ups is $\sim{}38$ per cent lower than in the \citetalias{Yates+13} set-up at low metallicities (\ie{}$Z\leq{}0.001$). This is predominantly caused by lower carbon yields from AGB stars. The minimum core mass and efficiency assumed for the third dredge up in the \citet{Karakas+02} models used in \binaryc{} are low compared to those from \citet{Marigo01} used by \citetalias{Yates+13}. This is because they are not explicitly calibrated to match carbon star luminosity functions (Z. Osborn, priv. comm.). Conversely, nitrogen is boosted by up to $\sim{}76$ per cent in the mixed \binaryc{} model compared to \citetalias{Yates+13} at low metallicities. This is due to enhanced $^{14}$N ejection from CE, novae, AGB stars, and SNe-Ia. However, this is no longer the case once metallicities reach the value of $Z\sim{}0.004$ seen in Fig.~\ref{fig:yields_comp}.}

There is little difference in the ejection rates from SNe-II at this metallicity, with the exception of a dip in the $\alpha$ element ejection rate at $\sim{}5$ Myr in the \citet{Portinari+98} single-star yields (used by \citetalias{Yates+13}) compared to \binaryc{}. This can be seen for the heavier $\alpha$ elements at $Z=0.004$ in Fig.~\ref{fig:yields_comp} but is also present for lighter $\alpha$ elements at lower metallicities. The impact of this is discussed in Section \ref{sec:General galaxy properties}. The late-time `SN-II group' ejection from the \binaryc{} mixed population is predominantly from WR winds, along with more minor contributions from accretion-induced collapse SNe (SNe-AIC) and the SNe-IIa of binary systems (see \citealt{Zapartas+17}).

However, there is a clear difference in the predicted SN-Ia group ejection rates for many elements. For example, N (mostly $^{14}$N) is ejected much more significantly after $\sim{}2$ Gyr in the \binaryc{} yields than those used by \citetalias{Yates+13} due to the inclusion of novae. Conversely, the total amount of Fe ejected by SNe-Ia is lower in \binaryc{} by $\sim{}1.0$ dex at all times, as discussed above. The H and He ejection seen for the `SN-Ia group' in \binaryc{} is predominantly from novae, with an additional contribution of accreted He from SNe-ELD.

Versions of Figs.~\ref{fig:binaryc_all_yields_z0004}, \ref{fig:mass_and_SNrate_comp}, and \ref{fig:yields_comp} for other metallicities considered in \binaryc{} are available online in the Supplementary Material.

\section{Dust model} \label{sec:Dust model}

The dust model implemented here is based on that developed for an earlier version of \lgal{} by \citet[][hereafter \citetalias{Vijayan+19}]{Vijayan+19}. This includes the production of four main dust types (silicates, carbon, iron oxides, and silicon carbides), their distribution across the diffuse and molecular components within the ISM, and their destruction by various mechanisms.

In this work, we extend that model to include the survival and gradual destruction of dust in the hot CGM and Ejecta components surrounding galaxies, as well as the spatial distribution of dust within the radially-resolved ISM. The partitioning of dust into diffuse and molecular gas components in the ISM is now also fully integrated with our HI-H$_{2}$ partitioning model.

We model the transport of dust into the CGM in the same way as for metals; (a) partial ejection of newly-formed dust from stars and SNe in the stellar disc, (b) entrainment in subsequent galactic outflows, and (c) full ejection of newly-formed dust from stars and SNe in the bulge and stellar halo. We find that allowing dust to escape galaxies, particularly via direct enrichment from disc SNe, has an important impact on the dust masses and therefore grain growth rates inside galaxies at high $z$ (see Sections \ref{sec:Dust scaling realtions} \& \ref{sec:Dust rates}). The key dust production and destruction processes modelled in \lgal{} are outlined below.

\subsection{\correc{Dust production and growth}} \label{sec:Dust production}
In this work, dust is assumed to be produced by \correc{three} sources: winds from asymptotic giant branch (AGB) stars, type Ia supernovae (SNe-Ia), and type II supernovae (SNe-II). \correc{In addition, the dust mass can increase through in-situ grain growth in the ISM. These processes are described in detail below.}

\subsubsection{Dust production by AGB winds} \label{sec:Dust production by AGB winds}
For AGB stars, we use the mass- and metallicity-dependent dust yields presented by \citet{Ferrarotti&Gail06}. The total dust production rate is then given by,
\begin{equation}\label{eqn:AGB_dust_ejection_rate}
\dot{M}\sub{d,AGB}(t) = \int^{7\Msun}_{0.85\Msun}M\sub{d,AGB}(M,Z)\,\psi{}(t-\tau\sub{M})\,\phi{}(M)\,\tn{d}M\ ,
\end{equation}
where $M\sub{d,AGB}(M,Z)$ is the dust mass ejected by one AGB star of mass $M$ and birth metallicity $Z$, $\psi{}(t-\tau\sub{M})$ is the star formation rate (SFR) at the birth of a star with lifetime $\tau\sub{M}$, and $\phi{}(M)$ is the IMF which is assumed to be constant across all time and space. The mass- and metallicity-dependent lifetimes from \citet{Portinari+98} are used for $\tau\sub{M}$ here. We note that dust is assumed to be formed at the end of the AGB star's life, and that we ensure the dust mass produced does not exceed the mass of the constituent chemical elements initially ejected by the AGB stars.

\subsubsection{Dust production by SNe} \label{sec:Dust production by SNe}
For dust production by SNe-Ia and SNe-II, we follow the method recommended by \citet{Zhukovska+08}, whereby the mass of dust formed is assumed to be proportional to the mass of the ``key element'' ejected by the SNe (Si for silicates and silicon carbides, C for carbon dust, and Fe for iron dust). The total dust production rate from SNe-Ia + SNe-II is therefore given by,
\begin{equation}\label{eqn:SN_dust_ejection_rate}
\dot{M}\sub{d,SN}(t) = \mathlarger{\sum}_{n} \dot{M}_{m\tn{,SN}}\,\eta_{n\tn{,SN}}\,\frac{A_{n}}{A_{m}}\ ,
\end{equation}
where $n$ is each of the four dust types considered, $m$ is the corresponding key element, $\dot{M}_{m\tn{,SN}}$ is the ejection rate of key element $m$ (which directly depends on the yields calculated by \binaryc{}, see Section \ref{sec:Binary stellar population model}), $\eta_{n\tn{,SN}}$ is the dust condensation efficiency parameter for dust type $n$ (see table 1 of \citetalias{Vijayan+19}) \correc{which accounts for the destruction of newly-formed dust by the SN reverse shock}, and $A_{n}$ and $A_{m}$ are the atomic weights of dust type $n$ and its corresponding key element $m$, respectively. For AGB stars and SNe, the fraction of dust ejected that goes into the molecular or diffuse components of the ISM is set such that their dust mass ratio, $M\sub{dust,clouds}/M\sub{dust,diff}$, is unchanged.

\begin{figure*}
\centering
 \includegraphics[angle=0,width=1.\linewidth]{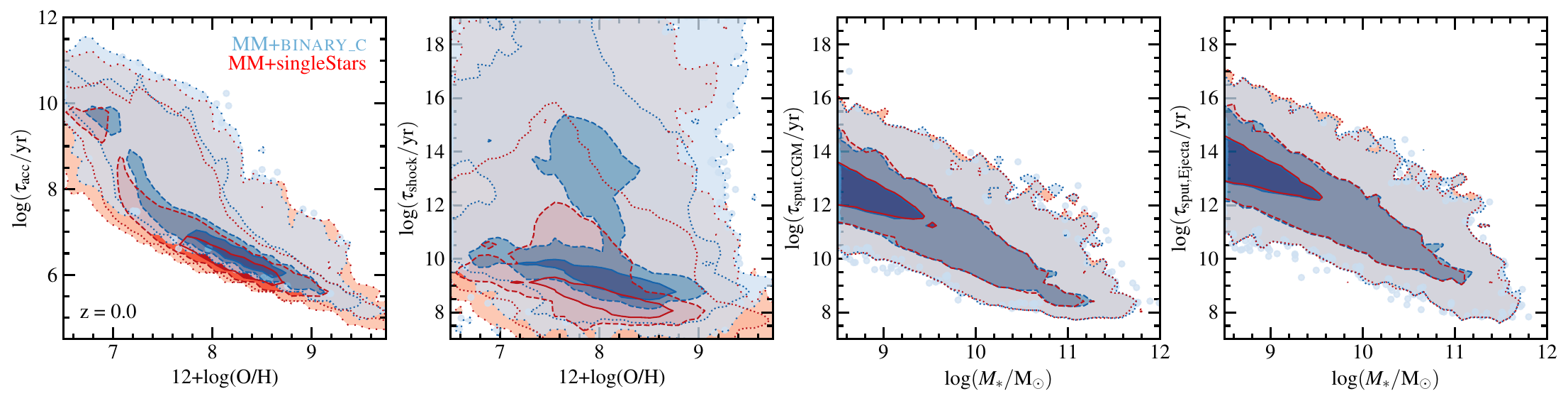}
 \caption{\textit{First panel:} Dust accretion timescale for grain growth in molecular clouds as a function of ISM oxygen abundance for individual radial rings. \textit{Second panel:} Dust destruction timescale for SN shocks in the ISM versus ISM oxygen abundance for individual rings. \textit{Third panel:} Dust destruction timescale for thermal sputtering in the CGM as a function of galaxy stellar mass. \textit{Fourth panel:} Dust destruction timescale for thermal sputtering in the Ejecta phase as a function of galaxy stellar mass. In all cases, the MM+\binaryc{} (blue) and MM+singleStars (red) versions of \lgal{} are shown, run on \textsc{Millennium} and \textsc{Millennium-II} combined. Contours represent the 1-4 $\sigma$ spread in the distributions.}
 \label{fig:timescales}
\end{figure*}

\subsubsection{\correc{Grain growth in the ISM}} \label{sec:Dust production by grain growth}
Growth of dust grains through accretion in the ISM is assumed to only take place inside molecular clouds. In \lgal{}, the H$_{2}$ fraction in the ISM is calculated at each timestep (for each radial ring) following the model developed by \citet{Fu+10,Fu+13}. This model uses the metallicity- and density-dependent H$_{2}$ partitioning formalism of \citet{Krumholz+09}, as discussed in section 2.2.3 of \citealt{Henriques+20}. \correc{We note that a metallicity floor of $Z=0.01$ is assumed when calculating the gas clumping factor in this formalism, in order to stimulate H2 formation at very early times.}

At the start of each timestep, we assume that the fraction of dust in molecular clouds is the same as the H$_{2}$ fraction, thereby ensuring that the chemical composition of diffuse and molecular components within each radial ring is initially the same. This is reasonable given that the expected lifetime of a molecular cloud ($\sim{}10$ Myr) is comparable to the timestep widths in \lgal{} ($<20$ Myr).

Once the total mass of molecular material is known, the maximum amount of each element in molecular clouds available for accretion onto dust grains is calculated. This is known as the maximum condensation fraction, $f\sub{max}$. For the refractory elements Mg, Si, Ca, and Fe, we assume $f\sub{max}=1.0$. For the volatile elements N, Ne, and S, we assume $f\sub{max}=0.0$. For C, we assume an $f\sub{max}$ equal to the fraction of C in the molecular gas phase that is not in CO molecules. In essentially all cases, the CO fraction for carbon is set to 0.3 (leading to $f\sub{max,C}=0.7$), in line with observations of molecular clouds in the Milky Way \citep{Irvine+87,vanDishoeck&Blake98}. However, it can be lower in environments where the oxygen abundance is particularly low.

The maximum condensation fraction for O, $f\sub{max,O}$, can then be determined at each timestep from the abundance of other refractory elements left available for grain growth (excluding oxygen atoms in CO), assuming that O forms only into silicate and iron oxide dust. To calculate this, we follow \citet{Zhukovska+08} such that
\begin{equation}\label{eqn:f_max,O}
f\sub{max,O} = \frac{M\sub{O,silicates,max}+M\sub{O,FeO,max}+M\sub{O,dust,clouds}}{M\sub{O,clouds}}\ ,
\end{equation}
where $M\sub{O,dust,clouds}$ is the mass of O already in dust in molecular clouds and $M\sub{O,clouds}$ is the total mass of O in molecular clouds.

Once the maximum condensation fractions are known, the \textit{actual} amount of each refractory element that accretes onto dust grains in that timestep can be calculated. To do this, we need to define (a) the timescale for the exchange of material between molecular and diffuse gas (\ie{}the effective molecular cloud lifetime), $\tau\sub{ex}$, and (b) the timescale for accretion of gas-phase elements onto existing dust grains, $\tau\sub{acc}$. For the exchange timescale, we assume $\tau\sub{ex}=10$ Myr (see \citealt{Zhukovska+08}). For the accretion timescale, we assume,
\begin{equation}\label{eqn:tau_acc}
\tau\sub{acc} = \tau\sub{acc,0}\,\frac{M\sub{clouds}}{M\sub{dust,clouds}}\ .
\end{equation}
This accretion timescale formalism differs from that presented by \citet{Asano+13} and commonly used in other cosmological simulations (\eg{}\citealt{deBennassuti+14,Popping+17,Triani+20,Graziani+20}), in that the efficiency of grain growth here depends on the mass of dust present in molecular clouds, rather than on gas-phase properties such as the ISM metallicity and density. This means that our efficiency parameter, $\tau\sub{acc,0}$, takes on a slightly different meaning to other works. Nonetheless, as in those works (but not \citetalias{Vijayan+19}), we tune our $\tau\sub{acc,0}$ to obtain $\tn{log}(\tau\sub{acc}/\tn{yr})\sim{}6.08$ at $Z = \Zsun = 0.0134$ and $n\sub{mol}\sim{}250\,\tn{cm}^{-3}$, as expected from the \citet{Asano+13} formalism. This returns $\tau\sub{acc,0}=5\times{}10^{3}\,\tn{yr}$, which we assume to be fixed across space and time. This calibration leads to a wider dynamic range of accretion timescales than found in most other simulations. For example, at fixed H$_{2}$ density, it produces shorter $\tau\sub{acc}$ in dust-rich environments and longer $\tau\sub{acc}$ in dust-poor environments than seen in \citet{Popping+17}. This has an important impact on the dust-to-metal ratios predicted for galaxies (see Section~\ref{sec:Dust scaling realtions}).

The first panel of Fig.~\ref{fig:timescales} shows the relation between ISM oxygen abundance and $\tau\sub{acc}$ at $z=0$ for the two versions of \lgal{} we consider in this work. These versions are defined in detail in Section \ref{sec:Results}.

The fraction of the available mass of a given chemical element that accretes onto dust grains within one timestep is then given by,
\begin{equation}\label{eqn:accretion_fraction}
f\sub{acc} = 1-e^{-\Delta{}t/\tau\sub{acc}}\ ,
\end{equation}
where $\Delta{}t$ is the timestep width and a constant $\tau\sub{acc}$ is implicitly assumed across the timestep. We also simultaneously solve for the fraction of material in molecular clouds that isn't yet exchanged with diffuse gas, given by
\begin{equation}\label{eqn:exchange_fraction}
f\sub{ex} = e^{-\Delta{}t/(1-\mu)\tau\sub{ex}}\ ,
\end{equation}
where $\mu=M\sub{H2}/(M\sub{HI}+M\sub{H2})$ is the molecular gas fraction. In the limit where $\tau\sub{acc}\gg{}\tau\sub{ex}$, the dust fractions in the molecular and diffuse phases are identical after grain growth. However, for cases where $\tau\sub{acc}\lesssim{}\tau\sub{ex}$ (\ie{}in dustier environments), the dust fraction in molecular clouds will exceed that present in the surrounding diffuse ISM after grain growth (see also section 3.2 of \citetalias{Vijayan+19}).

This model returns values for $f\sub{C,dust,clouds}$ of typically $0.6-0.7$ at $z=0$. There is a scatter down to lower values at high metallicity in early-type galaxies (ETGs), which have had most of their interstellar dust removed or destroyed by $z=0$. For oxygen, $f\sub{O,dust,clouds}$ is always $\lesssim{}0.2$, which can has a noticeable impact on measured gas-phase oxygen abundances in galaxies (see Section~\ref{sec:General galaxy properties}).

\correc{We note that the grain growth efficiency could also be affected by other properties, such as the grain size distribution \citep{Aoyama+20} and charge distribution \citep{Glatzle+22}. Such dependencies will be investigated in \lgal{} in future works.}

\subsection{Dust destruction} \label{sec:Dust destruction}
\subsubsection{\correc{Dust destruction by SN shocks and astration}} \label{sec:Dust destruction by SN shocks}
Following \citet{McKee89}, dust in the ISM is assumed to be destroyed via sputtering in SN-induced shocks. The destruction timescale for this process is
\begin{equation}\label{eqn:shock_dest_timescale}
\tau\sub{shock} = \frac{M\sub{cold}}{M\sub{cleared}f\sub{SN}R\sub{SN}}\ ,
\end{equation}
where $M\sub{cold}$ is the mass of cold gas within each ISM radial ring, $f\sub{SN}=0.36$ is the expected fraction of all SNe that interact with the ISM (see section 4 of \citealt{McKee89}), and $R\sub{SN}$ is the combined rate of SNe that explode within each radial ring at each timestep. $M\sub{cleared}$ represents the mass of cold gas assumed to be cleared of dust by destruction from a single SN. In \citetalias{Vijayan+19}, this value was fixed at $1200\Msun$. In this work, we instead follow the findings of \citet{Hu+19b} from their high-resolution hydrodynamical simulations by allowing $M\sub{cleared}$ to vary between 1180 and $1660\Msun$ depending on the composition of the dust, with the lower limit corresponding to $M\sub{C,dust}/(M\sub{C,dust}+M\sub{Si,dust})=1$ and the upper limit corresponding to $M\sub{C,dust}/(M\sub{C,dust}+M\sub{Si,dust})=0$. For typical dust-to-gas ratios of $\sim{}0.001$ at $z=0$ \citep{DeVis+19}, this leads to a typical dust-destruction rate of $1.18-1.66\Msun$ per SN, which is consistent with the upper limit of $3\Msun$ estimated by \citet{Ferrara&Peroux21}. We then compute the mass of dust in the ISM that is destroyed per timestep as
\begin{equation}\label{eqn:shock_dest_mass}
\Delta{}M\sub{shock} = (1-e^{-\Delta{}t/\tau\sub{shock}})M\sub{dust}\ .
\end{equation}

The second panel of Fig. \ref{fig:timescales} shows $\tau\sub{shock}$ as a function of overall gas-phase oxygen abundance for regions of the ISM within model galaxies at $z=0$. Most regions exhibit destruction timescales around 1 Gyr. This is in reasonable agreement with the values predicted for the solar neighbourhood by the hydrodynamical simulations of \citet{Slavin+15} and \citet{Hu+19b}. The scatter up to very large values of $\tau\sub{shock}$ is predominantly caused by regions with low current SFR and therefore low SN-II rates. At fixed metallicity, $\tau\sub{shock}$ is longer in MM+\binaryc{} than in MM+singleStars due to lower overall SN rates (see Section \ref{sec:Dust rates}).

\correc{In addition to SN shocks, dust is also destroyed in the ISM through astration. In \lgal{}, all dust present in the molecular gas used to form stars is assumed to be destroyed, with it's constituent elements becoming incorporated into the newly-formed stars.}

\subsubsection{Dust destruction by thermal sputtering} \label{sec:Dust destruction by thermal sputtering}
We also consider the survival of dust in galactic outflows and its subsequent gradual destruction via thermal sputtering in the CGM and Ejecta phase surrounding galaxies. Dust in extragalactic hot gas has been observed out to large distances, with a spatial distribution similar to the CGM itself (\eg{}\citealt{Menard+10,Peek+15}). Our model for thermal sputtering is based on that of \citet{Tsai&Mathews95}, which has also been used in other cosmological simulations (\eg{}\citealt{McKinnon+17,Popping+17,Triani+20}). The thermal-sputtering timescale depends on the hot gas density ($\rho\sub{hot}$) and temperature ($T\sub{hot}$) by
\begin{equation}\label{eqn:sput_dest_timescale}
\tau\sub{sput} = 0.17\,\tn{Gyr}\left(\frac{a}{a_{0}}\right)\left(\frac{10^{-27}\tn{g cm}^{-3}}{\rho\sub{hot}}\right)\left[\left(\frac{T\sub{crit}}{T\sub{hot}}\right)^{\omega}+1\right]\ ,
\end{equation}
where the initial grain radius at each timestep is assumed to be $a_{0} = 0.1\mu\tn{m}$, the factor controlling the sputtering timescale in the low-temperature regime is fixed at $\omega{}=2.5$, the critical temperature above which the sputtering rate becomes roughly constant is $T\sub{crit}=2\times{}10^{6}\,\tn{K}$, and the hot gas is assumed to be at the virial temperature given by $T\sub{hot}\equiv{}T\sub{vir} = 0.5\mumH{}V\sub{vir}^{2}/k\sub{B}$. The constants $\mumH$ and $k\sub{B}$ are the mean atomic mass and the Boltzmann constant, respectively. The dust-grain radius after sputtering, $a$, is calculated as
\begin{equation}\label{eqn:grain_radius}
\frac{\tn{d}a}{\tn{d}t} = -c\sub{sput}\left(\frac{\rho\sub{hot}}{m\sub{p}}\right)\left[\left(\frac{T\sub{crit}}{T\sub{hot}}\right)^{\omega}+1\right]^{-1}\ ,
\end{equation}
where $c\sub{sput}=3.2\times{}10^{-18}\ \tn{cm}^{4}\,\tn{s}^{-1}$, and $m\sub{p}$ is the mass of a proton. The mass of dust destroyed through sputtering per timestep is then given by
\begin{equation}\label{eqn:sput_dest_mass}
\Delta{}M\sub{sput} = \frac{M\sub{dust}}{\tau\sub{sput}/3}\,\Delta{}t\ ,
\end{equation}
where the factor of three arises from the grain mass ($m\sub{g}$) being related to the grain radius and density ($\rho\sub{g}$) by $m\sub{g} = (4/3)\pi{}a^{3}\rho\sub{g}$, as pointed out by \citet{McKinnon+17}.

When assessing how to measure $\rho\sub{hot}$ for use in Eqn.~\ref{eqn:sput_dest_timescale}, we note that assuming a uniform density of $\rho\sub{hot}=M\sub{hot}/(4/3)\pi{}R\sub{vir}^{3}$ (\ie{}the single isothermal sphere, or SIS, approximation) would likely under-estimate the true average in the CGM, as both gas and dust are observed to exhibit higher densities in the core than the outskirts (\eg{}\citealt{Vikhlinin+06,Menard+10}). Therefore, we approximate the gas density profile for Eqn.~\ref{eqn:sput_dest_timescale} with a $\beta$ model given by $\beta{}(r) = [1+(r/r\sub{c})^{2}]^{-3\beta\sub{s}/2}$ \citep{Cavaliere&Fusco-Femiano76}. Such a profile is commonly used to approximate the hot-gas density distribution in galaxy groups and clusters in observations (\eg{}\citealt{Reiprich&Boehringer02}) and semi-analytic simulations \citep{Yates+17}. Here, we assume the canonical shape parameter of $\beta\sub{s} = 2/3$ and that the core radius, $r\sub{c}$, equals the scale length of the dark-matter halo assuming an NFW profile \citep{Navarro+97}. The core density is then obtained from $\rho\sub{hot,0} = M\sub{hot} / \int_{0}^{R\sub{vir}} 4\pi{}r^{2}\beta(r)\tn{d}r$, and the density in Eqn.~\ref{eqn:sput_dest_timescale} as
\begin{equation}\label{eqn:ave_rhoHot}
\bar{\rho}\sub{hot} = \rho\sub{hot,0} \frac{\int_{0}^{R\sub{vir}} r^{2}\beta(r)^{2}\tn{d}r}{\int_{0}^{R\sub{vir}} r^{2}\beta(r)\tn{d}r}\ .
\end{equation}
We find that this assumption leads to shorter dust sputtering timescales in the CGM surrounding lower-mass galaxies compared to the simpler SIS approximation. However, we caution that the degree of this difference varies depending on the assumed gas density profile. Therefore, we take our preferred model as a reasonable intermediate case, which allows additional system-dependent flexibility in the CGM sputtering timescale.

For the Ejecta component surrounding haloes, we assume the simpler SIS density when calculating $\tau\sub{sput}$, as this is likely to better represent the lower ambient densities found in the intergalactic medium (IGM).

The third and fourth panels of Fig.~\ref{fig:timescales} show $\tau\sub{sput}$ for the CGM and Ejecta components, respectively, as a function of stellar mass for galaxies of all types at $z=0$. Sputtering timescales are much shorter in high-mass systems because of their higher core densities and virial temperatures. At $\tn{log}(M_{*}/\Msun) \gtrsim{} 11.0$, $\tau\sub{sput}$ tends to be shorter than the galaxy-averaged $\tau\sub{shock}$ within the ISM, meaning that thermal sputtering outside galaxies dominates the dust destruction in such systems (see Section \ref{sec:Dust rates}).

\begin{figure*}
 \centering
 \begin{tabular}{@{}c@{}c@{}c@{}}
\includegraphics[width=0.3275\linewidth]{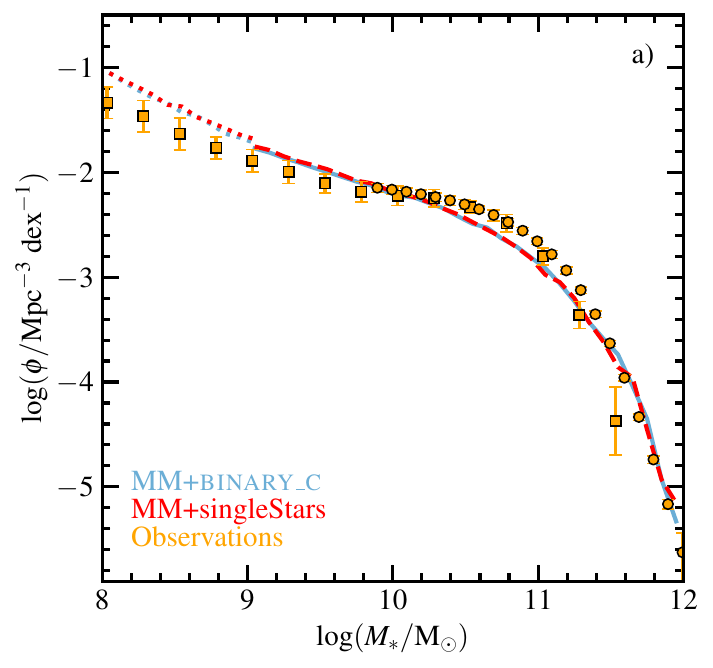} &
 \includegraphics[width=0.3275\linewidth]{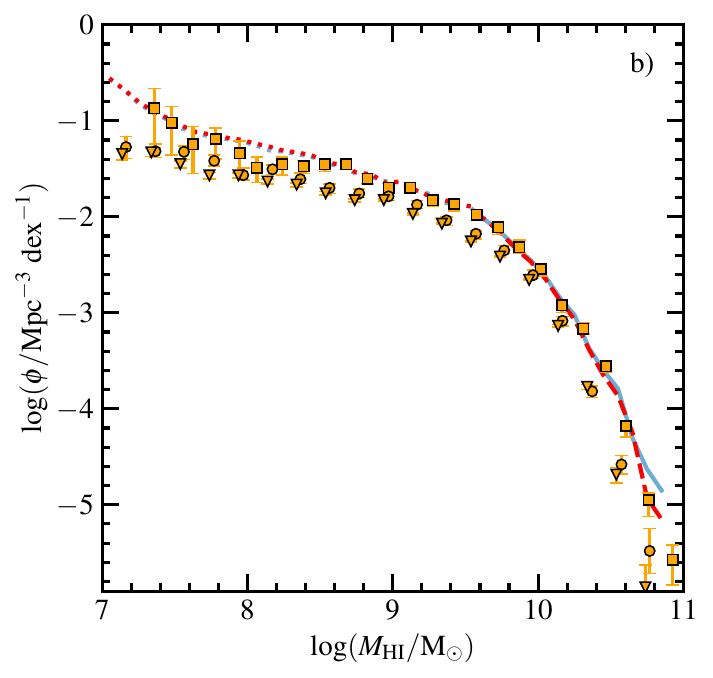} &
 \includegraphics[width=0.335\linewidth]{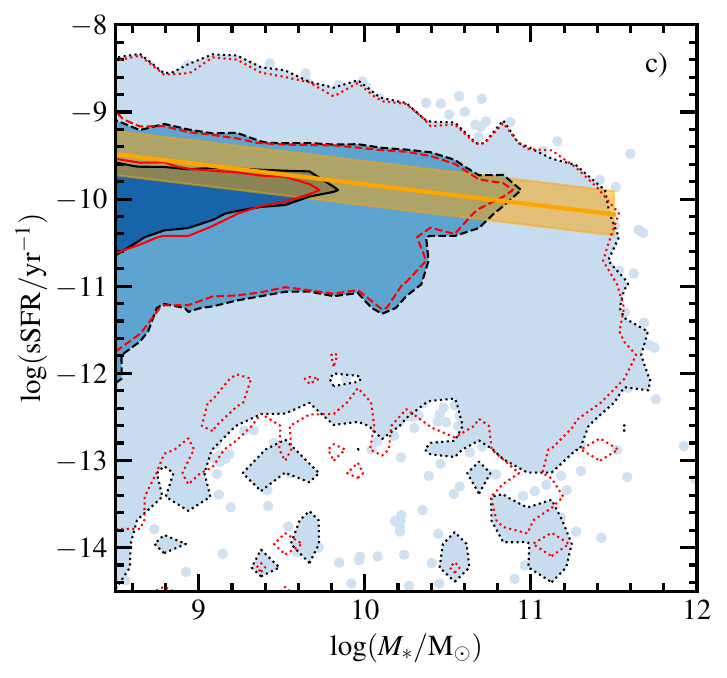} \\
 \includegraphics[width=0.325\linewidth]{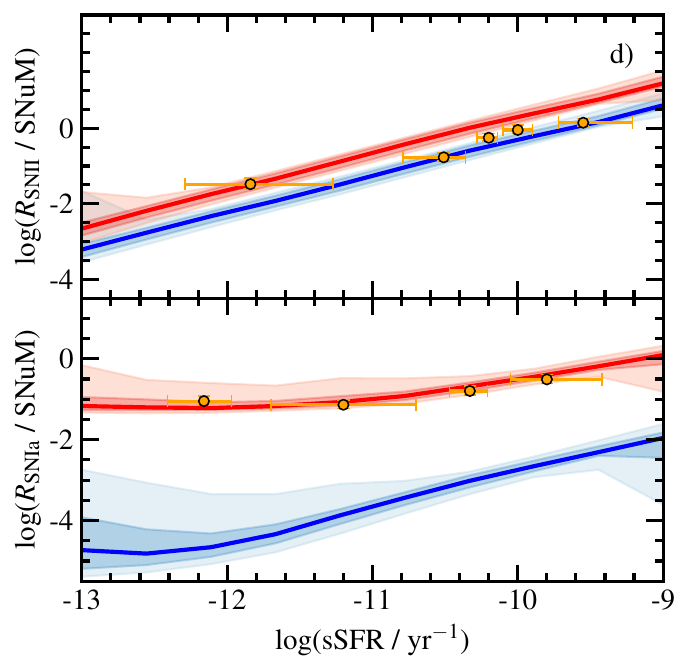} &
 \includegraphics[width=0.33\linewidth]{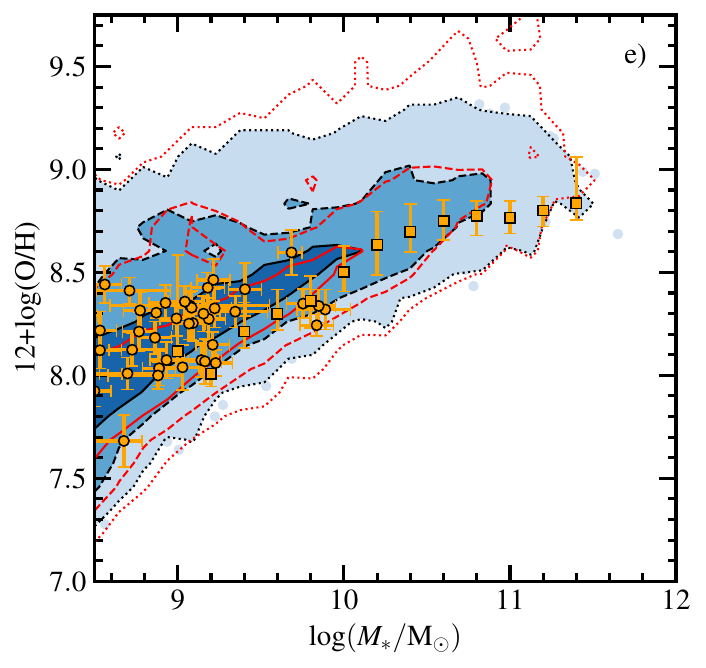} &
\includegraphics[width=0.345\linewidth]{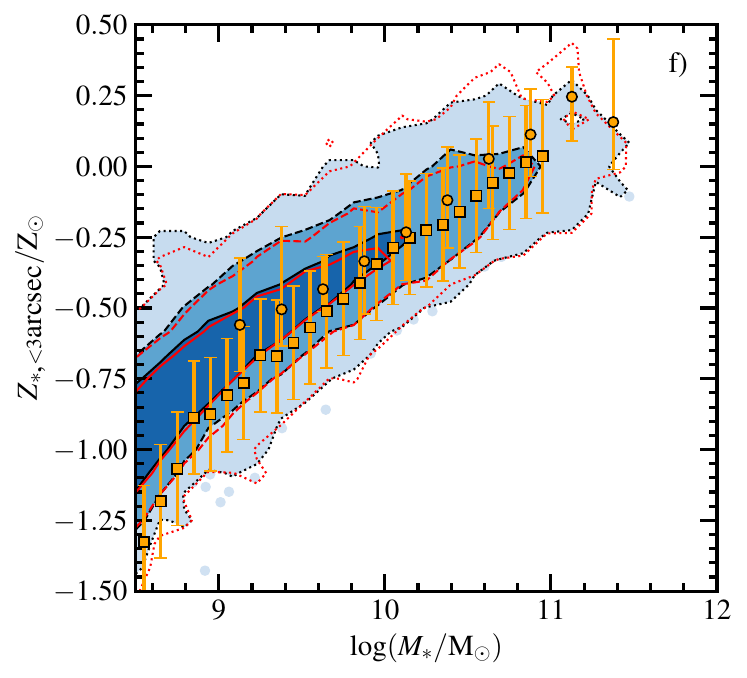} \\
 \end{tabular}
 \caption{Key scaling relations at $z=0$. \textit{Panel a)} The galaxy stellar mass function (SMF). Solid and dotted lines represent \lgal{} run on \textsc{Millennium} and \textsc{Millennium-II}, respectively. Observations are from \citet{Baldry+08,Baldry+12,Li&White09} (squares) and \citet{D'Souza+15} (circles). \textit{Panel b)} The galaxy HI mass function (HIMF). Observations are from \citet{Zwaan+05} (squares), \citet{Haynes+11} (circles), and \citet{Jones+18} (triangles). \textit{Panel c)} The galaxy stellar mass -- sSFR relation, with observations from \citet{Elbaz+07}. \textit{Panel d)} sSFR -- $R\sub{SN}$ relation for SNe-II (top) and SNe-Ia (bottom) at $z=0$ for all galaxies, with observations from \citet{Graur+17a}. \textit{Panel e)} The galaxy stellar mass -- gas-phase metallicity relation for star-forming disc galaxies. Observations are from \citet{Yates+20} (circles) and \citet{Yates+21a} (squares). \textit{Panel f)} The galaxy stellar mass -- stellar metallicity relation for star-forming disc galaxies, with metallicity measured within $\sim{}3$ arcsec. Observations are from \citet{Yates+21a} (circles) and \citet{Zahid+17} (squares). In all cases, the MM+\binaryc{} (blue) and MM+singleStars (red) versions of \lgal{} are shown. Overall, stellar masses, gas masses, and star formation rates are largely unaffected by switching from our single-star to \binaryc{} set-up, however gas-phase metallicities are higher at $z=0$ by $\sim{}0.1$ dex on average.}
 \label{fig:gen_gal_props}
\end{figure*}

\section{Results} \label{sec:Results}

To assess how binary stars affect galaxy evolution, we compare two versions of the \lgal{} MM simulation described in Section \ref{sec:L-Galaxies}: (a) a version using the mixed stellar populations from \binaryc{} presented in Section \ref{sec:Binary stellar populations} (hereafter, MM+\binaryc{}), and (b) a version using the single-star stellar populations plus SNe-Ia formalism presented in Section \ref{sec:Comparing single star and binary population yields} (hereafter, MM+singleStars). These two versions are chosen to best compare the new and old \lgal{} simulations, and our new dust model is included in both. Unless stated otherwise, in this section we combine model galaxies from runs on both \textsc{Millennium} and \textsc{Millennium-II}, with each galaxy weighted by the inverse of the effective volume of the underlying N-body simulation, considering only ``resolved'' systems above $\tn{log}(M_{*}/\Msun) = 8.0$ from \textsc{Millennium} and $\tn{log}(M_{*}/\Msun) = 7.0$ from \textsc{Millennium-II}. For plotting, we use representative sub-volumes of the \textsc{Millennium} and \textsc{Millennium-II} boxes, containing a total of $\sim{}495,000$ model galaxies by $z=0$.

\subsection{General galaxy properties} \label{sec:General galaxy properties}
We first present the key galaxy scaling relations used to assess the success of cosmological-scale simulations. Fig.~\ref{fig:gen_gal_props} shows the stellar mass function (SMF), HI mass function (HIMF), $M_{*}$ -- sSFR relation, SN rates, and mass -- metallicity relations (MZ$\sub{g}$R and MZ$_{*}$R) for galaxies at $z=0$. Observational data (orange) is shown alongside the MM+\binaryc{} (blue) and MM+singleStars (red) versions of \lgal{}.

The main conclusion from this comparison is that the use of binary stellar populations does not appear to significantly affect the general properties of galaxies at low $z$. For example, panels (a), (b), and (c) of Fig.~\ref{fig:gen_gal_props} show that the SMF, HIMF, and galaxy sSFRs are essentially unchanged when switching from the \citetalias{Yates+13} set-up to the \binaryc{} set-up in \lgal{} (see Section~\ref{sec:Comparing single star and binary population yields}). This is chiefly due to the similar mass ejection rates from massive stars in these two set-ups (as shown in Fig.~\ref{fig:mass_and_SNrate_comp}), which determines the total SN feedback energy in the simulation. A similar conclusion was also drawn by \citet{DeDonder&Vanbeveren04}, who found that binaries have a negligible impact on the star formation history of Milky-Way-type galaxies in one-zone GCE models with no outflows.

Panel (d) shows that SN-II rates are also in reasonable agreement between both the \lgal{} versions and observations. However, SN-Ia rates are severely under-estimated in MM+\binaryc{}, as expected from the yield analysis in Section~\ref{sec:Comparing single star and binary population yields}. This has an impact on the iron abundances of galaxies, which is discussed in Section~\ref{sec:Radial profiles}. Conversely, MM+singleStars is in good agreement with the observed SN-Ia rate from the LOSS survey \citet{Graur+17a}, primarily because the fraction of SN-Ia systems per stellar population (\ie{}the $A\sub{SNIa}$ parameter) is scaled to match global chemical properties at $z=0$.

Panel (e) shows the MZ$\sub{g}$R for star-forming disc galaxies at $z=0$, where gas-phase metallicity is calculated as the SFR-weighted average ISM oxygen abundance (excluding oxygen in dust) in units of 12+log(O/H) for \lgal{}, where $\tn{O/H} = N\sub{O}/N\sub{H} = (M\sub{O}/M\sub{H}) \cdot{} (A\sub{H}/A\sub{O}$) and $A\sub{H}$ and $A\sub{O}$ are the atomic weights of H and O, respectively. This method best matches the emission-line methods used for the observations to which we compare.

There is a slight but systematic difference in gas-phase metallicity between the two versions of \lgal{}, especially for low-mass galaxies. MM+\binaryc{} returns higher 12+log(O/H) than MM+singleStars by $\sim{}0.1$ dex on average, and is in marginally better agreement with the latest metallicity observations (although see \citealt{Cameron+23}). This difference is predominantly caused by differences in the SN-II yield tables assumed, rather than binary effects. In the \citet{Portinari+98} yields used in MM+singleStars, there is a drop in the O ejection rate from low-metallicity stars of mass $\sim{}50-60\Msun$ due to the build-up of large CO cores, which leads to larger remnant masses and therefore a lower ejected yield (see \citealt{Portinari+98}, section 5). This effect can also be seen as a dip in the single-star burst yields in Fig.~\ref{fig:yields_comp} for most heavy elements $\sim{}5$ Myr after star formation. In the \citet{Chieffi&Limongi04} yields used in MM+\binaryc{}, this effect is weaker, leading to higher O enrichment in low-metallicity environments, and consequently higher 12+log(O/H) in the ISM by $z=0$ in low-mass galaxies.

It is interesting to note that this boost in 12+log(O/H) at low mass in MM+\binaryc{} is almost exactly offset by the drop in 12+log(O/H) we find when accounting for the depletion of oxygen onto dust grains. This means that MM+\binaryc{} maintains the good agreement with the observed MZ$\sub{g}$R at $z=0$ that was found for previous versions of \lgal{} that did not account for dust depletion.

We also note that metallicities are not converged between \textsc{Millennium} and \textsc{Millennium-II}. For example, gas-phase metallicities at $z=0$ are, on average, $\sim{}0.1-0.3$ dex higher at fixed stellar mass when \lgal{} is run on the higher-resolution \textsc{Millennium-II}. This is due to the known effect of higher SFRs and gas masses in lower-mass galaxies for \textsc{Millennium-II} \citep{Henriques+20,Yates+21b}, which in turn leads to higher metal and dust abundances at larger radii in the ISM.

Finally, panel (f) shows the MZ$_{*}$R, where $Z_{*}$ is the total mass-weighted stellar metallicity measured within a 3 arcsec aperture centred on each galaxy for both observations and \lgal{}. These metallicities are normalised to the bulk metallicity of the Sun as measured by \citet{Asplund+09}. Interestingly, higher stellar metallicities are not seen in MM+\binaryc{} within 3 arcsec. This illustrates that the higher oxygen abundances discussed above are predominantly at larger galactocentric radii in $z=0$ galaxies, where the metallicity of the star-forming gas is lower.

\begin{figure}
 \centering
\includegraphics[width=0.77\linewidth]{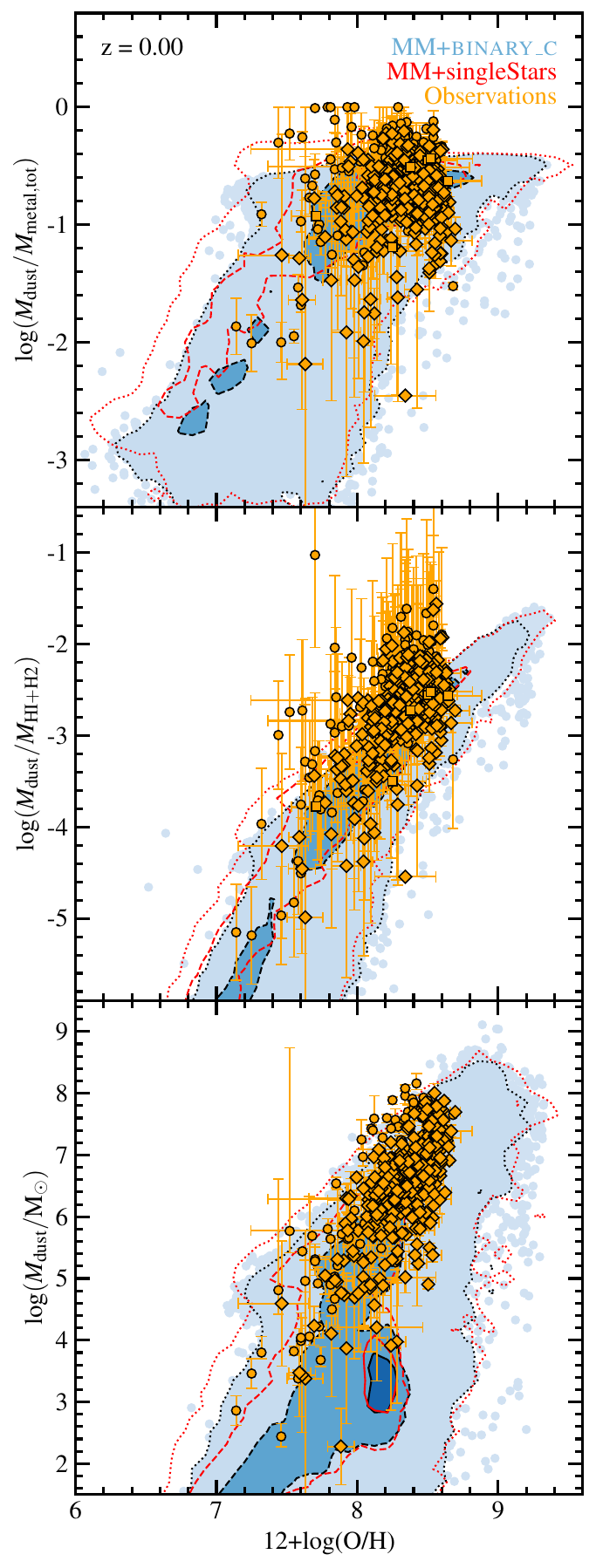}
 \caption{Relation between SFR-weighted metallicity and the DTM ratio (top panel), DTG ratio (middle panel), and $M\sub{dust}$ (bottom panel). In all cases, the MM+\binaryc{} (blue) and MM+singleStars (red) versions of \lgal{} are shown, run on \textsc{Millennium} and \textsc{Millennium-II} combined, with metallicity measured as the SFR-weighted oxygen abundance. Orange points represent observational data from \citet{Remy-Ruyer+15} (circles), DustPedia \citep{DeVis+19,Casasola+22} (diamonds), and the collection of \citet{Peroux&Howk20} (squares). There is a good match between both \lgal{} versions and observations at $z=0$.}
 \label{fig:dust_scaling_relations_z0}
\end{figure}

\begin{figure*}
 \centering
 \includegraphics[width=0.99\linewidth]{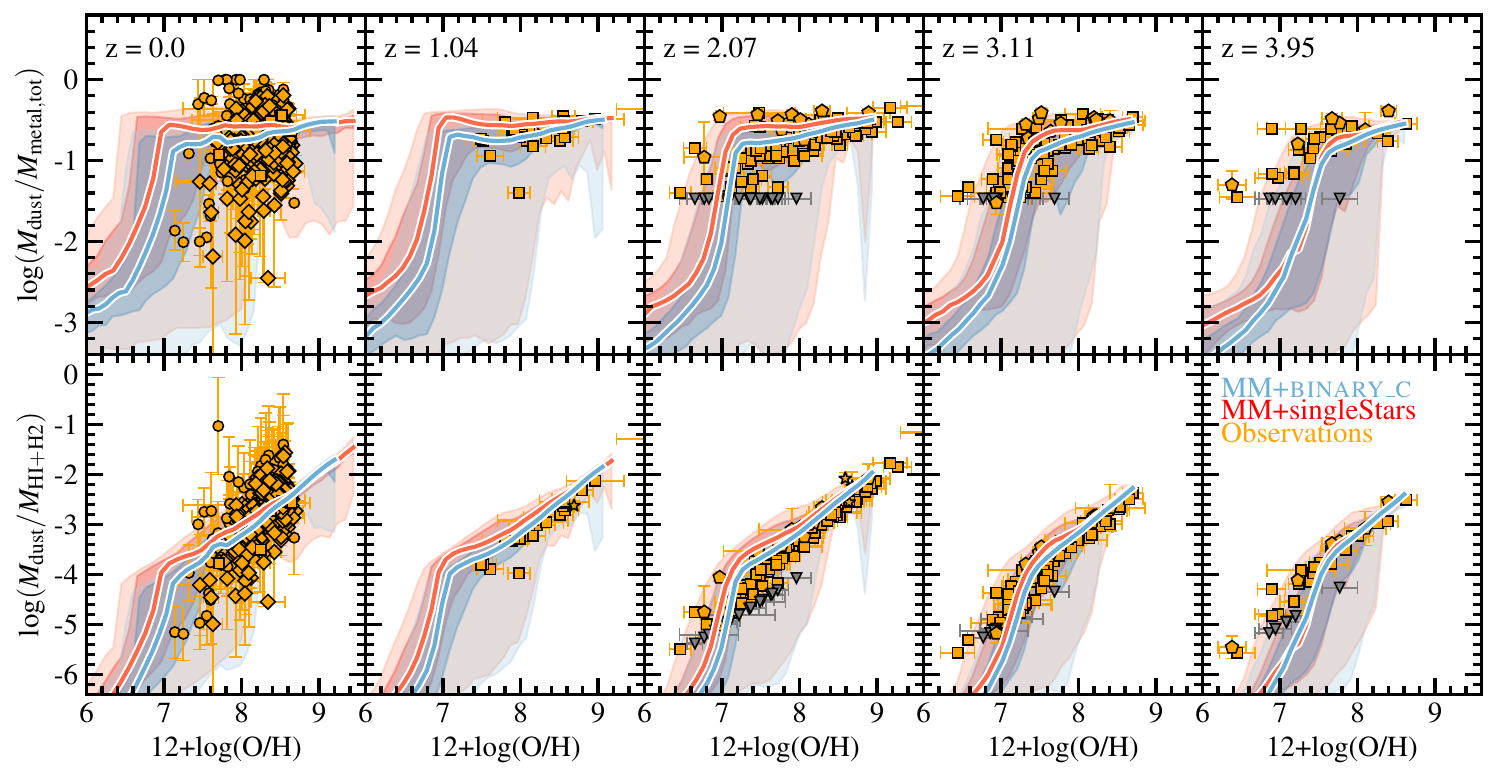} \\
 \includegraphics[width=0.99\linewidth]{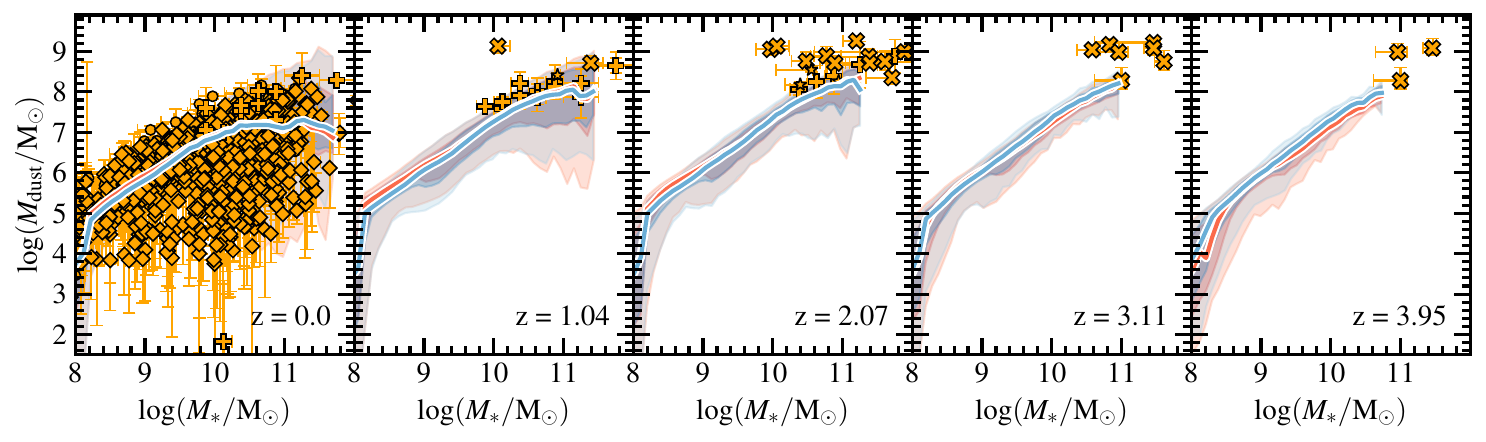}
 \caption{Evolution of key dust scaling relations from $z\sim{}0$ to $\sim{}4$ for MM+\binaryc{} (blue) and MM+singleStars (red). \textit{Top row:} Metallicity -- DTM relation. \textit{Middle row:} Metallicity -- DTG relation. \textit{Bottom row:} The stellar mass -- dust mass relation. In all cases, the median relation is shown (solid lines) along with the 1- and 2-$\sigma$ spread (shaded regions), with metallicity measured as the mass-weighted oxygen abundance. Observational data is from \citet[][circles]{Remy-Ruyer+15}, the DustPedia data set \citep[][diamonds]{DeVis+19}, the collection of \citet[][squares, with upper limits as grey triangles]{Peroux&Howk20}, \citet[][pentagons]{Heintz+23a}, \citet[][stars]{Shapley+20}, \citet[][pluses]{Santini+14}, and \citet[][crosses, considering only those galaxies with photometric redshift errors $< 0.5z$]{daCunha+15}, binned in redshift to match the \lgal{} outputs $\pm{}0.5z$ (see labels). Both versions of \lgal{} match observations well back to $z\sim{}4$, although dust masses may be under-estimated in metal-poor galaxies at $\gtrsim{}3$.}
 \label{fig:dtm_and_dtg_evos}
\end{figure*}

\subsection{Dust scaling relations} \label{sec:Dust scaling realtions}
We now consider the impact of binary stellar evolution on the dust properties of galaxies at low and high $z$. Fig.~\ref{fig:dust_scaling_relations_z0} shows the dust-to-metal ratio ($\tn{DTM} = M\sub{dust}/M\sub{metal,tot}$, top panel), dust-to-gas ratio ($\tn{DTG} = M\sub{dust}/M\sub{HI+H2}$, middle panel), and dust mass ($M\sub{dust}$, bottom panel) in the ISM of galaxies at $z=0$ as a function of their SFR-weighted average gas-phase oxygen abundance for the MM+singleStars (red) and MM+\binaryc{} (blue) versions of \lgal{}.

There is a very good correspondence between both versions of \lgal{} and observational data (orange points) in Fig.~\ref{fig:dust_scaling_relations_z0}. \correc{At lower metallicities, galaxies undergo an increase in their DTM, DTG, and $M\sub{dust}$ over time in \lgal{}, due to the increasing efficiency of grain growth as the dust mass in molecular clouds increases.} At higher metallicities, a typical value of $\tn{log(DTM)} \sim{} -0.6$ is found for star-forming systems at $z=0$, along with a significant scatter down to low DTM which is also seen in the DustPedia observations (\citealt{DeVis+19}, diamonds). In \lgal{}, this scatter at high metallicity contains mostly quenched galaxies, which have undergone net dust destruction at late times.

\correc{When comparing the two versions of \lgal{},} we find that MM+\binaryc{} returns slightly higher 12+log(O/H) at fixed DTM, DTG, and $M\sub{dust}$ by $z=0$. This is due to low-mass galaxies being more oxygen rich in MM+\binaryc{} (see Section~\ref{sec:General galaxy properties}), which has a greater effect on the gas than the dust in our model due to the lower maximum condensation factor for oxygen than other refractory elements (see Section~\ref{sec:Dust production by grain growth}). This effect shifts the low-mass galaxy population right and down in the top panel and right in the middle and bottom panels.

In Fig.~\ref{fig:dtm_and_dtg_evos}, we show the evolution back to $z\sim{}4$ of the median DTM and DTG ratios as a function of gas-phase oxygen abundance (top two rows) and median $M\sub{dust}$ as a function of galaxy $M_{*}$ (bottom row). A range of observational data is included for comparison (see figure caption). A similar observational trend is seen at high $z$ when only considering H$_{2}$ mass in the DTG ratio \citep{Saintonge+13,Popping+23}. We note that oxygen abundances for the QSO and GRB DLA data from \citet[][squares]{Peroux&Howk20} and \citet[][pentagons]{Heintz+23a} were derived from their measured metal-to-hydrogen mass ratios as follows: $12+\tn{log(O/H)} = 12 + \tn{[M/H]} + \tn{log}(M\sub{O,\textnormal{\astrosun}}/M\sub{H,\textnormal{\astrosun}}) + \tn{log}(A\sub{H}/A\sub{O})$. This approach assumes the fraction of all metal mass that is oxygen in these systems is the same as in the Sun, \ie{}$(M\sub{O}/M\sub{Z})\sub{\astrosun} = 0.43$ from \citet{Asplund+09}, which may under-estimate the true 12+log(O/H) at high $z$ by $\sim{}0.1$ dex. In order to better compare to these absorption-line-based measurements, we plot mass-weighted average gas-phase oxygen abundances for \lgal{} in the top two rows of Fig.~\ref{fig:dtm_and_dtg_evos}.

We find that both \lgal{} versions match observations well back to $z\sim{}4$, although the slightly lower DTM and DTG ratios found in MM+\binaryc{} (blue) at low $z$ appear to match the data better. An evolution in both the DTM and DTG ratios for metal-poor galaxies is also found, similar to that seen in GRB samples (\eg{}\citealt{Wiseman+17a}; \citealt{Heintz+23a}). This is a significant success of the new \lgal{} simulation, as previous cosmological-scale simulations have struggled to simultaneously match these dust properties across this redshift regime \citep{Ginolfi+18,Popping&Peroux22}. The key reason for this success is the shorter accretion timescales returned by our grain-growth formalism in dustier environments, which allow more efficient dust production at later times. Nonetheless, DTM ratios in metal-poor galaxies at $z\gtrsim{}4$ appear under-estimated in \lgal{} compared to absorption-line measurements from QSO and GRB DLAs (squares and pentagons). This is due to the inefficiency of grain growth in dust-poor environments in our model, which suggests either the need for enhanced dust production from other sources \correc{or a more detailed modelling of the dependence of the dust production rate on environmental properties \citep{Graziani+20}.} A similar result has also been found for the CROC simulations \citep{Esmerian&Gnedin22,Esmerian&Gnedin23} using a simpler dust model in post-processing. However, we note that the difficulty in detecting galaxies in this regime leaves open the possibility that current observational samples are biased towards dustier systems.

The sharp increase in DTM ratio seen at low metallicity in \lgal{} is driven by the transition between SN-dominated and grain-growth-dominated dust production. At low metallicities, dust is predominantly produced by SNe and AGB stars. However, once the metallicity, and thus in-situ dust mass, begins to increase, grain growth becomes increasingly efficient, causing a rapid increase in the DTM ratio until a rough equilibrium between grain growth and dust destruction is reached. This transition can also be seen in the grain growth accretion timescales of galaxies in Fig.~\ref{fig:timescales}. A similar effect is found in the \textsc{DustyGadget} hydrodynamical simulation developed by \citet{Graziani+20}, but is not present in most other cosmological-scale simulations \citep{Popping&Peroux22}. Other simulations tend to assume longer accretion timescales at low H$_{2}$ densities, leading to less efficient grain growth at late times (see Section~\ref{sec:Dust production by grain growth}). Such a sharp transition in DTM ratio with metallicity is unconfirmed in observations, however the upper limits obtained via absorption-line measurements (\citealt{Peroux&Howk20}, grey triangles) suggest that such a drop is not ruled out.

Finally, the bottom row of Fig.~\ref{fig:dtm_and_dtg_evos} illustrates that both versions of \lgal{} return very similar $M_{*}$ -- $M\sub{dust}$ relations at all redshifts. This indicates that net dust production from AGB stars and SNe is largely independent of the effects of binary stellar evolution. Both versions of \lgal{} are also in marginal agreement with the dust masses measured by \citet{Santini+14} and \citet{daCunha+15} back to $z\sim{}4$. We would expect these observational data to lie in the upper envelope of the model relation, as they include the most star-forming, dust-rich systems.

\begin{figure}
 \centering
 \includegraphics[width=1.0\linewidth]{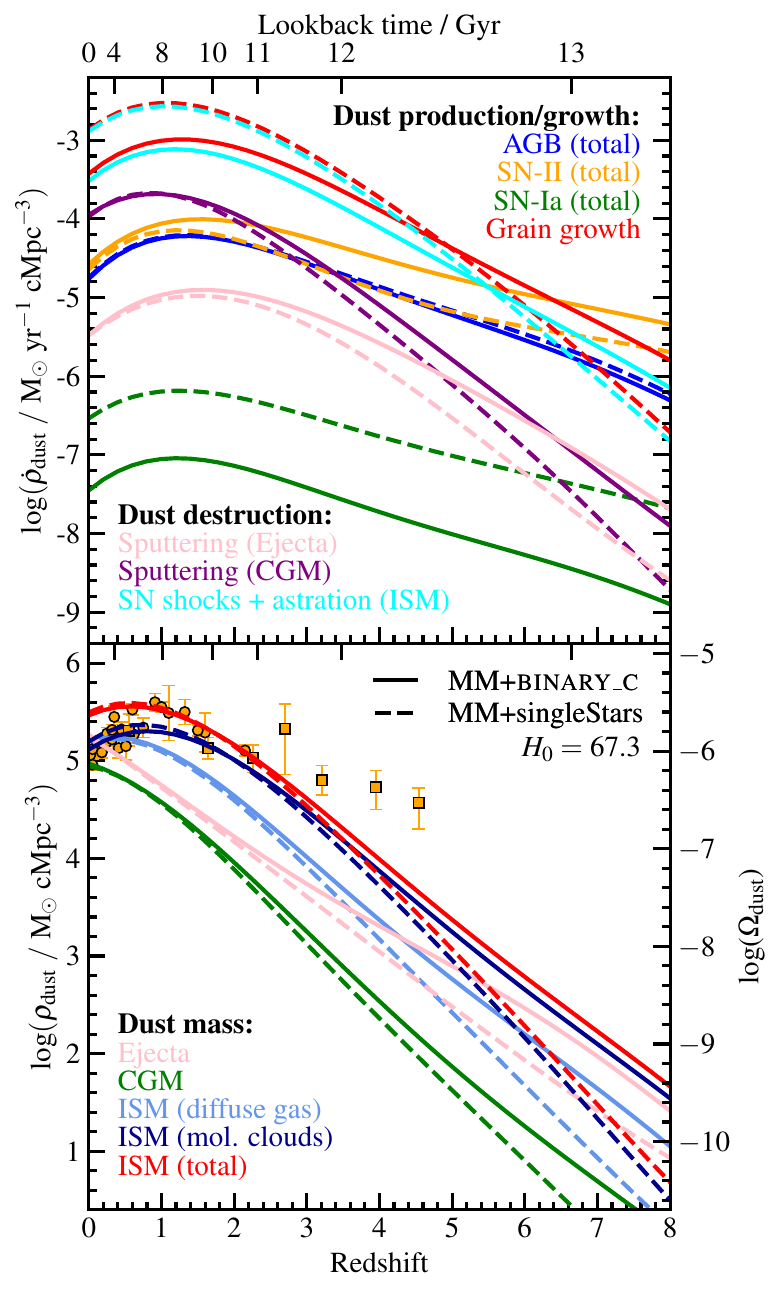}
 \caption{\textit{Top panel:} Evolution of various cosmic dust production and destruction rates from $z=8$ to 0 for the MM+\binaryc{} (solid lines) and MM+singleStars (dashed lines) versions of \lgal{} run on \textsc{Millennium-II}. \textit{Bottom panel:} Evolution of the dust mass density in various galaxy and halo components for the same \lgal{} versions. Observational data is taken from \citet{Dunne+11,Beeston+18,Driver+18,Pozzi+20} (orange circles) and \citet{Peroux&Howk20} (orange squares). In both panels, values at discreet output redshifts have been fit with 5th order polynomials. Dust evolution in the two versions of \lgal{} differs significantly at high $z$, predominantly due to differences in ISM metallicity.}
 \label{fig:dust_evos}
\end{figure}

\subsection{Dust production and destruction rates} \label{sec:Dust rates}
The top panel in Fig.~\ref{fig:dust_evos} shows the evolution of various cosmic dust \correc{growth} and destruction rate densities for MM+\binaryc{} (solid lines) and MM+singleStars (dashed lines) run on \textsc{Millennium-II}. This higher-resolution N-body simulation is preferred here, as it resolves more of the low-mass galaxies which dominate cosmic densities at high $z$ (see \citealt{Yates+21b}). For this reason, there is also no stellar mass cut applied for this analysis. 

We can see significant differences in the evolution of some rates between the two versions of \lgal{}. For example, the evolution of the cosmic destruction rate in the ISM (cyan), which is dominated by SN shock destruction ($\dot{\rho}\sub{shock}$), is flatter in MM+\binaryc{} than in MM+singleStars. This is because, at high $z$, $\dot{\rho}\sub{shock}$ is driven by its dependence on $M\sub{dust}$ (see Eqn.~\ref{eqn:shock_dest_mass}), which is higher at this epoch in MM+\binaryc{} due to higher dust and metal yields from low-metallicity SNe-II (see Section~\ref{sec:Comparing single star and binary population yields}). Whereas at low $z$, $\dot{\rho}\sub{shock}$ is limited by the dust destruction timescale (\ie{}$\tau\sub{shock}$), which is always longer in MM+\binaryc{} due to lower SN rates (see Fig.~\ref{fig:timescales}).

A similar trend is seen for the cosmic grain growth rate ($\dot{\rho}\sub{acc}$, red). At high $z$, there are more galaxies with high metal and dust masses in MM+\binaryc{}, leading to shorter accretion timescales (see Eqn.~\ref{eqn:tau_acc}) and thus higher $\dot{\rho}\sub{acc}$. At low $z$, there is less mass \textit{available} for accretion onto dust grains in MM+\binaryc{}, due to the lower dust destruction rates described above. In other words, grain growth saturates sooner in MM+\binaryc{} than MM+singleStars. Indeed, when turning-off dust destruction in \lgal{}, we find that $\dot{\rho}\sub{acc}$ is very similar between MM+\binaryc{} and MM+singleStars, indicating that grain growth is limited by $\tau\sub{shock}$ at low $z$. In addition, accretion timescales are slightly longer in MM+\binaryc{} by $z=0$, due to lower DTG ratios (see Fig.~\ref{fig:timescales}). This coupled evolution between the main dust production and destruction mechanisms nicely demonstrates the rapid cycling of dust in the ISM occurring in \lgal{}.

The cosmic SN-II dust production rate (orange) is higher at all epochs in MM+\binaryc{} due to its dependence on the mass of metals ejected by SNe-II, which is higher than in MM+singleStars. Conversely, the SN-Ia dust production rate (green) is significantly lower in MM+\binaryc{} due to the under-estimation of the SN-Ia rate in \binaryc{} (see Section~\ref{sec:Comparing single star and binary population yields}).

We also find that the \textit{total} dust production and destruction rates in MM+\binaryc{} also exceed those in MM+singleStars above $z\sim{}4.4$ and $\sim{}5.4$, respectively.

In both versions of \lgal{}, grain growth overtakes SNe-II as the dominant dust production mechanism at $z\sim{}6.4$. This ``crossover redshift'' is later than predicted by most other cosmological simulations (\eg{}\citealt{Popping+17,Vijayan+19,Graziani+20,Dayal+22,Parente+23}), likely due to our less-efficient early grain growth. A notable exception is \textsc{DustySAGE} \citep{Triani+20}, in which a crossover redshift of $z\sim{}2.2$ was found, due to lower grain growth efficiencies and a metallicity-dependent dust destruction efficiently (see \citealt{Popping&Peroux22}).

The bottom panel of Fig.~\ref{fig:dust_evos} compares \lgal{} to observational measurements of the cosmic dust mass density, $\rho\sub{dust}$, from SED fitting (orange circles, \citealt{Dunne+11,Beeston+18,Driver+18,Pozzi+20}) and neutral gas absorption lines in QSO-DLAs (orange squares, \citealt{Peroux&Howk20}). At low $z$, \correc{the total cosmic dust mass begins to level-off in both versions of \lgal{}, but does not decrease. This is because dust continues to build-up in the CGM and ejecta reservoirs surrounding galaxies, with the latter actually becoming the dominant dust reservoir by $z=0$. However, there \textit{is} a net decrease in the ISM dust mass at late times, due to the combination of decreased cosmic star formation (thus decreased stellar dust production) and decreased ISM gas reservoirs (thus decreased grain growth), as discussed by \citet{Yates+21b}.} This leads to a drop in $\rho\sub{dust}$ within molecular clouds (dark blue) by a factor of $\sim{}1.6$ from $z\sim{}1$ to 0. However, this factor is less than the factor of $3-4$ observed from the attenuation measurements using SED fitting.

To help reconcile this difference, \citet{Parente+23} have proposed allowing more efficient SMBH growth, and thus stronger AGN feedback, in simulations, to reduce star formation at late times and thus cause a stronger drop in $\rho\sub{dust}$. However, in our case, this could also lead to an under-estimation of dust masses at $z=0$, which match observations well (Fig. \ref{fig:dust_scaling_relations_z0}). Alternatively, an \textit{increase} in star-formation at $z\gtrsim{}1$ would cause a similar drop in $\rho\sub{dust}$ by boosting dust production at earlier times. The increased gas-cooling rates this requires would also stimulate dust production while leaving unaffected the good match \lgal{} has to observed DTG ratios back to $z\sim{}4$ (M. Parente, priv. comm.). Indeed, larger gas reservoirs in resolved subhaloes at high $z$ are already required in semi-analytic simulations to better match the observed cosmic H\textsc{i} density (see \citealt{Yates+21b}), as well as the cosmic SFR density from sub-mm and radio observations. Therefore, further investigation into the viability of such a modification in \lgal{} will be the focus of future work.

Finally, at high $z$, \lgal{} returns significantly lower $\rho\sub{dust}$ than is implied from DLA observations (orange  squares). This discrepancy could be even larger if very-dust-rich DLAs, which are typically missed from optically-selected QSO samples, are also considered \citep{Krogager+19}. Part of this discrepancy could be mitigated by the increased gas cooling and star formation discussed above. Increased dust production, including from additional \correc{binary} sources such as CE ejection and novae, \correc{or superluminous SNe \citep{Chen+23},} could also help. However, a residual offset in the cosmic metal mass density between \lgal{} and QSO-DLAs was also found by \citet{Yates+21b}, even when accounting for the simulation's gas deficit in resolved subhaloes. That work determined that this remaining difference could be reconciled if current high-$z$ DLA metallicity samples are biased towards galaxies with $\logMm{}\gtrsim{}7.5$. It is possible that such a mass bias could also play a role in an over-estimation of $\rho\sub{dust}$ here.

\begin{figure*}
 \centering
    \includegraphics[width=0.99\linewidth]{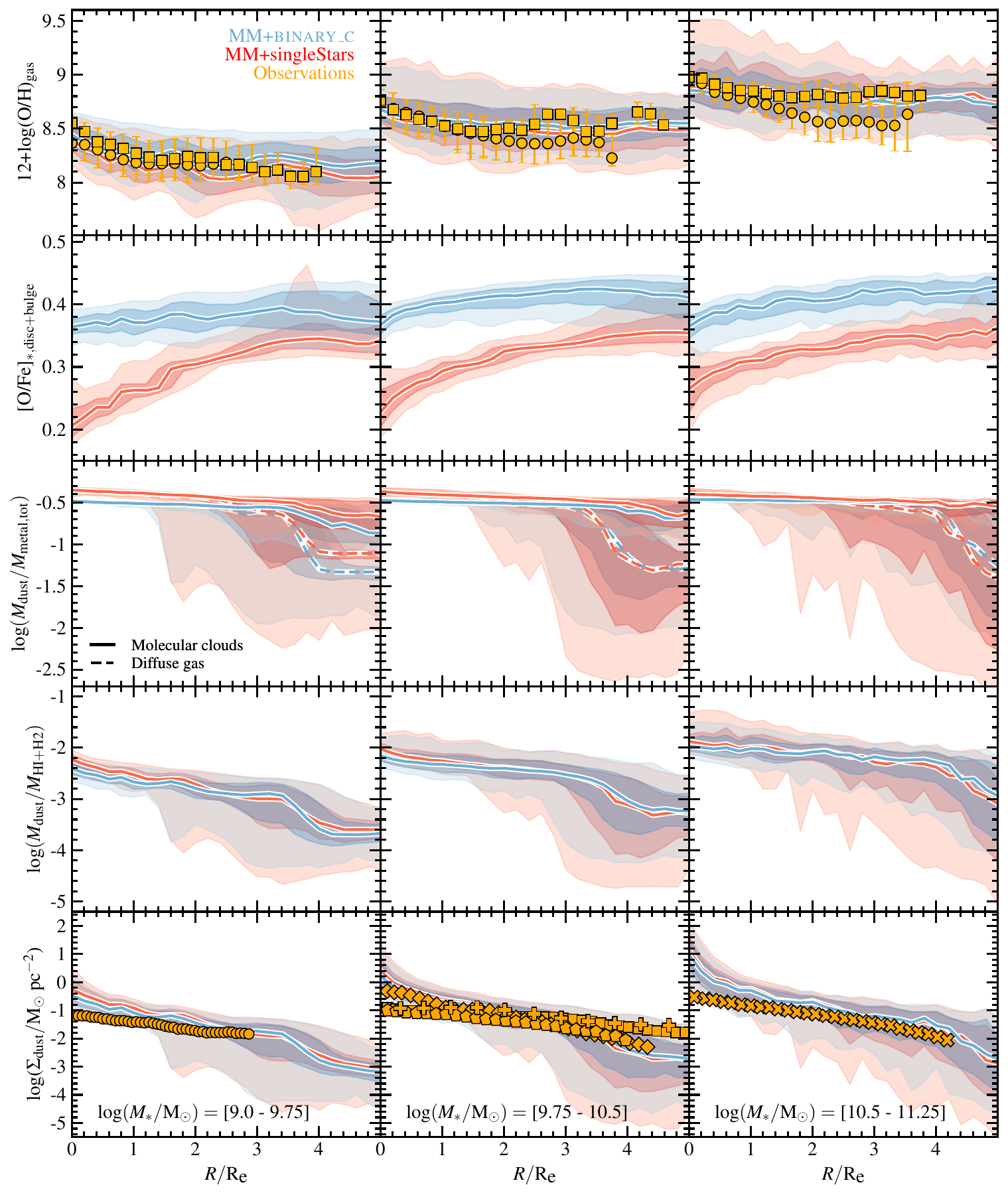}
 \caption{Stacked radial profiles for star-forming galaxies at $z=0$ from the MM+\binaryc{} (blue) and MM+singleStars (red) versions of \lgal{} run on \textsc{Millennium}, separated into three mass bins. \textit{First row:} Gas-phase oxygen abundance (excluding oxygen depleted onto dust). Observational data is from \citet{Yates+21a} (circles) and \citet{Erroz-Ferrer+19}; Easeman et al. (in prep.) (squares). \textit{Second row:} Oxygen enhancement in disc + bulge stars normalised to the solar photospheric value of $(M\sub{O}/M\sub{Fe})\sub{\astrosun} = 4.44$ from \citet{Asplund+09}. \textit{Third row:} The DTM ratio in the ISM for molecular clouds (solid lines) and diffuse gas (dashed lines). \textit{Fourth row:} The DTG ratio. \textit{Fifth row:} Dust mass surface density, with observational data of six galaxies from the DustPedia dataset of \citet{Casasola+17}. Both versions of \lgal{} match observational data well, and predict a significant drop in DTM and DTG ratios at large radii.}
 \label{fig:radial_profiles_z0_twoModels}
\end{figure*}

\subsection{Radial profiles} \label{sec:Radial profiles}
We now turn to studying the spatial distributions of dust and metals in the discs of massive star-forming galaxies at $z=0$. For this, we select model galaxies with $\logMm{} \geq 9.0$, $M\sub{bulge}/M_{*\tn{,tot}} < 0.3$, and $\tn{log}(\tn{sSFR/yr}^{-1}) \geq{} -10.9$ from the MM+\binaryc{} and MM+singleStars versions of \lgal{} run on \textsc{Millennium}. This larger N-body simulation is preferred here, as it provides better statistics while also being able to comfortably resolve higher-mass galaxies. The effective radius, $R\sub{e}$, is measured as the half light radius in the r-band for \lgal{}, to best match that measured in the observations to which we compare. \correc{We also note that ``local'' processes, such as star formation, dust production and destruction, and SN feedback, are calculated independently in each radial ring in \lgal{}. This means that radial profiles are allowed to evolve in an unparameterised way (\ie{}we do not impose any particular radial distribution function). However, we do assume an exponential radial profile for the accretion of gas and dust onto galaxy discs (see \citealt[][section 2.2.1]{Henriques+20}).}

The top row of Fig.~\ref{fig:radial_profiles_z0_twoModels} shows the median ISM gas-phase oxygen abundance profiles (excluding oxygen in dust) for star-forming disc galaxies at $z=0$ in three bins of stellar mass. The observational data shown (orange points) are from the MaNGA sample of \citet{Yates+21a} and the MUSE sample of \citep{Erroz-Ferrer+19} with spaxel-based metallicities re-calculated by Easeman et al. (in prep.). In both cases, the \citet{Dopita+16} strong-line metallicity diagnostic is used, as this best reproduces electron-temperature metallicities in low-mass galaxies, super-solar metallicities in very high mass galaxies, and has a negligible secondary dependence on ionisation parameter (see \citealt{Yates+21a}, section 3.2).

We find very good agreement between both \lgal{} models and the high-resolution MUSE data, in all mass bins. In \citet{Yates+21a}, we reported that the \lgal{} profiles are too flat at large radii in low-mass galaxies compared to this MUSE sample. However, this discrepancy is now greatly reduced by the careful re-analysis and removal of diffuse ionised gas (DIG) dominated spaxels from these data (Easeman et al. in prep.). This provides a purer sample of H\textsc{II} region measurements, which return higher metallicities at large radii in low-mass galaxies.

At high mass, the two versions of \lgal{} also exhibit similar 12+log(O/H) profiles to each other. However, at low mass, MM+\binaryc{} returns higher 12+log(O/H) (by up to $\sim{}0.15$ dex), particularly at large galactocentric radii. Similar trends are seen at large radii in the stellar metallicity profiles (not shown). This extra oxygen is a consequence of the smaller remnant masses predicted by the yield sets used in \binaryc{} at low metallicity (see Section \ref{sec:General galaxy properties}). We note that the strongest suppression of oxygen from the \citet{Portinari+98} yields used in MM+singleStars occurs below $Z\sim{}0.004$, which is roughly equivalent to $\tn{12+log(O/H)} < 8.4$.

Conversely, the gas-phase [Fe/H] profiles (not shown) exhibit an under-abundance of iron in MM+\binaryc{} compared to MM+singleStars, particularly at low radii. This is due to the lower SN-Ia rates in MM+\binaryc{}. The combination of these two factors (namely, more oxygen and less iron in the MM+\binaryc{} model) is most clearly seen in the second row of Fig.~\ref{fig:radial_profiles_z0_twoModels}, which shows the median azimuthally-averaged [O/Fe] profiles for bulge plus (thick + thin) disc stars of all ages. These profiles are flatter in MM+\binaryc{}, with a normalisation that is higher than expected in, for example, star-forming disc galaxies of Milky Way mass (see \eg{}\citealt{Hayden+15,Rojas-Arriagada+19}). This result suggests that measuring the $\alpha$ enhancements in the centres of disc galaxies could provide useful constraints on stellar yields.

The third row of Fig.~\ref{fig:radial_profiles_z0_twoModels} shows the DTM profiles for molecular clouds (solid lines) and diffuse gas (dashed lines) in the ISM. In both \lgal{} versions, the diffuse gas DTM ratio drops at larger radii. This indicates that the common assumption of a universal value (see \eg{}\citealt{Remy-Ruyer+15}) does not hold on sub-galactic scales. We find that the degree of this drop depends on (a) the efficiency of grain growth, and (b) the amount of dust that is able to re-accrete onto the ISM after being ejected by galactic outflows. A similar conclusion was made by \citet{DeVis+19} when interpreting DustPedia data. In the extreme case where no dust is allowed to survive in the CGM, values for the DTM and DTG in the outer disc can become very low in \lgal{}, as re-accretion cannot supplement the low in-situ dust formation rates at these radii. Central DTM ratios are also lower in the MM+\binaryc{} model than the MM+singleStars model, due to lower dust masses. This dust deficit is what drives the lower global DTM ratios in the MM+\binaryc{} model seen in Figs.~\ref{fig:dust_scaling_relations_z0} and \ref{fig:dtm_and_dtg_evos}.

\begin{figure}
 \centering
 \includegraphics[width=0.85\linewidth]{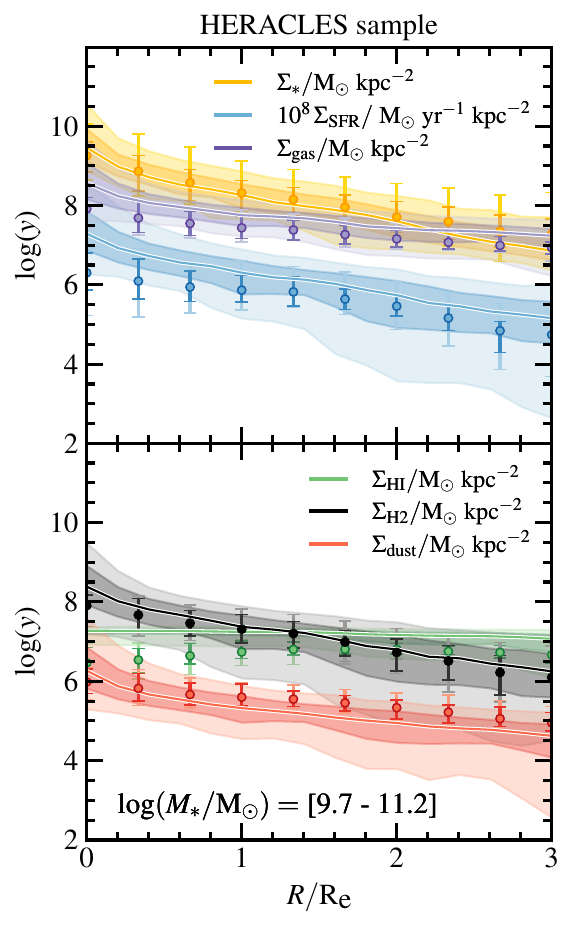}
 \caption{Radial profiles from the HERACLES sample \citep[][points]{Abdurrouf+22a} and star-forming disc galaxies within the same mass range [$9.7 \leq{} \logMm < 11.2$] from the MM+\binaryc{} version of \lgal{} (lines) run on \textsc{Millennium}. MM+\binaryc{} matches the observed profiles very well, albeit with a slight excess of SFR at low radii.}
 \label{fig:profs_HERACLES}
\end{figure}

We also note that the apparent difference between the average molecular and diffuse DTM profiles at large radii is due to averaging effects when stacking profiles. In \lgal{}, the few galaxies which have molecular gas at large radii are relatively evolved and therefore dust rich, whereas galaxies with only diffuse gas at large radii are relatively dust poor.

The fourth row of Fig.~\ref{fig:radial_profiles_z0_twoModels} shows DTG profiles for the same model disc galaxies. There is a clear decrease in DTG ratio with radius for all galaxies, including intermediate-mass systems which only exhibit the commonly-assumed universal DTG ratio of $\sim{}0.01$ in their innermost regions. There are also slightly lower DTG ratios at low radii in MM+\binaryc{} than MM+singleStars, due to lower central dust masses. The radial profiles of general properties such as H\textsc{i}, H$_{2}$, SFR, and $M_{*}$ densities are very similar in both versions of \lgal{}, reflecting their similar global values shown in Fig.~\ref{fig:gen_gal_props}.

Finally, the fifth row of Fig.~\ref{fig:radial_profiles_z0_twoModels} shows dust mass surface density, $\Sigma{}\sub{dust}$, profiles for \lgal{}. These are compared to six DustPedia galaxies from the sample of \citet{Casasola+17} for which effective radii could be obtained from the literature, namely NGC 2403 (circles), NGC 5457 (squares), NGC 5194 (diamonds), NGC 628 (pentagons), NGC 4736 (pluses), and HGC 5055 (crosses). Spatially resolved properties were obtained by the DustPedia team for these galaxies using multi-wavelength photometric maps.

There is a very good qualitative agreement between the observed $\Sigma{}\sub{dust}$ profiles and those from \lgal{} beyond $R/R\sub{e} \sim{} 1$. The flatter inner $\Sigma{}\sub{dust}$ profiles found in most DustPedia galaxies appear in conflict with the steeper central oxygen abundance profiles seen in MaNGA and MUSE data (top row of Fig.~\ref{fig:radial_profiles_z0_twoModels}). This could suggest either enhanced dust destruction (possibly by AGN feedback \correc{or older stellar populations in the bulge}) or suppressed star formation and thus dust production in galaxy centres.

A further comparison to spatially resolved observations is shown in Fig.~\ref{fig:profs_HERACLES} where we compare the MM+\binaryc{} model at $z=0$ (lines) to stacked radial profiles from the ten HERACLES/THINGS galaxies presented by \citet[][points]{Abdurrouf+22a}. Similar results are found for the MM+singleStars model. There is remarkable agreement in both slope and normalisation for the $\Sigma_{*}$, $\Sigma\sub{H2}$, and $\Sigma\sub{dust}$ profiles between the observations and \lgal{}. However, the central SFR and H\textsc{i} densities are slightly over-predicted in \lgal{} compared to these observations. As for the comparison to DustPedia galaxies above, this could be due to $\sim{}50$\% of the HERACLES sample hosting AGN, noting that most of their AGN hosts have flattened SFR profiles. Direct suppression of star formation via ISM gas removal from AGN feedback is not currently modelled in \lgal{}, which instead only allows AGN to indirectly suppress star formation by reducing further cooling onto the disc.

\section{Conclusions} \label{sec:Conclusions}
In this work, we implement models for (a) binary stellar evolution (BSE) using the \binaryc{} code and (b) dust production and destruction into the cosmological semi-analytic galaxy evolution simulation, \lgal{}. This opens up a whole new range of studies for cosmological-scale simulations, including comparisons to the growing plethora of observations of chemical abundances and dust properties in galaxies at low and high $z$. We hope that this work will serve as a template for implementing BSE sub-grid models into other galaxy evolution simulations in future.

When comparing stellar populations produced by \binaryc{} to those from a simpler single-star plus SNe-Ia set-up (Section \ref{sec:Comparing single star and binary population yields}), we find that the total mass ejected by massive stars is relatively unchanged. This is due to the similar assumptions made about explodability in both set-ups. However, the \textit{metal} mass ejected by massive stars is slightly larger in \binaryc{}. This is partly due to the more efficient mass loss in binary systems, but mostly due to the different SN-II yields assumed between the set-ups. \textsc{Binary\_c} also predicts significant ejection of light alpha elements (especially C) during the common envelope phase -- a process which is not possible when considering only single stars. However, the ejection of heavier elements such as iron from SNe-Ia is significantly under-predicted by \binaryc{}, due to an under-estimation of the SN-Ia rate. This issue is common to stellar evolution codes which attempt to model this process from first principles.

Our model for dust production and destruction is based on that developed by \citetalias{Vijayan+19}, but with the following key improvements: the inclusion of dust survival in the hot gas surrounding galaxies and its gradual destruction via thermal sputtering, the radial resolution of dust within the ISM, improvements to the SN shock destruction modelling, and full synchronisation with the \lgal{} HI-H$_{2}$ partitioning model. The new dust model also has important differences to those commonly used in other cosmological simulations, for example by allowing a wider range of accretion timescales during grain growth (Section \ref{sec:Dust model}).

With these models in place, we obtain the following results from two versions of \lgal{}, one including binary stars (MM+\binaryc{}) and one including only single stars (MM+singleStars):
\begin{itemize}
    \item The inclusion of binary stars into \lgal{} does not have a significant impact on key galaxy properties such as stellar mass, gas mass, and star formation rate. This is because the amount of mass and energy ejected by massive stars remains relatively unchanged compared to the single-star-only approximation. The mass ejected by massive stars determines the strength of the SN feedback in \lgal{}, which in turn drives gas ejection rates and consequently star formation and stellar mass growth (Section \ref{sec:General galaxy properties}).
    \item MM+\binaryc{} exhibits higher gas-phase oxygen abundances in low-mass galaxies than MM+singleStars. This is predominantly due to differences in the low-metallicity SN-II yields assumed, rather than binary effects. This additional oxygen is typically found in the outer disc of star-forming galaxies by $z=0$ (see Section \ref{sec:General galaxy properties}).
    \item \correc{Binary effects do impact the ejection of some lighter elements. For example, CE ejection, novae, and SNe-Ia help boost nitrogen yields at very early or very late times at low metallicities.}
    \item \lgal{} matches observed dust properties back to $z\sim{}4$, with slightly better agreement for MM+\binaryc{}. An evolution in the DTM and DTG ratios of metal-poor galaxies is also present, in qualitative agreement with that seen in GRB-DLA samples. These successes are largely due to the shorter accretion timescales in dustier environments predicted by our grain growth formalism, thus allowing more efficient dust production at late times (Section \ref{sec:Dust scaling realtions}).
    \item At high $z$, dust production and destruction rates are significantly higher in MM+\binaryc{} than MM+singleStars. This is due to higher dust masses and shorter accretion timescales at early times. These rates then drop below those found in MM+singleStars at low $z$, due to longer destruction timescales. In general, there is a coupled evolution between the main dust production and destruction mechanisms in \lgal{} as the key timescales and dust reservoirs change over time (see Section \ref{sec:Dust rates}).
    \item Grain growth overtakes SNe-II as the dominant dust production mechanism at $z\sim{}6.4$ in both versions of \lgal{}. This suggests that the balance between the main sources of dust production and destruction in galaxies is relatively independent of the details of stellar evolution (Section \ref{sec:Dust rates}).
    \item On sub-galactic scales, there is very good overall agreement between \lgal{} and observed dust and metal radial profiles at $z=0$. However, dust observations suggest slightly flatter $\Sigma{}\sub{dust}$ profiles in the centres of galaxies (\ie{}$\lesssim{}1 R\sub{e}$), in contrast to both \lgal{} and observed O/H profiles. This could suggest either enhanced dust destruction or suppressed dust production at low radii, perhaps associated with the presence of AGN \correc{or stellar bulge} (Section \ref{sec:Radial profiles}).
    \item The DTG ratio in the diffuse ISM drops by $\sim{}0.8$ dex at large radii (\ie{}$\gtrsim{}3 R\sub{e}$) on average in \lgal{}, particularly in low-mass (low-metallicity) systems and when binary stars are included. This demonstrates that the assumption of a constant DTM ratio does not hold on sub-galactic scales. The degree of this drop in DTM depends on the grain growth efficiency in molecular clouds and the amount of dust that is able to survive in the CGM and re-accrete onto galaxy discs at large radii (Section \ref{sec:Radial profiles}).
\end{itemize}

There are a number of avenues of further investigation made possible by this work. For example, the effect of modifying the standard parameters in \binaryc{} away from their default values will be studied, in particular the effect of more complex prescriptions for the fate of massive stars, SN-Ia progenitor evolution, and higher-order multiple systems. The evolution of more exotic chemical species, such as s- and r-process elements, individual isotopes, and rare elements of interest to Milky Way studies, will also be presented in follow-up papers. The potential for enhanced gas cooling at higher $z$ to reconcile a number of inconsistencies between \lgal{} and observations will also be investigated, as well as additional dust production and destruction mechanisms \correc{and the dependence of H$_{2}$ formation on the dust mass}. Finally, we highlight that the choice of stellar yields always has a significant impact on the chemical evolution of galaxies seen in simulations. In this work, we have chosen yield sets that are commonly used in the literature, to facilitate a more meaningful comparison between old and new versions of \lgal{}, and to other simulations. However, the inclusion of more detailed chemical yields will be the main the focus of future work.

\section*{Acknowledgements}
\correc{The authors would like to thank the referee for their helpful and insightful comments,} as well as Kasper Heintz, Chiaki Kobayashi, Zara Osborn, Massimiliano Parente, and Patricia Schady for valuable discussions during the undertaking of this work. We also thank Abdurro'uf, Viviana Casasola, Bethan Easeman, and Kasper Heintz for sharing their observational data. APV acknowledges support from the Carlsberg Foundation (grant no CF20-0534). The Cosmic Dawn Center (DAWN) is funded by the Danish National Research Foundation under grant No. 140.

\section*{Data Availability}
\correc{The full source code for the \lgal{} versions used in this work are publicly available via \href{https://github.com/LGalaxiesPublicRelease/LGalaxies2020_PublicRepository/tree/Yates2023}{GitHub} (see the \href{https://lgalaxiespublicrelease.github.io/index.html\#DOWNLOADS}{\lgal{} website} for more details, including installation and running instructions). Associated \lgal{} output catalogues can be directly obtained via \href{https://doi.org/10.5281/zenodo.10019005}{Zenodo}. The \binaryc{} ensembles used (in JSON format), including the \textsc{python} script used to generate \lgal{} yield tables from them, are also available via the \lgal{} GitHub and Zenodo repositories. Additional simulation and observation data products derived by the authors and presented here can be obtained upon request.}


\bibliographystyle{mnras}
\bibliography{robyates.bib,collaborators.bib}

\begin{thebibliography}{}
\makeatletter
\relax
\def\mn@urlcharsother{\let\do\@makeother \do\$\do\&\do\#\do\^\do\_\do\%\do\~}
\def\mn@doi{\begingroup\mn@urlcharsother \@ifnextchar [ {\mn@doi@}
  {\mn@doi@[]}}
\def\mn@doi@[#1]#2{\def\@tempa{#1}\ifx\@tempa\@empty \href
  {http://dx.doi.org/#2} {doi:#2}\else \href {http://dx.doi.org/#2} {#1}\fi
  \endgroup}
\def\mn@eprint#1#2{\mn@eprint@#1:#2::\@nil}
\def\mn@eprint@arXiv#1{\href {http://arxiv.org/abs/#1} {{\tt arXiv:#1}}}
\def\mn@eprint@dblp#1{\href {http://dblp.uni-trier.de/rec/bibtex/#1.xml}
  {dblp:#1}}
\def\mn@eprint@#1:#2:#3:#4\@nil{\def\@tempa {#1}\def\@tempb {#2}\def\@tempc
  {#3}\ifx \@tempc \@empty \let \@tempc \@tempb \let \@tempb \@tempa \fi \ifx
  \@tempb \@empty \def\@tempb {arXiv}\fi \@ifundefined
  {mn@eprint@\@tempb}{\@tempb:\@tempc}{\expandafter \expandafter \csname
  mn@eprint@\@tempb\endcsname \expandafter{\@tempc}}}

\bibitem[\protect\citeauthoryear{{Abdurro'uf,}, {Lin}, {Hirashita},
  {Morishita}, {Tacchella}, {Akiyama}, {Takeuchi}  \& {Wu}}{{Abdurro'uf,}
  et~al.}{2022}]{Abdurrouf+22a}
{Abdurro'uf,} {Lin} Y.-T.,  {Hirashita} H.,  {Morishita} T.,  {Tacchella} S.,
  {Akiyama} M.,  {Takeuchi} T.~T.,   {Wu} P.-F.,  2022, \mn@doi [\apj]
  {10.3847/1538-4357/ac439a}, \href
  {https://ui.adsabs.harvard.edu/abs/2022ApJ...926...81A} {926, 81}

\bibitem[\protect\citeauthoryear{{Angulo} \& {Hilbert}}{{Angulo} \&
  {Hilbert}}{2015}]{Angulo&Hilbert15}
{Angulo} R.~E.,  {Hilbert} S.,  2015, \mn@doi [MNRAS] {10.1093/mnras/stv050},
  \href {http://adsabs.harvard.edu/abs/2015MNRAS.448..364A} {448, 364}

\bibitem[\protect\citeauthoryear{{Aoyama}, {Hirashita}  \& {Nagamine}}{{Aoyama}
  et~al.}{2020}]{Aoyama+20}
{Aoyama} S.,  {Hirashita} H.,   {Nagamine} K.,  2020, \mn@doi [\mnras]
  {10.1093/mnras/stz3253}, \href
  {https://ui.adsabs.harvard.edu/abs/2020MNRAS.491.3844A} {491, 3844}

\bibitem[\protect\citeauthoryear{{Arellano-C{\'o}rdova}
  et~al.,}{{Arellano-C{\'o}rdova} et~al.}{2022}]{Arellano-Cordova+22}
{Arellano-C{\'o}rdova} K.~Z.,  et~al., 2022, \mn@doi [\apjl]
  {10.3847/2041-8213/ac9ab2}, \href
  {https://ui.adsabs.harvard.edu/abs/2022ApJ...940L..23A} {940, L23}

\bibitem[\protect\citeauthoryear{{Arrigoni}, {Trager}, {Somerville}  \&
  {Gibson}}{{Arrigoni} et~al.}{2010}]{Arrigoni+10a}
{Arrigoni} M.,  {Trager} S.~C.,  {Somerville} R.~S.,   {Gibson} B.~K.,  2010,
  \mn@doi [\mnras] {10.1111/j.1365-2966.2009.15924.x}, \href
  {https://ui.adsabs.harvard.edu/abs/2010MNRAS.402..173A} {402, 173}

\bibitem[\protect\citeauthoryear{{Artale}, {Mapelli}, {Bouffanais}, {Giacobbo},
  {Pasquato}  \& {Spera}}{{Artale} et~al.}{2020}]{Artale+20}
{Artale} M.~C.,  {Mapelli} M.,  {Bouffanais} Y.,  {Giacobbo} N.,  {Pasquato}
  M.,   {Spera} M.,  2020, \mn@doi [\mnras] {10.1093/mnras/stz3190}, \href
  {https://ui.adsabs.harvard.edu/abs/2020MNRAS.491.3419A} {491, 3419}

\bibitem[\protect\citeauthoryear{{Asano}, {Takeuchi}, {Hirashita}  \&
  {Inoue}}{{Asano} et~al.}{2013}]{Asano+13}
{Asano} R.~S.,  {Takeuchi} T.~T.,  {Hirashita} H.,   {Inoue} A.~K.,  2013,
  \mn@doi [Earth, Planets and Space] {10.5047/eps.2012.04.014}, \href
  {https://ui.adsabs.harvard.edu/abs/2013EP&S...65..213A} {65, 213}

\bibitem[\protect\citeauthoryear{{Asplund}, {Grevesse}, {Sauval}  \&
  {Scott}}{{Asplund} et~al.}{2009}]{Asplund+09}
{Asplund} M.,  {Grevesse} N.,  {Sauval} A.~J.,   {Scott} P.,  2009, \mn@doi
  [\araa] {10.1146/annurev.astro.46.060407.145222}, \href
  {http://adsabs.harvard.edu/abs/2009ARA%26A..47..481A} {47, 481}

\bibitem[\protect\citeauthoryear{{Ayromlou}, {Nelson}, {Yates}, {Kauffmann},
  {Renneby}  \& {White}}{{Ayromlou} et~al.}{2021}]{Ayromlou+21a}
{Ayromlou} M.,  {Nelson} D.,  {Yates} R.~M.,  {Kauffmann} G.,  {Renneby} M.,
  {White} S. D.~M.,  2021, \mn@doi [\mnras] {10.1093/mnras/staa4011}, \href
  {https://ui.adsabs.harvard.edu/abs/2021MNRAS.502.1051A} {502, 1051}

\bibitem[\protect\citeauthoryear{{Baldry}, {Glazebrook}  \& {Driver}}{{Baldry}
  et~al.}{2008}]{Baldry+08}
{Baldry} I.~K.,  {Glazebrook} K.,   {Driver} S.~P.,  2008, \mn@doi [\mnras]
  {10.1111/j.1365-2966.2008.13348.x}, \href
  {https://ui.adsabs.harvard.edu/abs/2008MNRAS.388..945B} {388, 945}

\bibitem[\protect\citeauthoryear{{Baldry} et~al.,}{{Baldry}
  et~al.}{2012}]{Baldry+12}
{Baldry} I.~K.,  et~al., 2012, \mn@doi [\mnras]
  {10.1111/j.1365-2966.2012.20340.x}, \href
  {https://ui.adsabs.harvard.edu/abs/2012MNRAS.421..621B} {421, 621}

\bibitem[\protect\citeauthoryear{{Beeston} et~al.,}{{Beeston}
  et~al.}{2018}]{Beeston+18}
{Beeston} R.~A.,  et~al., 2018, \mn@doi [\mnras] {10.1093/mnras/sty1460}, \href
  {https://ui.adsabs.harvard.edu/abs/2018MNRAS.479.1077B} {479, 1077}

\bibitem[\protect\citeauthoryear{{Boylan-Kolchin}, {Springel}, {White},
  {Jenkins}  \& {Lemson}}{{Boylan-Kolchin} et~al.}{2009}]{Boylan-Kolchin+09}
{Boylan-Kolchin} M.,  {Springel} V.,  {White} S.~D.~M.,  {Jenkins} A.,
  {Lemson} G.,  2009, \mn@doi [MNRAS] {10.1111/j.1365-2966.2009.15191.x}, \href
  {http://adsabs.harvard.edu/abs/2009MNRAS.398.1150B} {398, 1150}

\bibitem[\protect\citeauthoryear{{Brinchmann}}{{Brinchmann}}{2023}]{Brinchmann23}
{Brinchmann} J.,  2023, \mn@doi [\mnras] {10.1093/mnras/stad1704}, \href
  {https://ui.adsabs.harvard.edu/abs/2023MNRAS.525.2087B} {525, 2087}

\bibitem[\protect\citeauthoryear{{Brinchmann}, {Charlot}, {Kauffmann},
  {Heckman}, {White}  \& {Tremonti}}{{Brinchmann} et~al.}{2013}]{Brinchmann+13}
{Brinchmann} J.,  {Charlot} S.,  {Kauffmann} G.,  {Heckman} T.,  {White} S.
  D.~M.,   {Tremonti} C.,  2013, \mn@doi [\mnras] {10.1093/mnras/stt551}, \href
  {https://ui.adsabs.harvard.edu/abs/2013MNRAS.432.2112B} {432, 2112}

\bibitem[\protect\citeauthoryear{{Cameron}, {Katz}  \& {Rey}}{{Cameron}
  et~al.}{2023}]{Cameron+23}
{Cameron} A.~J.,  {Katz} H.,   {Rey} M.~P.,  2023, \mn@doi [\mnras]
  {10.1093/mnrasl/slad046}, \href
  {https://ui.adsabs.harvard.edu/abs/2023MNRAS.522L..89C} {522, L89}

\bibitem[\protect\citeauthoryear{{Casasola} et~al.,}{{Casasola}
  et~al.}{2017}]{Casasola+17}
{Casasola} V.,  et~al., 2017, \mn@doi [\aap] {10.1051/0004-6361/201731020},
  \href {https://ui.adsabs.harvard.edu/abs/2017A&A...605A..18C} {605, A18}

\bibitem[\protect\citeauthoryear{{Casasola} et~al.,}{{Casasola}
  et~al.}{2022}]{Casasola+22}
{Casasola} V.,  et~al., 2022, \mn@doi [\aap] {10.1051/0004-6361/202245043},
  \href {https://ui.adsabs.harvard.edu/abs/2022A&A...668A.130C} {668, A130}

\bibitem[\protect\citeauthoryear{{Cavaliere} \& {Fusco-Femiano}}{{Cavaliere} \&
  {Fusco-Femiano}}{1976}]{Cavaliere&Fusco-Femiano76}
{Cavaliere} A.,  {Fusco-Femiano} R.,  1976, \aap, \href
  {https://ui.adsabs.harvard.edu/abs/1976A&A....49..137C} {49, 137}

\bibitem[\protect\citeauthoryear{{Chabrier}}{{Chabrier}}{2003}]{Chabrier03}
{Chabrier} G.,  2003, \mn@doi [\pasp] {10.1086/376392}, \href
  {http://adsabs.harvard.edu/abs/2003PASP..115..763C} {115, 763}

\bibitem[\protect\citeauthoryear{{Chen} et~al.,}{{Chen} et~al.}{2021}]{Chen+23}
{Chen} T.~W.,  et~al., 2021, \mn@doi [arXiv e-prints]
  {10.48550/arXiv.2109.07942}, \href
  {https://ui.adsabs.harvard.edu/abs/2021arXiv210907942C} {p. arXiv:2109.07942}

\bibitem[\protect\citeauthoryear{{Chieffi} \& {Limongi}}{{Chieffi} \&
  {Limongi}}{2004}]{Chieffi&Limongi04}
{Chieffi} A.,  {Limongi} M.,  2004, \mn@doi [\apj] {10.1086/392523}, \href
  {https://ui.adsabs.harvard.edu/abs/2004ApJ...608..405C} {608, 405}

\bibitem[\protect\citeauthoryear{{Chruslinska}, {Belczynski}, {Klencki}  \&
  {Benacquista}}{{Chruslinska} et~al.}{2018}]{Chruslinska+18}
{Chruslinska} M.,  {Belczynski} K.,  {Klencki} J.,   {Benacquista} M.,  2018,
  \mn@doi [\mnras] {10.1093/mnras/stx2923}, \href
  {https://ui.adsabs.harvard.edu/abs/2018MNRAS.474.2937C} {474, 2937}

\bibitem[\protect\citeauthoryear{{Claeys}, {Pols}, {Izzard}, {Vink}  \&
  {Verbunt}}{{Claeys} et~al.}{2014}]{Claeys+14}
{Claeys} J.~S.~W.,  {Pols} O.~R.,  {Izzard} R.~G.,  {Vink} J.,   {Verbunt}
  F.~W.~M.,  2014, \mn@doi [\aap] {10.1051/0004-6361/201322714}, \href
  {https://ui.adsabs.harvard.edu/abs/2014A&A...563A..83C} {563, A83}

\bibitem[\protect\citeauthoryear{{Collacchioni}, {Cora}, {Lagos}  \&
  {Vega-Mart{\'\i}nez}}{{Collacchioni} et~al.}{2018}]{Collacchioni+18}
{Collacchioni} F.,  {Cora} S.~A.,  {Lagos} C. D.~P.,   {Vega-Mart{\'\i}nez}
  C.~A.,  2018, \mn@doi [\mnras] {10.1093/mnras/sty2347}, \href
  {https://ui.adsabs.harvard.edu/abs/2018MNRAS.481..954C} {481, 954}

\bibitem[\protect\citeauthoryear{{C{\^o}t{\'e}} et~al.,}{{C{\^o}t{\'e}}
  et~al.}{2018}]{Cote+18}
{C{\^o}t{\'e}} B.,  et~al., 2018, \mn@doi [\apj] {10.3847/1538-4357/aaad67},
  \href {https://ui.adsabs.harvard.edu/abs/2018ApJ...855...99C} {855, 99}

\bibitem[\protect\citeauthoryear{{C{\^o}t{\'e}} et~al.,}{{C{\^o}t{\'e}}
  et~al.}{2019}]{Cote+19}
{C{\^o}t{\'e}} B.,  et~al., 2019, \mn@doi [\apj] {10.3847/1538-4357/ab10db},
  \href {https://ui.adsabs.harvard.edu/abs/2019ApJ...875..106C} {875, 106}

\bibitem[\protect\citeauthoryear{{Curti} et~al.,}{{Curti}
  et~al.}{2023}]{Curti+23}
{Curti} M.,  et~al., 2023, \mn@doi [\mnras] {10.1093/mnras/stac2737}, \href
  {https://ui.adsabs.harvard.edu/abs/2023MNRAS.518..425C} {518, 425}

\bibitem[\protect\citeauthoryear{\DA{Cunha} {Da}{da}~Cunha et~al.,}{\DA{Cunha}
  {Da}{da}~Cunha et~al.}{2015}]{daCunha+15}
\DA{Cunha} {Da}{da}~Cunha E.,  et~al., 2015, \mn@doi [\apj]
  {10.1088/0004-637X/806/1/110}, \href
  {https://ui.adsabs.harvard.edu/abs/2015ApJ...806..110D} {806, 110}

\bibitem[\protect\citeauthoryear{\DE{Bennassuti} {De}{de}~Bennassuti,
  {Schneider}, {Valiante}  \& {Salvadori}}{\DE{Bennassuti} {De}{de}~Bennassuti
  et~al.}{2014}]{deBennassuti+14}
\DE{Bennassuti} {De}{de}~Bennassuti M.,  {Schneider} R.,  {Valiante} R.,
  {Salvadori} S.,  2014, \mn@doi [\mnras] {10.1093/mnras/stu1962}, \href
  {https://ui.adsabs.harvard.edu/abs/2014MNRAS.445.3039D} {445, 3039}

\bibitem[\protect\citeauthoryear{{D'Souza}, {Vegetti}  \&
  {Kauffmann}}{{D'Souza} et~al.}{2015}]{D'Souza+15}
{D'Souza} R.,  {Vegetti} S.,   {Kauffmann} G.,  2015, \mn@doi [\mnras]
  {10.1093/mnras/stv2234}, \href
  {https://ui.adsabs.harvard.edu/abs/2015MNRAS.454.4027D} {454, 4027}

\bibitem[\protect\citeauthoryear{{Dav{\'e}}, {Angl{\'e}s-Alc{\'a}zar},
  {Narayanan}, {Li}, {Rafieferantsoa}  \& {Appleby}}{{Dav{\'e}}
  et~al.}{2019}]{Dave+19}
{Dav{\'e}} R.,  {Angl{\'e}s-Alc{\'a}zar} D.,  {Narayanan} D.,  {Li} Q.,
  {Rafieferantsoa} M.~H.,   {Appleby} S.,  2019, \mn@doi [\mnras]
  {10.1093/mnras/stz937}, \href
  {https://ui.adsabs.harvard.edu/abs/2019MNRAS.486.2827D} {486, 2827}

\bibitem[\protect\citeauthoryear{{Dayal} et~al.,}{{Dayal}
  et~al.}{2022}]{Dayal+22}
{Dayal} P.,  et~al., 2022, \mn@doi [\mnras] {10.1093/mnras/stac537}, \href
  {https://ui.adsabs.harvard.edu/abs/2022MNRAS.512..989D} {512, 989}

\bibitem[\protect\citeauthoryear{{De Donder} \& {Vanbeveren}}{{De Donder} \&
  {Vanbeveren}}{2004}]{DeDonder&Vanbeveren04}
{De Donder} E.,  {Vanbeveren} D.,  2004, \mn@doi [\nar]
  {10.1016/j.newar.2004.07.001}, \href
  {https://ui.adsabs.harvard.edu/abs/2004NewAR..48..861D} {48, 861}

\bibitem[\protect\citeauthoryear{{De Lucia}, {Tornatore}, {Frenk}, {Helmi},
  {Navarro}  \& {White}}{{De Lucia} et~al.}{2014}]{DeLucia+14}
{De Lucia} G.,  {Tornatore} L.,  {Frenk} C.~S.,  {Helmi} A.,  {Navarro} J.~F.,
   {White} S. D.~M.,  2014, \mn@doi [\mnras] {10.1093/mnras/stu1752}, \href
  {https://ui.adsabs.harvard.edu/abs/2014MNRAS.445..970D} {445, 970}

\bibitem[\protect\citeauthoryear{{De Marco} \& {Izzard}}{{De Marco} \&
  {Izzard}}{2017}]{DeMarco&Izzard17}
{De Marco} O.,  {Izzard} R.~G.,  2017, \mn@doi [\pasa] {10.1017/pasa.2016.52},
  \href {https://ui.adsabs.harvard.edu/abs/2017PASA...34....1D} {34, e001}

\bibitem[\protect\citeauthoryear{{De Vis} et~al.,}{{De Vis}
  et~al.}{2019}]{DeVis+19}
{De Vis} P.,  et~al., 2019, \mn@doi [\aap] {10.1051/0004-6361/201834444}, \href
  {https://ui.adsabs.harvard.edu/abs/2019A&A...623A...5D} {623, A5}

\bibitem[\protect\citeauthoryear{{Dopita}, {Kewley}, {Sutherland}  \&
  {Nicholls}}{{Dopita} et~al.}{2016}]{Dopita+16}
{Dopita} M.~A.,  {Kewley} L.~J.,  {Sutherland} R.~S.,   {Nicholls} D.~C.,
  2016, \mn@doi [\apss] {10.1007/s10509-016-2657-8}, \href
  {http://adsabs.harvard.edu/abs/2016Ap%26SS.361...61D} {361, 61}

\bibitem[\protect\citeauthoryear{{Driver} et~al.,}{{Driver}
  et~al.}{2018}]{Driver+18}
{Driver} S.~P.,  et~al., 2018, \mn@doi [\mnras] {10.1093/mnras/stx2728}, \href
  {https://ui.adsabs.harvard.edu/abs/2018MNRAS.475.2891D} {475, 2891}

\bibitem[\protect\citeauthoryear{{Dunne} et~al.,}{{Dunne}
  et~al.}{2011}]{Dunne+11}
{Dunne} L.,  et~al., 2011, \mn@doi [\mnras] {10.1111/j.1365-2966.2011.19363.x},
  \href {https://ui.adsabs.harvard.edu/abs/2011MNRAS.417.1510D} {417, 1510}

\bibitem[\protect\citeauthoryear{{Eales} et~al.,}{{Eales}
  et~al.}{2012}]{Eales+12}
{Eales} S.,  et~al., 2012, \mn@doi [\apj] {10.1088/0004-637X/761/2/168}, \href
  {https://ui.adsabs.harvard.edu/abs/2012ApJ...761..168E} {761, 168}

\bibitem[\protect\citeauthoryear{{Elbaz} et~al.,}{{Elbaz}
  et~al.}{2007}]{Elbaz+07}
{Elbaz} D.,  et~al., 2007, \mn@doi [\aap] {10.1051/0004-6361:20077525}, \href
  {http://adsabs.harvard.edu/abs/2007A%26A...468...33E} {468, 33}

\bibitem[\protect\citeauthoryear{{Eldridge}, {Stanway}, {Xiao}, {McClelland},
  {Taylor}, {Ng}, {Greis}  \& {Bray}}{{Eldridge} et~al.}{2017}]{Eldridge+17}
{Eldridge} J.~J.,  {Stanway} E.~R.,  {Xiao} L.,  {McClelland} L.~A.~S.,
  {Taylor} G.,  {Ng} M.,  {Greis} S.~M.~L.,   {Bray} J.~C.,  2017, \mn@doi
  [\pasa] {10.1017/pasa.2017.51}, \href
  {https://ui.adsabs.harvard.edu/abs/2017PASA...34...58E} {34, e058}

\bibitem[\protect\citeauthoryear{{Emerick}, {Bryan}  \& {Mac Low}}{{Emerick}
  et~al.}{2020}]{Emerick+20a}
{Emerick} A.,  {Bryan} G.~L.,   {Mac Low} M.-M.,  2020, \mn@doi [\apj]
  {10.3847/1538-4357/ab6efc}, \href
  {https://ui.adsabs.harvard.edu/abs/2020ApJ...890..155E} {890, 155}

\bibitem[\protect\citeauthoryear{{Erroz-Ferrer} et~al.,}{{Erroz-Ferrer}
  et~al.}{2019}]{Erroz-Ferrer+19}
{Erroz-Ferrer} S.,  et~al., 2019, \mn@doi [\mnras] {10.1093/mnras/stz194},
  \href {https://ui.adsabs.harvard.edu/abs/2019MNRAS.484.5009E} {484, 5009}

\bibitem[\protect\citeauthoryear{{Ertl}, {Janka}, {Woosley}, {Sukhbold}  \&
  {Ugliano}}{{Ertl} et~al.}{2016}]{Ertl+16}
{Ertl} T.,  {Janka} H.~T.,  {Woosley} S.~E.,  {Sukhbold} T.,   {Ugliano} M.,
  2016, \mn@doi [\apj] {10.3847/0004-637X/818/2/124}, \href
  {https://ui.adsabs.harvard.edu/abs/2016ApJ...818..124E} {818, 124}

\bibitem[\protect\citeauthoryear{{Esmerian} \& {Gnedin}}{{Esmerian} \&
  {Gnedin}}{2022}]{Esmerian&Gnedin22}
{Esmerian} C.~J.,  {Gnedin} N.~Y.,  2022, \mn@doi [\apj]
  {10.3847/1538-4357/ac9612}, \href
  {https://ui.adsabs.harvard.edu/abs/2022ApJ...940...74E} {940, 74}

\bibitem[\protect\citeauthoryear{{Esmerian} \& {Gnedin}}{{Esmerian} \&
  {Gnedin}}{2023}]{Esmerian&Gnedin23}
{Esmerian} C.~J.,  {Gnedin} N.~Y.,  2023, \mn@doi [arXiv e-prints]
  {10.48550/arXiv.2308.11723}, \href
  {https://ui.adsabs.harvard.edu/abs/2023arXiv230811723E} {p. arXiv:2308.11723}

\bibitem[\protect\citeauthoryear{{Farmer}, {Laplace}, {Ma}, {de Mink}  \&
  {Justham}}{{Farmer} et~al.}{2023}]{Farmer+23}
{Farmer} R.,  {Laplace} E.,  {Ma} J.-z.,  {de Mink} S.~E.,   {Justham} S.,
  2023, \mn@doi [\apj] {10.3847/1538-4357/acc315}, \href
  {https://ui.adsabs.harvard.edu/abs/2023ApJ...948..111F} {948, 111}

\bibitem[\protect\citeauthoryear{{Ferrara} \& {Peroux}}{{Ferrara} \&
  {Peroux}}{2021}]{Ferrara&Peroux21}
{Ferrara} A.,  {Peroux} C.,  2021, \mn@doi [\mnras] {10.1093/mnras/stab761},
  \href {https://ui.adsabs.harvard.edu/abs/2021MNRAS.503.4537F} {503, 4537}

\bibitem[\protect\citeauthoryear{{Ferrara}, {Viti}  \& {Ceccarelli}}{{Ferrara}
  et~al.}{2016}]{Ferrara+16}
{Ferrara} A.,  {Viti} S.,   {Ceccarelli} C.,  2016, \mn@doi [\mnras]
  {10.1093/mnrasl/slw165}, \href
  {https://ui.adsabs.harvard.edu/abs/2016MNRAS.463L.112F} {463, L112}

\bibitem[\protect\citeauthoryear{{Ferrarotti} \& {Gail}}{{Ferrarotti} \&
  {Gail}}{2006}]{Ferrarotti&Gail06}
{Ferrarotti} A.~S.,  {Gail} H.~P.,  2006, \mn@doi [\aap]
  {10.1051/0004-6361:20041198}, \href
  {https://ui.adsabs.harvard.edu/abs/2006A&A...447..553F} {447, 553}

\bibitem[\protect\citeauthoryear{{Fryer}, {Belczynski}, {Wiktorowicz},
  {Dominik}, {Kalogera}  \& {Holz}}{{Fryer} et~al.}{2012}]{Fryer+12}
{Fryer} C.~L.,  {Belczynski} K.,  {Wiktorowicz} G.,  {Dominik} M.,  {Kalogera}
  V.,   {Holz} D.~E.,  2012, \mn@doi [\apj] {10.1088/0004-637X/749/1/91}, \href
  {https://ui.adsabs.harvard.edu/abs/2012ApJ...749...91F} {749, 91}

\bibitem[\protect\citeauthoryear{{Fu}, {Guo}, {Kauffmann}  \& {Krumholz}}{{Fu}
  et~al.}{2010}]{Fu+10}
{Fu} J.,  {Guo} Q.,  {Kauffmann} G.,   {Krumholz} M.~R.,  2010, \mn@doi [MNRAS]
  {10.1111/j.1365-2966.2010.17342.x}, \href
  {http://adsabs.harvard.edu/abs/2010MNRAS.409..515F} {409, 515}

\bibitem[\protect\citeauthoryear{{Fu} et~al.,}{{Fu} et~al.}{2013}]{Fu+13}
{Fu} J.,  et~al., 2013, \mn@doi [\mnras] {10.1093/mnras/stt1117}, \href
  {https://ui.adsabs.harvard.edu/abs/2013MNRAS.434.1531F} {434, 1531}

\bibitem[\protect\citeauthoryear{{Galliano}, {Galametz}  \& {Jones}}{{Galliano}
  et~al.}{2018}]{Galliano+18}
{Galliano} F.,  {Galametz} M.,   {Jones} A.~P.,  2018, \mn@doi [\araa]
  {10.1146/annurev-astro-081817-051900}, \href
  {https://ui.adsabs.harvard.edu/abs/2018ARA&A..56..673G} {56, 673}

\bibitem[\protect\citeauthoryear{{Giacobbo}, {Mapelli}  \& {Spera}}{{Giacobbo}
  et~al.}{2018}]{Giacobbo+18}
{Giacobbo} N.,  {Mapelli} M.,   {Spera} M.,  2018, \mn@doi [\mnras]
  {10.1093/mnras/stx2933}, \href
  {https://ui.adsabs.harvard.edu/abs/2018MNRAS.474.2959G} {474, 2959}

\bibitem[\protect\citeauthoryear{{Ginolfi}, {Graziani}, {Schneider}, {Marassi},
  {Valiante}, {Dell'Agli}, {Ventura}  \& {Hunt}}{{Ginolfi}
  et~al.}{2018}]{Ginolfi+18}
{Ginolfi} M.,  {Graziani} L.,  {Schneider} R.,  {Marassi} S.,  {Valiante} R.,
  {Dell'Agli} F.,  {Ventura} P.,   {Hunt} L.~K.,  2018, \mn@doi [\mnras]
  {10.1093/mnras/stx2572}, \href
  {https://ui.adsabs.harvard.edu/abs/2018MNRAS.473.4538G} {473, 4538}

\bibitem[\protect\citeauthoryear{{Glatzle}, {Graziani}  \& {Ciardi}}{{Glatzle}
  et~al.}{2022}]{Glatzle+22}
{Glatzle} M.,  {Graziani} L.,   {Ciardi} B.,  2022, \mn@doi [\mnras]
  {10.1093/mnras/stab3459}, \href
  {https://ui.adsabs.harvard.edu/abs/2022MNRAS.510.1068G} {510, 1068}

\bibitem[\protect\citeauthoryear{{G{\"o}tberg}, {de Mink}, {Groh}, {Leitherer}
  \& {Norman}}{{G{\"o}tberg} et~al.}{2019}]{Goetberg+19}
{G{\"o}tberg} Y.,  {de Mink} S.~E.,  {Groh} J.~H.,  {Leitherer} C.,   {Norman}
  C.,  2019, \mn@doi [\aap] {10.1051/0004-6361/201834525}, \href
  {https://ui.adsabs.harvard.edu/abs/2019A&A...629A.134G} {629, A134}

\bibitem[\protect\citeauthoryear{{Graur}, {Bianco}, {Huang}, {Modjaz},
  {Shivvers}, {Filippenko}, {Li}  \& {Eldridge}}{{Graur}
  et~al.}{2017}]{Graur+17a}
{Graur} O.,  {Bianco} F.~B.,  {Huang} S.,  {Modjaz} M.,  {Shivvers} I.,
  {Filippenko} A.~V.,  {Li} W.,   {Eldridge} J.~J.,  2017, \mn@doi [\apj]
  {10.3847/1538-4357/aa5eb8}, \href
  {https://ui.adsabs.harvard.edu/abs/2017ApJ...837..120G} {837, 120}

\bibitem[\protect\citeauthoryear{{Graziani}, {Schneider}, {Ginolfi}, {Hunt},
  {Maio}, {Glatzle}  \& {Ciardi}}{{Graziani} et~al.}{2020}]{Graziani+20}
{Graziani} L.,  {Schneider} R.,  {Ginolfi} M.,  {Hunt} L.~K.,  {Maio} U.,
  {Glatzle} M.,   {Ciardi} B.,  2020, \mn@doi [\mnras] {10.1093/mnras/staa796},
  \href {https://ui.adsabs.harvard.edu/abs/2020MNRAS.494.1071G} {494, 1071}

\bibitem[\protect\citeauthoryear{{Greggio}}{{Greggio}}{2005}]{Greggio+05}
{Greggio} L.,  2005, \mn@doi [\aap] {10.1051/0004-6361:20052926}, \href
  {https://ui.adsabs.harvard.edu/abs/2005A&A...441.1055G} {441, 1055}

\bibitem[\protect\citeauthoryear{{Guo} et~al.,}{{Guo} et~al.}{2011}]{Guo+11}
{Guo} Q.,  et~al., 2011, \mn@doi [MNRAS] {10.1111/j.1365-2966.2010.18114.x},
  \href {http://adsabs.harvard.edu/abs/2011MNRAS.413..101G} {413, 101}

\bibitem[\protect\citeauthoryear{{Gutcke}, {Pakmor}, {Naab}  \&
  {Springel}}{{Gutcke} et~al.}{2021}]{Gutcke+20}
{Gutcke} T.~A.,  {Pakmor} R.,  {Naab} T.,   {Springel} V.,  2021, \mn@doi
  [\mnras] {10.1093/mnras/staa3875}, \href
  {https://ui.adsabs.harvard.edu/abs/2021MNRAS.tmp..177G} {}

\bibitem[\protect\citeauthoryear{{Hayden} et~al.,}{{Hayden}
  et~al.}{2015}]{Hayden+15}
{Hayden} M.~R.,  et~al., 2015, \mn@doi [\apj] {10.1088/0004-637X/808/2/132},
  \href {https://ui.adsabs.harvard.edu/abs/2015ApJ...808..132H} {808, 132}

\bibitem[\protect\citeauthoryear{{Haynes} et~al.,}{{Haynes}
  et~al.}{2011}]{Haynes+11}
{Haynes} M.~P.,  et~al., 2011, \mn@doi [\aj] {10.1088/0004-6256/142/5/170},
  \href {https://ui.adsabs.harvard.edu/abs/2011AJ....142..170H} {142, 170}

\bibitem[\protect\citeauthoryear{{Heintz} et~al.,}{{Heintz}
  et~al.}{2023}]{Heintz+23a}
{Heintz} K.~E.,  et~al., 2023, \mn@doi [arXiv e-prints]
  {10.48550/arXiv.2308.14812}, \href
  {https://ui.adsabs.harvard.edu/abs/2023arXiv230814812H} {p. arXiv:2308.14812}

\bibitem[\protect\citeauthoryear{{Hendriks} \& {Izzard}}{{Hendriks} \&
  {Izzard}}{2023}]{Hendriks+23}
{Hendriks} D.~D.,  {Izzard} R.~G.,  2023, \mn@doi [Journal of Open Source
  Software] {10.21105/joss.04642}, 8, 4642

\bibitem[\protect\citeauthoryear{{Henriques}, {White}, {Thomas}, {Angulo},
  {Guo}, {Lemson}, {Springel}  \& {Overzier}}{{Henriques}
  et~al.}{2015}]{Henriques+15}
{Henriques} B.~M.~B.,  {White} S.~D.~M.,  {Thomas} P.~A.,  {Angulo} R.,  {Guo}
  Q.,  {Lemson} G.,  {Springel} V.,   {Overzier} R.,  2015, \mn@doi [MNRAS]
  {10.1093/mnras/stv705}, \href
  {http://adsabs.harvard.edu/abs/2015MNRAS.451.2663H} {451, 2663}

\bibitem[\protect\citeauthoryear{{Henriques}, {Yates}, {Fu}, {Guo},
  {Kauffmann}, {Srisawat}, {Thomas}  \& {White}}{{Henriques}
  et~al.}{2020}]{Henriques+20}
{Henriques} B. M.~B.,  {Yates} R.~M.,  {Fu} J.,  {Guo} Q.,  {Kauffmann} G.,
  {Srisawat} C.,  {Thomas} P.~A.,   {White} S. D.~M.,  2020, \mn@doi [\mnras]
  {10.1093/mnras/stz3233}, \href
  {https://ui.adsabs.harvard.edu/abs/2020MNRAS.491.5795H} {491, 5795}

\bibitem[\protect\citeauthoryear{{Hopkins} et~al.,}{{Hopkins}
  et~al.}{2018}]{Hopkins+18}
{Hopkins} P.~F.,  et~al., 2018, \mn@doi [\mnras] {10.1093/mnras/sty1690}, \href
  {https://ui.adsabs.harvard.edu/abs/2018MNRAS.480..800H} {480, 800}

\bibitem[\protect\citeauthoryear{{Hou}, {Aoyama}, {Hirashita}, {Nagamine}  \&
  {Shimizu}}{{Hou} et~al.}{2019}]{Hou+19}
{Hou} K.-C.,  {Aoyama} S.,  {Hirashita} H.,  {Nagamine} K.,   {Shimizu} I.,
  2019, \mn@doi [\mnras] {10.1093/mnras/stz121}, \href
  {https://ui.adsabs.harvard.edu/abs/2019MNRAS.485.1727H} {485, 1727}

\bibitem[\protect\citeauthoryear{{Hough}, {Rennehan}, {Kobayashi}, {Loubser},
  {Dav{\'e}}, {Babul}  \& {Cui}}{{Hough} et~al.}{2023}]{Hough+23}
{Hough} R.~T.,  {Rennehan} D.,  {Kobayashi} C.,  {Loubser} S.~I.,  {Dav{\'e}}
  R.,  {Babul} A.,   {Cui} W.,  2023, \mn@doi [\mnras]
  {10.1093/mnras/stad2394}, \href
  {https://ui.adsabs.harvard.edu/abs/2023MNRAS.tmp.2309H} {}

\bibitem[\protect\citeauthoryear{{Hu}, {Zhukovska}, {Somerville}  \&
  {Naab}}{{Hu} et~al.}{2019}]{Hu+19b}
{Hu} C.-Y.,  {Zhukovska} S.,  {Somerville} R.~S.,   {Naab} T.,  2019, \mn@doi
  [\mnras] {10.1093/mnras/stz1481}, \href
  {https://ui.adsabs.harvard.edu/abs/2019MNRAS.487.3252H} {487, 3252}

\bibitem[\protect\citeauthoryear{{Hurley}, {Pols}  \& {Tout}}{{Hurley}
  et~al.}{2000}]{Hurley+00}
{Hurley} J.~R.,  {Pols} O.~R.,   {Tout} C.~A.,  2000, \mn@doi [\mnras]
  {10.1046/j.1365-8711.2000.03426.x}, \href
  {https://ui.adsabs.harvard.edu/abs/2000MNRAS.315..543H} {315, 543}

\bibitem[\protect\citeauthoryear{{Hurley}, {Tout}  \& {Pols}}{{Hurley}
  et~al.}{2002}]{Hurley+02}
{Hurley} J.~R.,  {Tout} C.~A.,   {Pols} O.~R.,  2002, \mn@doi [\mnras]
  {10.1046/j.1365-8711.2002.05038.x}, \href
  {https://ui.adsabs.harvard.edu/abs/2002MNRAS.329..897H} {329, 897}

\bibitem[\protect\citeauthoryear{{Irvine}, {Goldsmith}  \&
  {Hjalmarson}}{{Irvine} et~al.}{1987}]{Irvine+87}
{Irvine} W.~M.,  {Goldsmith} P.~F.,   {Hjalmarson} A.,  1987, in {Hollenbach}
  D.~J.,  {Thronson} Harley~A. J.,  eds,  Interstellar Processes Vol. 134,
  Interstellar Processes. p.~561, \mn@doi{10.1007/978-94-009-3861-8_21}

\bibitem[\protect\citeauthoryear{{Iwamoto}, {Brachwitz}, {Nomoto}, {Kishimoto},
  {Umeda}, {Hix}  \& {Thielemann}}{{Iwamoto} et~al.}{1999}]{Iwamoto+99}
{Iwamoto} K.,  {Brachwitz} F.,  {Nomoto} K.,  {Kishimoto} N.,  {Umeda} H.,
  {Hix} W.~R.,   {Thielemann} F.-K.,  1999, \mn@doi [\apjs] {10.1086/313278},
  \href {https://ui.adsabs.harvard.edu/abs/1999ApJS..125..439I} {125, 439}

\bibitem[\protect\citeauthoryear{{Izzard} \& {Jermyn}}{{Izzard} \&
  {Jermyn}}{2022}]{Izzard&Jermyn22}
{Izzard} R.~G.,  {Jermyn} A.~S.,  2022, \mn@doi [\mnras]
  {10.1093/mnras/stac2899}, \href
  {https://ui.adsabs.harvard.edu/abs/2022MNRAS.tmp.2755I} {}

\bibitem[\protect\citeauthoryear{{Izzard}, {Tout}, {Karakas}  \&
  {Pols}}{{Izzard} et~al.}{2004}]{Izzard+04}
{Izzard} R.~G.,  {Tout} C.~A.,  {Karakas} A.~I.,   {Pols} O.~R.,  2004, \mn@doi
  [\mnras] {10.1111/j.1365-2966.2004.07446.x}, \href
  {https://ui.adsabs.harvard.edu/abs/2004MNRAS.350..407I} {350, 407}

\bibitem[\protect\citeauthoryear{{Izzard}, {Dray}, {Karakas}, {Lugaro}  \&
  {Tout}}{{Izzard} et~al.}{2006}]{Izzard+06}
{Izzard} R.~G.,  {Dray} L.~M.,  {Karakas} A.~I.,  {Lugaro} M.,   {Tout} C.~A.,
  2006, \mn@doi [\aap] {10.1051/0004-6361:20066129}, \href
  {https://ui.adsabs.harvard.edu/abs/2006A&A...460..565I} {460, 565}

\bibitem[\protect\citeauthoryear{{Izzard}, {Glebbeek}, {Stancliffe}  \&
  {Pols}}{{Izzard} et~al.}{2009}]{Izzard+09}
{Izzard} R.~G.,  {Glebbeek} E.,  {Stancliffe} R.~J.,   {Pols} O.~R.,  2009,
  \mn@doi [\aap] {10.1051/0004-6361/200912827}, \href
  {http://cdsads.u-strasbg.fr/abs/2009A%26A...508.1359I} {508, 1359}

\bibitem[\protect\citeauthoryear{{Izzard}, {Preece}, {Jofre}, {Halabi},
  {Masseron}  \& {Tout}}{{Izzard} et~al.}{2018}]{Izzard+18}
{Izzard} R.~G.,  {Preece} H.,  {Jofre} P.,  {Halabi} G.~M.,  {Masseron} T.,
  {Tout} C.~A.,  2018, \mn@doi [\mnras] {10.1093/mnras/stx2355}, \href
  {http://adsabs.harvard.edu/abs/2018MNRAS.473.2984I} {473, 2984}

\bibitem[\protect\citeauthoryear{{Jenkins}}{{Jenkins}}{2009}]{Jenkins09}
{Jenkins} E.~B.,  2009, \mn@doi [\apj] {10.1088/0004-637X/700/2/1299}, \href
  {http://adsabs.harvard.edu/abs/2009ApJ...700.1299J} {700, 1299}

\bibitem[\protect\citeauthoryear{{Jones}, {Haynes}, {Giovanelli}  \&
  {Moorman}}{{Jones} et~al.}{2018}]{Jones+18}
{Jones} M.~G.,  {Haynes} M.~P.,  {Giovanelli} R.,   {Moorman} C.,  2018,
  \mn@doi [\mnras] {10.1093/mnras/sty521}, \href
  {https://ui.adsabs.harvard.edu/abs/2018MNRAS.477....2J} {477, 2}

\bibitem[\protect\citeauthoryear{{Jones}, {Sanders}, {Roberts-Borsani},
  {Ellis}, {Laporte}, {Treu}  \& {Harikane}}{{Jones} et~al.}{2020}]{Jones+20}
{Jones} T.,  {Sanders} R.,  {Roberts-Borsani} G.,  {Ellis} R.~S.,  {Laporte}
  N.,  {Treu} T.,   {Harikane} Y.,  2020, \mn@doi [\apj]
  {10.3847/1538-4357/abb943}, \href
  {https://ui.adsabs.harvard.edu/abs/2020ApJ...903..150J} {903, 150}

\bibitem[\protect\citeauthoryear{{Karakas}, {Lattanzio}  \& {Pols}}{{Karakas}
  et~al.}{2002}]{Karakas+02}
{Karakas} A.~I.,  {Lattanzio} J.~C.,   {Pols} O.~R.,  2002, \mn@doi [\pasa]
  {10.1071/AS02013}, \href
  {https://ui.adsabs.harvard.edu/abs/2002PASA...19..515K} {19, 515}

\bibitem[\protect\citeauthoryear{{Kobayashi}, {Karakas}  \&
  {Lugaro}}{{Kobayashi} et~al.}{2020}]{Kobayashi+20}
{Kobayashi} C.,  {Karakas} A.~I.,   {Lugaro} M.,  2020, \mn@doi [\apj]
  {10.3847/1538-4357/abae65}, \href
  {https://ui.adsabs.harvard.edu/abs/2020ApJ...900..179K} {900, 179}

\bibitem[\protect\citeauthoryear{{Kobayashi} et~al.,}{{Kobayashi}
  et~al.}{2023}]{Kobayashi+23}
{Kobayashi} C.,  et~al., 2023, \mn@doi [\apjl] {10.3847/2041-8213/acad82},
  \href {https://ui.adsabs.harvard.edu/abs/2023ApJ...943L..12K} {943, L12}

\bibitem[\protect\citeauthoryear{{Krogager}, {Fynbo}, {M{\o}ller},
  {Noterdaeme}, {Heintz}  \& {Pettini}}{{Krogager} et~al.}{2019}]{Krogager+19}
{Krogager} J.-K.,  {Fynbo} J. P.~U.,  {M{\o}ller} P.,  {Noterdaeme} P.,
  {Heintz} K.~E.,   {Pettini} M.,  2019, \mn@doi [\mnras]
  {10.1093/mnras/stz1120}, \href
  {https://ui.adsabs.harvard.edu/abs/2019MNRAS.486.4377K} {486, 4377}

\bibitem[\protect\citeauthoryear{{Kroupa}}{{Kroupa}}{2001}]{Kroupa01}
{Kroupa} P.,  2001, \mn@doi [\mnras] {10.1046/j.1365-8711.2001.04022.x}, \href
  {http://adsabs.harvard.edu/abs/2001MNRAS.322..231K} {322, 231}

\bibitem[\protect\citeauthoryear{{Krumholz}, {McKee}  \&
  {Tumlinson}}{{Krumholz} et~al.}{2009}]{Krumholz+09}
{Krumholz} M.~R.,  {McKee} C.~F.,   {Tumlinson} J.,  2009, \mn@doi [\apj]
  {10.1088/0004-637X/693/1/216}, \href
  {https://ui.adsabs.harvard.edu/abs/2009ApJ...693..216K} {693, 216}

\bibitem[\protect\citeauthoryear{{Li} \& {White}}{{Li} \&
  {White}}{2009}]{Li&White09}
{Li} C.,  {White} S. D.~M.,  2009, \mn@doi [\mnras]
  {10.1111/j.1365-2966.2009.15268.x}, \href
  {https://ui.adsabs.harvard.edu/abs/2009MNRAS.398.2177L} {398, 2177}

\bibitem[\protect\citeauthoryear{{Li}, {Narayanan}  \& {Dav{\'e}}}{{Li}
  et~al.}{2019}]{Li+19}
{Li} Q.,  {Narayanan} D.,   {Dav{\'e}} R.,  2019, \mn@doi [\mnras]
  {10.1093/mnras/stz2684}, \href
  {https://ui.adsabs.harvard.edu/abs/2019MNRAS.490.1425L} {490, 1425}

\bibitem[\protect\citeauthoryear{{Limongi} \& {Chieffi}}{{Limongi} \&
  {Chieffi}}{2018}]{Limongi&Chieffi18}
{Limongi} M.,  {Chieffi} A.,  2018, \mn@doi [\apjs] {10.3847/1538-4365/aacb24},
  \href {https://ui.adsabs.harvard.edu/abs/2018ApJS..237...13L} {237, 13}

\bibitem[\protect\citeauthoryear{{Livne} \& {Arnett}}{{Livne} \&
  {Arnett}}{1995}]{Livne&Arnett95}
{Livne} E.,  {Arnett} D.,  1995, \mn@doi [\apj] {10.1086/176279}, \href
  {https://ui.adsabs.harvard.edu/abs/1995ApJ...452...62L} {452, 62}

\bibitem[\protect\citeauthoryear{{Maoz}, {Mannucci}, {Li}, {Filippenko}, {Della
  Valle}  \& {Panagia}}{{Maoz} et~al.}{2011}]{Maoz+11}
{Maoz} D.,  {Mannucci} F.,  {Li} W.,  {Filippenko} A.~V.,  {Della Valle} M.,
  {Panagia} N.,  2011, \mn@doi [\mnras] {10.1111/j.1365-2966.2010.16808.x},
  \href {https://ui.adsabs.harvard.edu/abs/2011MNRAS.412.1508M} {412, 1508}

\bibitem[\protect\citeauthoryear{{Maoz}, {Mannucci}  \& {Nelemans}}{{Maoz}
  et~al.}{2014}]{Maoz+14}
{Maoz} D.,  {Mannucci} F.,   {Nelemans} G.,  2014, \mn@doi [\araa]
  {10.1146/annurev-astro-082812-141031}, \href
  {https://ui.adsabs.harvard.edu/abs/2014ARA&A..52..107M} {52, 107}

\bibitem[\protect\citeauthoryear{{Marigo}}{{Marigo}}{2001}]{Marigo01}
{Marigo} P.,  2001, \mn@doi [\aap] {10.1051/0004-6361:20000247}, \href
  {https://ui.adsabs.harvard.edu/abs/2001A&A...370..194M} {370, 194}

\bibitem[\protect\citeauthoryear{{Martin}, {Kobulnicky}  \& {Heckman}}{{Martin}
  et~al.}{2002}]{Martin+02}
{Martin} C.~L.,  {Kobulnicky} H.~A.,   {Heckman} T.~M.,  2002, \mn@doi [\apj]
  {10.1086/341092}, \href
  {https://ui.adsabs.harvard.edu/abs/2002ApJ...574..663M} {574, 663}

\bibitem[\protect\citeauthoryear{{Matteucci}}{{Matteucci}}{2012}]{Matteucci+12}
{Matteucci} F.,  2012, {Chemical Evolution of Galaxies}.
Springer-Verlag, \mn@doi{10.1007/978-3-642-22491-1}

\bibitem[\protect\citeauthoryear{{Matteucci}, {Panagia}, {Pipino}, {Mannucci},
  {Recchi}  \& {Della Valle}}{{Matteucci} et~al.}{2006}]{Matteucci+06}
{Matteucci} F.,  {Panagia} N.,  {Pipino} A.,  {Mannucci} F.,  {Recchi} S.,
  {Della Valle} M.,  2006, \mn@doi [\mnras] {10.1111/j.1365-2966.2006.10848.x},
  \href {https://ui.adsabs.harvard.edu/abs/2006MNRAS.372..265M} {372, 265}

\bibitem[\protect\citeauthoryear{{Matteucci}, {Spitoni}, {Recchi}  \&
  {Valiante}}{{Matteucci} et~al.}{2009}]{Matteucci+09}
{Matteucci} F.,  {Spitoni} E.,  {Recchi} S.,   {Valiante} R.,  2009, \mn@doi
  [\aap] {10.1051/0004-6361/200911869}, \href
  {https://ui.adsabs.harvard.edu/abs/2009A&A...501..531M} {501, 531}

\bibitem[\protect\citeauthoryear{{Mattsson}}{{Mattsson}}{2015}]{Mattsson+15}
{Mattsson} L.,  2015, \mn@doi [arXiv e-prints] {10.48550/arXiv.1505.04758},
  \href {https://ui.adsabs.harvard.edu/abs/2015arXiv150504758M} {p.
  arXiv:1505.04758}

\bibitem[\protect\citeauthoryear{{McKee}}{{McKee}}{1989}]{McKee89}
{McKee} C.,  1989, in {Allamandola} L.~J.,  {Tielens} A.~G.~G.~M.,  eds,
  Interstellar Dust Vol. 135, Interstellar Dust. p.~431

\bibitem[\protect\citeauthoryear{{McKinnon}, {Torrey}, {Vogelsberger},
  {Hayward}  \& {Marinacci}}{{McKinnon} et~al.}{2017}]{McKinnon+17}
{McKinnon} R.,  {Torrey} P.,  {Vogelsberger} M.,  {Hayward} C.~C.,
  {Marinacci} F.,  2017, \mn@doi [\mnras] {10.1093/mnras/stx467}, \href
  {https://ui.adsabs.harvard.edu/abs/2017MNRAS.468.1505M} {468, 1505}

\bibitem[\protect\citeauthoryear{{M{\'e}nard}, {Scranton}, {Fukugita}  \&
  {Richards}}{{M{\'e}nard} et~al.}{2010}]{Menard+10}
{M{\'e}nard} B.,  {Scranton} R.,  {Fukugita} M.,   {Richards} G.,  2010,
  \mn@doi [\mnras] {10.1111/j.1365-2966.2010.16486.x}, \href
  {https://ui.adsabs.harvard.edu/abs/2010MNRAS.405.1025M} {405, 1025}

\bibitem[\protect\citeauthoryear{{Mennekens} \& {Vanbeveren}}{{Mennekens} \&
  {Vanbeveren}}{2016}]{Mennekens&Vanbeveren16}
{Mennekens} N.,  {Vanbeveren} D.,  2016, \mn@doi [\aap]
  {10.1051/0004-6361/201628193}, \href
  {https://ui.adsabs.harvard.edu/abs/2016A&A...589A..64M} {589, A64}

\bibitem[\protect\citeauthoryear{{Moe} \& {Di Stefano}}{{Moe} \& {Di
  Stefano}}{2017}]{Moe&DiStefano17}
{Moe} M.,  {Di Stefano} R.,  2017, \mn@doi [\apjs] {10.3847/1538-4365/aa6fb6},
  \href {https://ui.adsabs.harvard.edu/abs/2017ApJS..230...15M} {230, 15}

\bibitem[\protect\citeauthoryear{{Navarro}, {Frenk}  \& {White}}{{Navarro}
  et~al.}{1997}]{Navarro+97}
{Navarro} J.~F.,  {Frenk} C.~S.,   {White} S. D.~M.,  1997, \mn@doi [\apj]
  {10.1086/304888}, \href
  {https://ui.adsabs.harvard.edu/abs/1997ApJ...490..493N} {490, 493}

\bibitem[\protect\citeauthoryear{{Nomoto}, {Kobayashi}  \& {Tominaga}}{{Nomoto}
  et~al.}{2013}]{Nomoto+13}
{Nomoto} K.,  {Kobayashi} C.,   {Tominaga} N.,  2013, \mn@doi [\araa]
  {10.1146/annurev-astro-082812-140956}, \href
  {https://ui.adsabs.harvard.edu/abs/2013ARA&A..51..457N} {51, 457}

\bibitem[\protect\citeauthoryear{{Parente}, {Ragone-Figueroa}, {Granato},
  {Borgani}, {Murante}, {Valentini}, {Bressan}  \& {Lapi}}{{Parente}
  et~al.}{2022}]{Parente+22}
{Parente} M.,  {Ragone-Figueroa} C.,  {Granato} G.~L.,  {Borgani} S.,
  {Murante} G.,  {Valentini} M.,  {Bressan} A.,   {Lapi} A.,  2022, \mn@doi
  [\mnras] {10.1093/mnras/stac1913}, \href
  {https://ui.adsabs.harvard.edu/abs/2022MNRAS.515.2053P} {515, 2053}

\bibitem[\protect\citeauthoryear{{Parente}, {Ragone-Figueroa}, {Granato}  \&
  {Lapi}}{{Parente} et~al.}{2023}]{Parente+23}
{Parente} M.,  {Ragone-Figueroa} C.,  {Granato} G.~L.,   {Lapi} A.,  2023,
  \mn@doi [\mnras] {10.1093/mnras/stad907}, \href
  {https://ui.adsabs.harvard.edu/abs/2023MNRAS.521.6105P} {521, 6105}

\bibitem[\protect\citeauthoryear{{Peek}, {M{\'e}nard}  \& {Corrales}}{{Peek}
  et~al.}{2015}]{Peek+15}
{Peek} J.~E.~G.,  {M{\'e}nard} B.,   {Corrales} L.,  2015, \mn@doi [\apj]
  {10.1088/0004-637X/813/1/7}, \href
  {https://ui.adsabs.harvard.edu/abs/2015ApJ...813....7P} {813, 7}

\bibitem[\protect\citeauthoryear{{Peron}, {Libanore}, {Ravenni}, {Liguori}  \&
  {Artale}}{{Peron} et~al.}{2023}]{Peron+23}
{Peron} M.,  {Libanore} S.,  {Ravenni} A.,  {Liguori} M.,   {Artale} M.~C.,
  2023, arXiv e-prints, \href
  {https://ui.adsabs.harvard.edu/abs/2023arXiv230518003P} {p. arXiv:2305.18003}

\bibitem[\protect\citeauthoryear{{P{\'e}roux} \& {Howk}}{{P{\'e}roux} \&
  {Howk}}{2020}]{Peroux&Howk20}
{P{\'e}roux} C.,  {Howk} J.~C.,  2020, \mn@doi [\araa]
  {10.1146/annurev-astro-021820-120014}, \href
  {https://ui.adsabs.harvard.edu/abs/2020ARA&A..58..363P} {58, 363}

\bibitem[\protect\citeauthoryear{{Pillepich} et~al.,}{{Pillepich}
  et~al.}{2018}]{Pillepich+18a}
{Pillepich} A.,  et~al., 2018, \mn@doi [\mnras] {10.1093/mnras/stx2656}, \href
  {https://ui.adsabs.harvard.edu/abs/2018MNRAS.473.4077P} {473, 4077}

\bibitem[\protect\citeauthoryear{{Planck Collaboration} et~al.,}{{Planck
  Collaboration} et~al.}{2014}]{Planck14}
{Planck Collaboration} et~al., 2014, \mn@doi [\aap]
  {10.1051/0004-6361/201321591}, \href
  {http://adsabs.harvard.edu/abs/2014A%26A...571A..16P} {571, A16}

\bibitem[\protect\citeauthoryear{{Podsiadlowski}, {Joss}  \&
  {Hsu}}{{Podsiadlowski} et~al.}{1992}]{Podsiadlowski+92}
{Podsiadlowski} P.,  {Joss} P.~C.,   {Hsu} J.~J.~L.,  1992, \mn@doi [\apj]
  {10.1086/171341}, \href
  {https://ui.adsabs.harvard.edu/abs/1992ApJ...391..246P} {391, 246}

\bibitem[\protect\citeauthoryear{{Popping} \& {P{\'e}roux}}{{Popping} \&
  {P{\'e}roux}}{2022}]{Popping&Peroux22}
{Popping} G.,  {P{\'e}roux} C.,  2022, \mn@doi [\mnras]
  {10.1093/mnras/stac695}, \href
  {https://ui.adsabs.harvard.edu/abs/2022MNRAS.513.1531P} {513, 1531}

\bibitem[\protect\citeauthoryear{{Popping}, {Somerville}  \&
  {Galametz}}{{Popping} et~al.}{2017}]{Popping+17}
{Popping} G.,  {Somerville} R.~S.,   {Galametz} M.,  2017, \mn@doi [\mnras]
  {10.1093/mnras/stx1545}, \href
  {https://ui.adsabs.harvard.edu/abs/2017MNRAS.471.3152P} {471, 3152}

\bibitem[\protect\citeauthoryear{{Popping} et~al.,}{{Popping}
  et~al.}{2023}]{Popping+23}
{Popping} G.,  et~al., 2023, \mn@doi [\aap] {10.1051/0004-6361/202243817},
  \href {https://ui.adsabs.harvard.edu/abs/2023A&A...670A.138P} {670, A138}

\bibitem[\protect\citeauthoryear{{Portegies Zwart} \& {Verbunt}}{{Portegies
  Zwart} \& {Verbunt}}{1996}]{PortegiesZwart+96}
{Portegies Zwart} S.~F.,  {Verbunt} F.,  1996, \aap, \href
  {https://ui.adsabs.harvard.edu/abs/1996A&A...309..179P} {309, 179}

\bibitem[\protect\citeauthoryear{{Portinari}, {Chiosi}  \&
  {Bressan}}{{Portinari} et~al.}{1998}]{Portinari+98}
{Portinari} L.,  {Chiosi} C.,   {Bressan} A.,  1998, \aap, \href
  {https://ui.adsabs.harvard.edu/abs/1998A&A...334..505P} {334, 505}

\bibitem[\protect\citeauthoryear{{Pozzi}, {Calura}, {Zamorani}, {Delvecchio},
  {Gruppioni}  \& {Santini}}{{Pozzi} et~al.}{2020}]{Pozzi+20}
{Pozzi} F.,  {Calura} F.,  {Zamorani} G.,  {Delvecchio} I.,  {Gruppioni} C.,
  {Santini} P.,  2020, \mn@doi [\mnras] {10.1093/mnras/stz2724}, \href
  {https://ui.adsabs.harvard.edu/abs/2020MNRAS.491.5073P} {491, 5073}

\bibitem[\protect\citeauthoryear{{Reiprich} \& {B{\"o}hringer}}{{Reiprich} \&
  {B{\"o}hringer}}{2002}]{Reiprich&Boehringer02}
{Reiprich} T.~H.,  {B{\"o}hringer} H.,  2002, \mn@doi [\apj] {10.1086/338753},
  \href {https://ui.adsabs.harvard.edu/abs/2002ApJ...567..716R} {567, 716}

\bibitem[\protect\citeauthoryear{{R{\'e}my-Ruyer} et~al.,}{{R{\'e}my-Ruyer}
  et~al.}{2015}]{Remy-Ruyer+15}
{R{\'e}my-Ruyer} A.,  et~al., 2015, \mn@doi [\aap]
  {10.1051/0004-6361/201526067}, \href
  {https://ui.adsabs.harvard.edu/abs/2015A&A...582A.121R} {582, A121}

\bibitem[\protect\citeauthoryear{{Rojas-Arriagada}, {Zoccali}, {Schultheis},
  {Recio-Blanco}, {Zasowski}, {Minniti}, {J{\"o}nsson}  \&
  {Cohen}}{{Rojas-Arriagada} et~al.}{2019}]{Rojas-Arriagada+19}
{Rojas-Arriagada} A.,  {Zoccali} M.,  {Schultheis} M.,  {Recio-Blanco} A.,
  {Zasowski} G.,  {Minniti} D.,  {J{\"o}nsson} H.,   {Cohen} R.~E.,  2019,
  \mn@doi [\aap] {10.1051/0004-6361/201834126}, \href
  {https://ui.adsabs.harvard.edu/abs/2019A&A...626A..16R} {626, A16}

\bibitem[\protect\citeauthoryear{{Saintonge} et~al.,}{{Saintonge}
  et~al.}{2013}]{Saintonge+13}
{Saintonge} A.,  et~al., 2013, \mn@doi [\apj] {10.1088/0004-637X/778/1/2},
  \href {https://ui.adsabs.harvard.edu/abs/2013ApJ...778....2S} {778, 2}

\bibitem[\protect\citeauthoryear{{Sana} et~al.,}{{Sana} et~al.}{2012}]{Sana+12}
{Sana} H.,  et~al., 2012, \mn@doi [Science] {10.1126/science.1223344}, \href
  {https://ui.adsabs.harvard.edu/abs/2012Sci...337..444S} {337, 444}

\bibitem[\protect\citeauthoryear{{Sanders} et~al.,}{{Sanders}
  et~al.}{2021}]{Sanders+21}
{Sanders} R.~L.,  et~al., 2021, \mn@doi [\apj] {10.3847/1538-4357/abf4c1},
  \href {https://ui.adsabs.harvard.edu/abs/2021ApJ...914...19S} {914, 19}

\bibitem[\protect\citeauthoryear{{Sanders} et~al.,}{{Sanders}
  et~al.}{2023}]{Sanders+23}
{Sanders} R.~L.,  et~al., 2023, \mn@doi [\apj] {10.3847/1538-4357/aca9cc},
  \href {https://ui.adsabs.harvard.edu/abs/2023ApJ...943...75S} {943, 75}

\bibitem[\protect\citeauthoryear{{Santini} et~al.,}{{Santini}
  et~al.}{2014}]{Santini+14}
{Santini} P.,  et~al., 2014, \mn@doi [\aap] {10.1051/0004-6361/201322835},
  \href {https://ui.adsabs.harvard.edu/abs/2014A&A...562A..30S} {562, A30}

\bibitem[\protect\citeauthoryear{{Savage} \& {Sembach}}{{Savage} \&
  {Sembach}}{1996}]{Savage&Sembach96}
{Savage} B.~D.,  {Sembach} K.~R.,  1996, \mn@doi [\araa]
  {10.1146/annurev.astro.34.1.279}, \href
  {https://ui.adsabs.harvard.edu/abs/1996ARA&A..34..279S} {34, 279}

\bibitem[\protect\citeauthoryear{{Schaerer}, {Marques-Chaves}, {Barrufet},
  {Oesch}, {Izotov}, {Naidu}, {Guseva}  \& {Brammer}}{{Schaerer}
  et~al.}{2022}]{Schaerer+22}
{Schaerer} D.,  {Marques-Chaves} R.,  {Barrufet} L.,  {Oesch} P.,  {Izotov}
  Y.~I.,  {Naidu} R.,  {Guseva} N.~G.,   {Brammer} G.,  2022, \mn@doi [\aap]
  {10.1051/0004-6361/202244556}, \href
  {https://ui.adsabs.harvard.edu/abs/2022A&A...665L...4S} {665, L4}

\bibitem[\protect\citeauthoryear{{Schaye} et~al.,}{{Schaye}
  et~al.}{2015}]{Schaye+15}
{Schaye} J.,  et~al., 2015, \mn@doi [\mnras] {10.1093/mnras/stu2058}, \href
  {https://ui.adsabs.harvard.edu/abs/2015MNRAS.446..521S} {446, 521}

\bibitem[\protect\citeauthoryear{{Schneider}, {Graziani}, {Marassi}, {Spera},
  {Mapelli}, {Alparone}  \& {Bennassuti}}{{Schneider}
  et~al.}{2017}]{Schneider+17}
{Schneider} R.,  {Graziani} L.,  {Marassi} S.,  {Spera} M.,  {Mapelli} M.,
  {Alparone} M.,   {Bennassuti} M.~d.,  2017, \mn@doi [\mnras]
  {10.1093/mnrasl/slx118}, \href
  {https://ui.adsabs.harvard.edu/abs/2017MNRAS.471L.105S} {471, L105}

\bibitem[\protect\citeauthoryear{{Schneider}, {Podsiadlowski}  \&
  {M{\"u}ller}}{{Schneider} et~al.}{2021}]{Schneider+21}
{Schneider} F.~R.~N.,  {Podsiadlowski} P.,   {M{\"u}ller} B.,  2021, \mn@doi
  [\aap] {10.1051/0004-6361/202039219}, \href
  {https://ui.adsabs.harvard.edu/abs/2021A&A...645A...5S} {645, A5}

\bibitem[\protect\citeauthoryear{{Shamshiri}, {Thomas}, {Henriques}, {Tojeiro},
  {Lemson}, {Oliver}  \& {Wilkins}}{{Shamshiri} et~al.}{2015}]{Shamshiri+15}
{Shamshiri} S.,  {Thomas} P.~A.,  {Henriques} B.~M.,  {Tojeiro} R.,  {Lemson}
  G.,  {Oliver} S.~J.,   {Wilkins} S.,  2015, \mn@doi [\mnras]
  {10.1093/mnras/stv883}, \href
  {https://ui.adsabs.harvard.edu/abs/2015MNRAS.451.2681S} {451, 2681}

\bibitem[\protect\citeauthoryear{{Shapley}, {Cullen}, {Dunlop}, {McLure},
  {Kriek}, {Reddy}  \& {Sanders}}{{Shapley} et~al.}{2020}]{Shapley+20}
{Shapley} A.~E.,  {Cullen} F.,  {Dunlop} J.~S.,  {McLure} R.~J.,  {Kriek} M.,
  {Reddy} N.~A.,   {Sanders} R.~L.,  2020, \mn@doi [\apjl]
  {10.3847/2041-8213/abc006}, \href
  {https://ui.adsabs.harvard.edu/abs/2020ApJ...903L..16S} {903, L16}

\bibitem[\protect\citeauthoryear{{Slavin}, {Dwek}  \& {Jones}}{{Slavin}
  et~al.}{2015}]{Slavin+15}
{Slavin} J.~D.,  {Dwek} E.,   {Jones} A.~P.,  2015, \mn@doi [\apj]
  {10.1088/0004-637X/803/1/7}, \href
  {https://ui.adsabs.harvard.edu/abs/2015ApJ...803....7S} {803, 7}

\bibitem[\protect\citeauthoryear{{Springel}, {White}, {Jenkins}
  et~al.}{{Springel} et~al.}{2005}]{Springel+05}
{Springel} V.,  {White} S.~D.~M.,  {Jenkins} A.,   et~al., 2005, \mn@doi
  [Nature] {10.1038/nature03597}, \href
  {http://adsabs.harvard.edu/abs/2005Natur.435..629S} {435, 629}

\bibitem[\protect\citeauthoryear{{Stanway}, {Hoskin}, {Lane}, {Brown},
  {Childs}, {Greis}  \& {Levan}}{{Stanway} et~al.}{2018}]{Stanway+18}
{Stanway} E.~R.,  {Hoskin} M.~J.,  {Lane} M.~A.,  {Brown} G.~C.,  {Childs}
  H.~J.~T.,  {Greis} S.~M.~L.,   {Levan} A.~J.,  2018, \mn@doi [\mnras]
  {10.1093/mnras/stx3305}, \href
  {https://ui.adsabs.harvard.edu/abs/2018MNRAS.475.1829S} {475, 1829}

\bibitem[\protect\citeauthoryear{{Strickland}, {Heckman}, {Colbert}, {Hoopes}
  \& {Weaver}}{{Strickland} et~al.}{2004}]{Strickland+04}
{Strickland} D.~K.,  {Heckman} T.~M.,  {Colbert} E. J.~M.,  {Hoopes} C.~G.,
  {Weaver} K.~A.,  2004, \mn@doi [\apj] {10.1086/383136}, \href
  {https://ui.adsabs.harvard.edu/abs/2004ApJ...606..829S} {606, 829}

\bibitem[\protect\citeauthoryear{{Sutherland} \& {Dopita}}{{Sutherland} \&
  {Dopita}}{1993}]{Sutherland&Dopita93}
{Sutherland} R.~S.,  {Dopita} M.~A.,  1993, \mn@doi [\apjs] {10.1086/191823},
  \href {https://ui.adsabs.harvard.edu/abs/1993ApJS...88..253S} {88, 253}

\bibitem[\protect\citeauthoryear{{Temmink}, {Pols}, {Justham}, {Istrate}  \&
  {Toonen}}{{Temmink} et~al.}{2023}]{Temmink+23}
{Temmink} K.~D.,  {Pols} O.~R.,  {Justham} S.,  {Istrate} A.~G.,   {Toonen} S.,
   2023, \mn@doi [\aap] {10.1051/0004-6361/202244137}, \href
  {https://ui.adsabs.harvard.edu/abs/2023A&A...669A..45T} {669, A45}

\bibitem[\protect\citeauthoryear{{Thielemann} et~al.,}{{Thielemann}
  et~al.}{2003}]{Thielemann+03}
{Thielemann} F.~K.,  et~al., 2003, in {Hillebrandt} W.,  {Leibundgut} B.,  eds,
  From Twilight to Highlight: The Physics of Supernovae. p.~331,
  \mn@doi{10.1007/10828549_46}

\bibitem[\protect\citeauthoryear{{Tinsley}}{{Tinsley}}{1980}]{Tinsley80}
{Tinsley} B.~M.,  1980, \mn@doi [\fcp] {10.48550/arXiv.2203.02041}, \href
  {https://ui.adsabs.harvard.edu/abs/1980FCPh....5..287T} {5, 287}

\bibitem[\protect\citeauthoryear{{Triani}, {Sinha}, {Croton}, {Pacifici}  \&
  {Dwek}}{{Triani} et~al.}{2020}]{Triani+20}
{Triani} D.~P.,  {Sinha} M.,  {Croton} D.~J.,  {Pacifici} C.,   {Dwek} E.,
  2020, \mn@doi [\mnras] {10.1093/mnras/staa446}, \href
  {https://ui.adsabs.harvard.edu/abs/2020MNRAS.493.2490T} {493, 2490}

\bibitem[\protect\citeauthoryear{{Trump} et~al.,}{{Trump}
  et~al.}{2023}]{Trump+23}
{Trump} J.~R.,  et~al., 2023, \mn@doi [\apj] {10.3847/1538-4357/acba8a}, \href
  {https://ui.adsabs.harvard.edu/abs/2023ApJ...945...35T} {945, 35}

\bibitem[\protect\citeauthoryear{{Tsai} \& {Mathews}}{{Tsai} \&
  {Mathews}}{1995}]{Tsai&Mathews95}
{Tsai} J.~C.,  {Mathews} W.~G.,  1995, \mn@doi [\apj] {10.1086/175943}, \href
  {https://ui.adsabs.harvard.edu/abs/1995ApJ...448...84T} {448, 84}

\bibitem[\protect\citeauthoryear{{Tumlinson} et~al.,}{{Tumlinson}
  et~al.}{2011}]{Tumlinson+11}
{Tumlinson} J.,  et~al., 2011, \mn@doi [Science] {10.1126/science.1209840},
  \href {https://ui.adsabs.harvard.edu/abs/2011Sci...334..948T} {334, 948}

\bibitem[\protect\citeauthoryear{\VAN{Dishoeck} {Van}{van}~Dishoeck \&
  {Blake}}{\VAN{Dishoeck} {Van}{van}~Dishoeck \&
  {Blake}}{1998}]{vanDishoeck&Blake98}
\VAN{Dishoeck} {Van}{van}~Dishoeck E.~F.,  {Blake} G.~A.,  1998, \mn@doi
  [\araa] {10.1146/annurev.astro.36.1.317}, \href
  {https://ui.adsabs.harvard.edu/abs/1998ARA&A..36..317V} {36, 317}

\bibitem[\protect\citeauthoryear{\VAN{de Voort} {Van}{van}~de Voort,
  {Quataert}, {Hopkins}, {Kere{\v{s}}}  \& {Faucher-Gigu{\`e}re}}{\VAN{de
  Voort} {Van}{van}~de Voort et~al.}{2015}]{van_de_Voort+15}
\VAN{de Voort} {Van}{van}~de Voort F.,  {Quataert} E.,  {Hopkins} P.~F.,
  {Kere{\v{s}}} D.,   {Faucher-Gigu{\`e}re} C.-A.,  2015, \mn@doi [\mnras]
  {10.1093/mnras/stu2404}, \href
  {https://ui.adsabs.harvard.edu/abs/2015MNRAS.447..140V} {447, 140}

\bibitem[\protect\citeauthoryear{{Vijayan}, {Clay}, {Thomas}, {Yates},
  {Wilkins}  \& {Henriques}}{{Vijayan} et~al.}{2019}]{Vijayan+19}
{Vijayan} A.~P.,  {Clay} S.~J.,  {Thomas} P.~A.,  {Yates} R.~M.,  {Wilkins}
  S.~M.,   {Henriques} B.~M.,  2019, \mn@doi [\mnras] {10.1093/mnras/stz1948},
  \href {https://ui.adsabs.harvard.edu/abs/2019MNRAS.489.4072V} {489, 4072}

\bibitem[\protect\citeauthoryear{{Vikhlinin}, {Kravtsov}, {Forman}, {Jones},
  {Markevitch}, {Murray}  \& {Van Speybroeck}}{{Vikhlinin}
  et~al.}{2006}]{Vikhlinin+06}
{Vikhlinin} A.,  {Kravtsov} A.,  {Forman} W.,  {Jones} C.,  {Markevitch} M.,
  {Murray} S.~S.,   {Van Speybroeck} L.,  2006, \mn@doi [\apj]
  {10.1086/500288}, \href
  {https://ui.adsabs.harvard.edu/abs/2006ApJ...640..691V} {640, 691}

\bibitem[\protect\citeauthoryear{{Vogelsberger}, {Genel}, {Sijacki}, {Torrey},
  {Springel}  \& {Hernquist}}{{Vogelsberger} et~al.}{2013}]{Vogelsberger+13}
{Vogelsberger} M.,  {Genel} S.,  {Sijacki} D.,  {Torrey} P.,  {Springel} V.,
  {Hernquist} L.,  2013, \mn@doi [\mnras] {10.1093/mnras/stt1789}, \href
  {https://ui.adsabs.harvard.edu/abs/2013MNRAS.436.3031V} {436, 3031}

\bibitem[\protect\citeauthoryear{Webbink}{Webbink}{1984}]{webbinkDoubleWhiteDwarfs1984}
Webbink R.~F.,  1984, \mn@doi [The Astrophysical Journal] {10.1086/161701},
  277, 355

\bibitem[\protect\citeauthoryear{{Wiseman}, {Schady}, {Bolmer}, {Kr{\"u}hler},
  {Yates}, {Greiner}  \& {Fynbo}}{{Wiseman} et~al.}{2017}]{Wiseman+17a}
{Wiseman} P.,  {Schady} P.,  {Bolmer} J.,  {Kr{\"u}hler} T.,  {Yates} R.~M.,
  {Greiner} J.,   {Fynbo} J.~P.~U.,  2017, \mn@doi [\aap]
  {10.1051/0004-6361/201629228}, \href
  {http://adsabs.harvard.edu/abs/2017A%26A...599A..24W} {599, A24}

\bibitem[\protect\citeauthoryear{{Yates}, {Henriques}, {Thomas}, {Kauffmann},
  {Johansson}  \& {White}}{{Yates} et~al.}{2013}]{Yates+13}
{Yates} R.~M.,  {Henriques} B.,  {Thomas} P.~A.,  {Kauffmann} G.,  {Johansson}
  J.,   {White} S. D.~M.,  2013, \mn@doi [\mnras] {10.1093/mnras/stt1542},
  \href {https://ui.adsabs.harvard.edu/abs/2013MNRAS.435.3500Y} {435, 3500}

\bibitem[\protect\citeauthoryear{{Yates}, {Thomas}  \& {Henriques}}{{Yates}
  et~al.}{2017}]{Yates+17}
{Yates} R.~M.,  {Thomas} P.~A.,   {Henriques} B. M.~B.,  2017, \mn@doi [\mnras]
  {10.1093/mnras/stw2361}, \href
  {https://ui.adsabs.harvard.edu/abs/2017MNRAS.464.3169Y} {464, 3169}

\bibitem[\protect\citeauthoryear{{Yates}, {Schady}, {Chen}, {Schweyer}  \&
  {Wiseman}}{{Yates} et~al.}{2020}]{Yates+20}
{Yates} R.~M.,  {Schady} P.,  {Chen} T.~W.,  {Schweyer} T.,   {Wiseman} P.,
  2020, \mn@doi [\aap] {10.1051/0004-6361/201936506}, \href
  {https://ui.adsabs.harvard.edu/abs/2020A&A...634A.107Y} {634, A107}

\bibitem[\protect\citeauthoryear{{Yates}, {Henriques}, {Fu}, {Kauffmann},
  {Thomas}, {Guo}, {White}  \& {Schady}}{{Yates} et~al.}{2021a}]{Yates+21a}
{Yates} R.~M.,  {Henriques} B. M.~B.,  {Fu} J.,  {Kauffmann} G.,  {Thomas}
  P.~A.,  {Guo} Q.,  {White} S. D.~M.,   {Schady} P.,  2021a, \mn@doi [\mnras]
  {10.1093/mnras/stab741}, \href
  {https://ui.adsabs.harvard.edu/abs/2021MNRAS.503.4474Y} {503, 4474}

\bibitem[\protect\citeauthoryear{{Yates}, {P{\'e}roux}  \& {Nelson}}{{Yates}
  et~al.}{2021b}]{Yates+21b}
{Yates} R.~M.,  {P{\'e}roux} C.,   {Nelson} D.,  2021b, \mn@doi [\mnras]
  {10.1093/mnras/stab2837}, \href
  {https://ui.adsabs.harvard.edu/abs/2021MNRAS.508.3535Y} {508, 3535}

\bibitem[\protect\citeauthoryear{{Zahid}, {Kudritzki}, {Conroy}, {Andrews}  \&
  {Ho}}{{Zahid} et~al.}{2017}]{Zahid+17}
{Zahid} H.~J.,  {Kudritzki} R.-P.,  {Conroy} C.,  {Andrews} B.,   {Ho} I.-T.,
  2017, \mn@doi [\apj] {10.3847/1538-4357/aa88ae}, \href
  {http://adsabs.harvard.edu/abs/2017ApJ...847...18Z} {847, 18}

\bibitem[\protect\citeauthoryear{{Zapartas} et~al.,}{{Zapartas}
  et~al.}{2017}]{Zapartas+17}
{Zapartas} E.,  et~al., 2017, \mn@doi [\aap] {10.1051/0004-6361/201629685},
  \href {https://ui.adsabs.harvard.edu/abs/2017A&A...601A..29Z} {601, A29}

\bibitem[\protect\citeauthoryear{{Zhukovska}, {Gail}  \&
  {Trieloff}}{{Zhukovska} et~al.}{2008}]{Zhukovska+08}
{Zhukovska} S.,  {Gail} H.~P.,   {Trieloff} M.,  2008, \mn@doi [\aap]
  {10.1051/0004-6361:20077789}, \href
  {https://ui.adsabs.harvard.edu/abs/2008A&A...479..453Z} {479, 453}

\bibitem[\protect\citeauthoryear{{Zwaan}, {Meyer}, {Staveley-Smith}  \&
  {Webster}}{{Zwaan} et~al.}{2005}]{Zwaan+05}
{Zwaan} M.~A.,  {Meyer} M.~J.,  {Staveley-Smith} L.,   {Webster} R.~L.,  2005,
  \mn@doi [\mnras] {10.1111/j.1745-3933.2005.00029.x}, \href
  {https://ui.adsabs.harvard.edu/abs/2005MNRAS.359L..30Z} {359, L30}

\makeatother
\end{thebibliography}







\bsp	
\label{lastpage}
\end{document}